\documentclass[10pt]{article}
\usepackage[utf8]{inputenc}
\usepackage{amsmath,setspace,geometry}
\usepackage{amsfonts}
\usepackage[shortlabels]{enumitem}
\usepackage{graphicx}
\usepackage{bbm}
\usepackage{xcolor}

\usepackage[colorlinks,citecolor=purple,urlcolor=blue,bookmarks=false,hypertexnames=true]{hyperref}
\usepackage[]{natbib}
\usepackage{bibunits}
\bibpunct[:]{(}{)}{,}{a}{}{,}
\makeatletter
\newcommand{\setbibunitsuffix}[1]{%
  \gdef\@extra@b@citeb{#1}%
  \gdef\@extra@binfo{#1}%
  \renewcommand\bibcite[2]{\global\@namedef{b@##1\@extra@binfo}{##2}}%
}
\makeatother
\usepackage[english]{babel}
\usepackage{float}
\usepackage{placeins}
\usepackage{subfig}
\usepackage{booktabs}
\usepackage{array}
\usepackage{pdfpages}
\usepackage{threeparttable}
\usepackage{lscape}

\title{Industry Dynamics with Cartels: The Case of the Container Shipping Industry\thanks{\href{mailto:suguru.otani@e.u-tokyo.ac.jp}{suguru.otani@e.u-tokyo.ac.jp}, Market Design Center, University of Tokyo
\\I thank my advisor Jeremy Fox for valuable advice. I also thank my committee members, Yunmi Kong, Maura Coughlin, and Arun Gopalakrishnan. I am especially grateful to Mitsuru Igami for his detailed comments and guidance. I benefited from extensive comments from Yinghua He, Koichiro Ito, Jihye Jeon, Myrto Kalouptsidi, Kei Kawai, Takuma Matsuda, Yuri Matsumura, Takeshi Murooka, Jun Nakabayashi, Masato Nishiwaki, Taisei Noda, Shosuke Noguchi, Isabelle Perrigne, Yuta Toyama, Kosuke Uetake, Naoki Wakamori, and Yasutora Watanabe, as well as from participants at the Summer Workshop on Economic Theory, the 18th Applied Econometrics Conference, the 18th Annual Economics Graduate Student Conference at Washington University in St. Louis, the 2024 Asia-Pacific Industrial Organization Conference, the Kanematsu Prize conference, and the Kanto regional meeting of the Japan Society of Logistics and Shipping Economics. This work won the 2023 Kanematsu Prize from the Research Institute for Economics and Business Administration (RIEB), Kobe University. This work was supported by JST ERATO Grant Number JPMJER2301 and JSPS KAKENHI Grant Number 24K22604. 
}
}
\author{Suguru Otani}

\date{
First version: October 27, 2023\\
Current version: \today
}

\setbibunitsuffix{@main}
\begin{document}

\maketitle

\begin{abstract}
    This paper studies how shipping conferences---explicit cartels---shaped container shipping through prices, entry, and investment from 1973--1990. I estimate a structural model to disentangle static pricing and internal allocation effects from dynamic entry and investment responses. \textcolor{black}{Estimated price wedges equal 30--70\% of mean observed freight rates, and conference rents encouraged entry and shipbuilding.} \textcolor{black}{Removing the conference regime barely changes consumer plus producer surplus but raises net social welfare across markets by reducing resource costs.} For fixed route quantity, \textcolor{black}{the capacity-proportional benchmark allocation} equalizes members' marginal costs, but alternative quota tilts produce different dynamic welfare rankings across markets.
\end{abstract}

\noindent\textbf{JEL Classification:} L41, L91, C73, D25.\\
\noindent\textbf{Keywords:} cartels; industry dynamics; entry and exit; investment; container shipping.

\newpage
\section{Introduction}

Do cartel profits support the development of a capital-intensive industry, or do they induce socially costly investment?
This question cannot be answered from prices alone.
Classic theory shows that market structure shapes incentives for cost-reducing investment \citep{spence1984cost}, while fixed setup costs and business stealing can make the private return to entry exceed its social value \citep{mankiw1986free}.
Higher markups transfer surplus to producers and contract output, but the resulting rents can also induce entry and investment in long-lived capacity.
That capacity may expand future output and lower marginal cost, while entry, continued operation, exit, and investment consume real resources.
Cartels can therefore change both the current allocation and the path of industry formation.
This paper quantifies these forces during the formation of global container shipping.

Container shipping provides a transparent setting because its institutions make both the cartel and its investment channel observable.
Under the long-standing principle of freedom of shipping, governments generally did not reserve a country's foreign-trade cargo for its own flag vessels or assign shippers to particular carriers.
For more than a century, however, many jurisdictions also allowed liner carriers serving the same routes to form explicit cartels.
These route-level cartels, known as shipping conferences, set common freight rates and allocated cargo among members.\footnote{Conference membership was route-specific: a carrier could belong to the conference on one route while operating outside the conference on another. Unlike conferences, modern alliances are global operational agreements.}
\textcolor{black}{The capacity-proportional benchmark studied here allocates each carrier's conference cargo in proportion to its route-assigned capacity.}
\textcolor{black}{In the model, this mapping links each carrier's revenue to its installed capacity and thus makes cartel organization directly relevant for shipbuilding incentives.}

The industry also experienced a transition from conference control to competition.
During the 1960s and 1970s, containerization expanded while carriers from developing countries entered the world shipping market.
The 1974 Code broadened access to conference membership and established trade-share principles \citep{UNLinerCode1974}.
Conference control subsequently weakened in two stages.
\textcolor{black}{In early 1980, Sea-Land withdrew from a transpacific market to set its rates independently \citep{FMC1980annual}.}
The Shipping Act of 1984 then required conferences on U.S.-related routes to allow individual members to set rates independently, making common tariffs harder to sustain.
Although the Act did not govern Asia--Europe routes, their rates also remained low after 1984.
Based on these institutional changes and price patterns, I distinguish a strong conference regime in 1973--1979, a weakened conference regime in 1980--1983, and a competitive regime from 1984 onward.
Historical accounts refer to the sharp and persistent rate decline surrounding this breakdown as the ``Container Crisis'' \citep{broeze2002globalisation}.
By focusing on the conference era, this paper complements \cite{jeon2022learning}, who shows that uncertainty about the demand process and strategic incentives amplify boom--bust investment in modern container shipping.

This transition brings a long-standing policy debate into focus: did conferences stabilize a young, capital-intensive industry, or did they protect carriers at shippers' expense?
\textcolor{black}{The welfare calculation reflects a broader industrial-policy trade-off between static distortions and dynamic gains from investment and industry development \citep[see][]{juhasz2023new}.}
Higher conference prices reduce current shipping demand, whereas the resulting profits may induce entry and shipbuilding.
The additional capacity can lower marginal cost, but entry, continued fleet operation, exit, and shipbuilding absorb resources.
Moreover, capacity-based cargo allocation changes how conference rents are distributed across capacity levels and hence which firms invest.
Which force dominates is an empirical question.

Despite the long policy debate, the literature has lacked a quantitative welfare analysis of competition and collusion in this industry.
Distinguishing these forces requires jointly recovering demand, marginal cost, the cartel price wedge, and the costs governing firms' dynamic choices.
I assemble the fragmented historical records into a unified dataset and estimate a structural model that uses route-year variation in prices, quantities, and regimes together with firm-market-year variation in entry, exit, and shipbuilding investment.

The data combine route-year and firm-market-year information.
The route-year data contain conference freight rates and shipping quantities for six directional routes: eastbound and westbound services in the transpacific, transatlantic, and Asia--Europe markets.
The firm-market-year data use ship-level records from the \textit{Containerisation International Yearbook} to measure each carrier's capacity, entry, exit, and shipbuilding investment in the three markets.\footnote{See \cite{matsuda2022unified,otani2025unified} for more information on data construction.}
The descriptive data cover 1966--1990.
\textcolor{black}{I define the conference segment to include conference members while collective rate setting was effective and the corresponding former members after it became ineffective.}
I treat non-conference quantities as exogenous because their price data are not available.

I first document the price decline.
\textcolor{black}{Descriptive regressions associate the common empirical boundaries in 1980 and 1984 with a 49\% fall in conference freight rates.}
This fact shows that the regime change is economically large, but price regressions alone cannot separate cartel price effects from changes in demand, marginal cost, entry, and shipbuilding investment.
\textcolor{black}{A structural model is therefore needed to disentangle these forces and quantify their welfare effects.}

The structural model has a static part and a dynamic part.
The static demand and supply system follows the empirical cartel literature \citep{porter1983study,igami2015market} in estimating route-year demand, marginal cost, and a regime-specific cartel price wedge.
The dynamic game builds on empirical models of industry evolution \citep{igami2017estimating} and uses the estimated static profits as payoffs.
Each year, firms choose whether to enter, exit, stay, or build larger ships, and the state is the number of firms by capacity class in each market.
Conference rules affect profits through two channels: the estimated cartel price wedge and \textcolor{black}{the capacity-proportional benchmark allocation}.
I estimate entry, exit, operating, and investment costs by matching observed firm decisions.
Firms are assumed to anticipate the weakening of the conference regime in 1980 but not its breakdown in 1984; therefore, the dynamic model uses decisions during 1973--1983 and treats 1984 as the terminal period.

The estimates imply large cartel price effects.
\textcolor{black}{The demand elasticity is about $-1.1$, compared with $-3.89$ for modern container shipping in \cite{jeon2022learning}; the lower responsiveness plausibly reflects shippers' limited access to close transport substitutes during the early container era.}
Conferences raised prices above marginal cost by \$1,807 per \textcolor{black}{twenty-foot equivalent unit (TEU)} during 1973--1979 and by \$998 per TEU during 1980--1983.
Across routes, these effects correspond to roughly 30--70\% of mean observed freight rates in the corresponding regimes.
\textcolor{black}{The dynamic estimates imply entry costs of about \$3--4 billion and large shipbuilding sunk costs.}
These costs matter because cartel rents encouraged entry and investment in large container ships.

The counterfactuals vary the strength of cartel control and allow entry and shipbuilding to respond.
\textcolor{black}{In the modeled conference segment, removing the conference regime over 1973--1983 raises consumer surplus by about 2--4\% and lowers producer surplus by about 17--27\%, while changing their sum by less than 1\%.}
\textcolor{black}{After operating, exit, entry, and shipbuilding costs are included, removal raises net social welfare by 2.7\% in the transpacific market, 4.4\% in Asia--Europe, and 64.3\% in the transatlantic market; the large transatlantic percentage reflects its low benchmark net social welfare after operating costs.}
\textcolor{black}{At the opposite extreme, hypothetical full collusion lowers net social welfare by 8.4\% in the transpacific market, 10.0\% in Asia--Europe, and 35.2\% in the transatlantic market, while leaving it positive in all three markets.}
\textcolor{black}{Evaluated at observed market states, the estimated conference price effect is 15.8\% of the full-collusion effect in the transpacific market, 14.7\% in Asia--Europe, and 31.4\% in the transatlantic market, so full collusion remains an upper bound rather than a close approximation to the estimated conference regime.}
These calculations end in 1983 and therefore do not credit possible subsequent returns to the accumulated capacity.

I also study the conferences' internal quota rule.
For a fixed route quantity, \textcolor{black}{the capacity-proportional benchmark allocation} equalizes marginal costs across conference members.
I compare this rule with alternatives that shift conference cargo toward smaller or larger carriers.
\textcolor{black}{In the model, these alternatives create a participation--upgrading trade-off: favoring small carriers encourages entry but weakens incentives to upgrade into large-capacity states, whereas favoring large carriers weakens entry but strengthens upgrading incentives.}
Favoring large carriers induces a costly transition to the largest ships in the transatlantic market but reduces smaller-carrier capacity and resource costs in the transpacific and Asia--Europe markets.
Thus, a quota rule that is statically efficient need not be dynamically optimal, and the preferred dynamic allocation is market-specific.

The broader policy implication is that competition authorities should evaluate coordination in capital-intensive industries dynamically. Current price effects can miss the entry and long-lived investment induced by cartel rents, while internal quota rules determine which firms receive those rents and expand. Where coordination is permitted, policy should assess both its pricing effects and its internal allocation rule: within-period shipping-cost minimization alone is not a sufficient efficiency defense, and the preferred rule may differ across markets.

\paragraph{Related literature}

This paper contributes to three strands of the literature: empirical cartel analysis, dynamic models of entry and investment, and shipping and containerization.
To the best of my knowledge, it is the first paper to combine explicit cartels, investment dynamics, and \textcolor{black}{an explicit capacity-proportional allocation benchmark} in a welfare analysis of container shipping.
The combination matters because conferences affected both current markups and the future market structure through entry and shipbuilding.

First, the paper contributes to the empirical literature on cartels.
Structural work has measured the welfare consequences of collusion in many industries and auction formats.\footnote{See the survey by \cite{asker2021}.}
Much of the empirical industrial organization literature instead studies tacit collusion or estimates conduct \citep{porter1983study,bresnahan1987competition,byrne2019learning,matsumura2023revisiting}.
Rather than estimating a conduct parameter, I study shipping conferences as explicit cartel institutions and use observed conference regimes to estimate price wedges.
Related case studies include sugar \citep{genesove1998testing,genesove2001rules}, cement \citep{roller2006workings}, bidding rings \citep{asker2010study}, gasoline \citep{clark2013collusion,clark2014effect}, coffee \citep{igami2015market}, beer \citep{miller2017understanding}, oil \citep{asker2019mis}, gynecologists \citep{ale2020trade}, vitamins \citep{igami2021measuring}, and U.S. local newspapers \citep{tiew2022flailing}.
Each of these papers studies a specific cartel environment.
My paper adds a setting in which the cartel is an industry institution, \textcolor{black}{the internal allocation benchmark is explicit within the model}, and dynamic investment responses are central to welfare.
This differs from many cartel applications in which the cartel episode is local to a subset of firms or markets and the main object is the static price effect.

The closest empirical approach is \cite{igami2015market}, who studies an international coffee agreement and compares collusion with Cournot-Nash competition in a homogeneous-good market.
\cite{miller2017understanding} use merger-induced conduct changes to distinguish Bertrand-Nash competition from collusion in the U.S. beer industry; \cite{clark2013collusion} and \cite{clark2014effect} compare collusive and competitive gasoline markets in Canada.
Like these papers, I exploit a regime change.
\textcolor{black}{Unlike a difference-in-differences setting, however, the empirical regime boundaries are imposed across all six routes, so the analysis uses no untreated-route comparison and instead relies on a structural model to separate cartel price effects from changes in demand, costs, entry, and investment.}
The model also lets me quantify counterfactual welfare, which cannot be recovered from the descriptive price decline alone.

Shipping conferences themselves have long been recognized as important cartels \citep{levenstein2006determines,asker2021}.
Historical studies examine particular UK shipping cartels in the nineteenth century \citep{morton1997entry,podolny1999social}.
\cite{wilson1991some} provide evidence on the Shipping Act of 1984 using quarterly data on freight rates and quantities for five selected commodities on the transpacific route.
Their evidence is valuable but does not use cross-sectional variation across U.S. and non-U.S. routes or firm-level investment behavior.
My contribution is to quantify markups, investment responses, and welfare for the six main container routes during the expansion of global containerization, where data limitations have previously made broad quantitative analysis difficult.

Second, the paper contributes to dynamic models of entry, exit, and investment.
In container shipping, ship size is both capacity and technology: larger ships lower marginal shipping costs but require large sunk investment.
New generations of ships, such as Panamax ships, therefore matter for both static costs and dynamic market structure.
In my model, firms decide each year whether to enter, exit, stay, or build larger ships, conditional on the distribution of firms across capacity classes.
This builds on empirical dynamic models such as \cite{igami2017estimating} and \cite{igami2020mergers}.
I complement these papers by adding explicit cartel price wedges and \textcolor{black}{an explicitly modeled internal allocation mechanism} that reallocates cartel output across firms and changes investment incentives.
The setting also tracks the early state of a new global industry, so the model can study how collusive profits shape market structure during industry formation.

Third, the paper connects to recent work on shipping, shipbuilding, and containerization.\footnote{The maritime economics literature includes work on core-theory predictions using two routes around 1983--1985 \citep{pirrong1992application}, market power and conference market share after the Act \citep{clyde1998market}, and conference theories such as monopolistic cartels, contestable markets, destructive competition, and empty cores \citep{sjostrom2013competition}. A separate international trade literature studies containerization as a technology affecting trade flows and adoption. \cite{rua2014diffusion} provides port-level evidence on diffusion, \cite{bernhofen2016estimating} study country-level containerization adoption for 1962--1990, and \cite{cocsar2018shipping} study modal choice between containerization and breakbulk shipping using Turkish export data. This paper is related to that work through the historical spread of containerization, but its object is the industrial organization of carriers rather than trade quantities, adoption, or modal choice.}
The closest paper is \cite{jeon2022learning}, who shows that uncertainty about the demand process and strategic incentives amplify investment cycles in container shipping.
She combines route-level price and quantity data from 1997--2014 with firm-level capacity and investment data from 2006--2014.
Both papers estimate constant-elasticity demand on six major directional routes and map capacity-dependent static profits into shipbuilding decisions.
The questions and periods are complementary: she studies demand learning, investment, and scrapping among major carriers in the modern industry, whereas I study explicit cartels, entry and exit, and \textcolor{black}{a capacity-proportional benchmark allocation} during the industry's formative decades.
Recent studies of modern liner alliances analyze the trade-off between economies of scale and market power using structural and Automatic Identification System (AIS)-based evidence \citep{li2024shippingalliances,petrose2025economies}.
Related work studies tanker cartel conduct \citep{asker2010leniency}, bulk shipping \citep{kalouptsidi2014aer,greenwood2015waves,brancaccio2020geography}, and tanker congestion \citep{bai2021congestion}.
Upstream shipbuilding studies analyze industrial policy in ship production \citep{kalouptsidi2017res,barwick2019china}.
These papers are closely related but generally use modern Clarksons data or post-2000 markets, whereas this paper studies the earlier container era and the transition from conferences to competition.
They also study markets with different competitive structures, so their estimates do not directly answer how liner conferences affected the container industry in the 1970s and 1980s.

\section{Data and Institutional Background}\label{sec:data_and_institiutional_background}

This section introduces the data and institutional changes and then documents the price decline that motivates the model.
A \textit{route} is directional, such as the eastbound transpacific route, whereas a \textit{market} combines both directions between the same pair of regions.
\textcolor{black}{Throughout, $i$ indexes carrier firms, $r$ directed routes, $m$ markets, and $t$ years; $\mathcal R_m$ denotes the two directed routes in market $m$.}
Because ships normally serve both directions, I define firm entry and exit at the market level.

\subsection{Data}\label{subsec:data}

I combine route-year data on prices and quantities with ship-level records aggregated to the firm-market-year level.
I treat an independent operator or corporate group as one firm.
Each route and market includes conference and non-conference carriers: conference carriers participated in the route agreement, whereas non-conference carriers operated outside it.
Reliable freight-rate series are unavailable for non-conference carriers, so the price analysis uses conference rates; capacity data cover both groups.
Appendix \ref{sec:data_details} provides additional data-construction details.

\paragraph{Route-level prices and quantities}

\begin{figure}[!htbp]
\begin{center}
  \subfloat[Price]{\includegraphics[width = 0.5\textwidth]
  {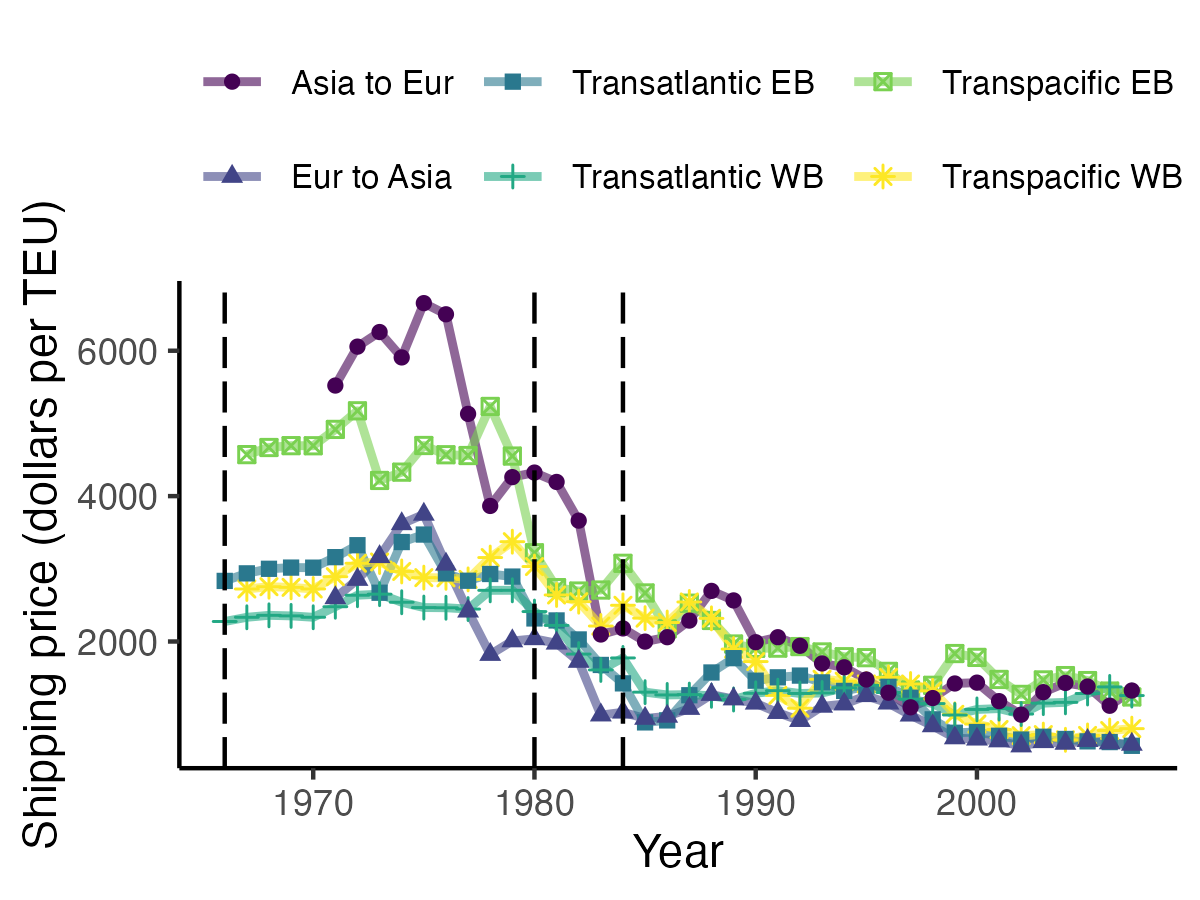}}
  \subfloat[Quantity]{\includegraphics[width = 0.5\textwidth]
  {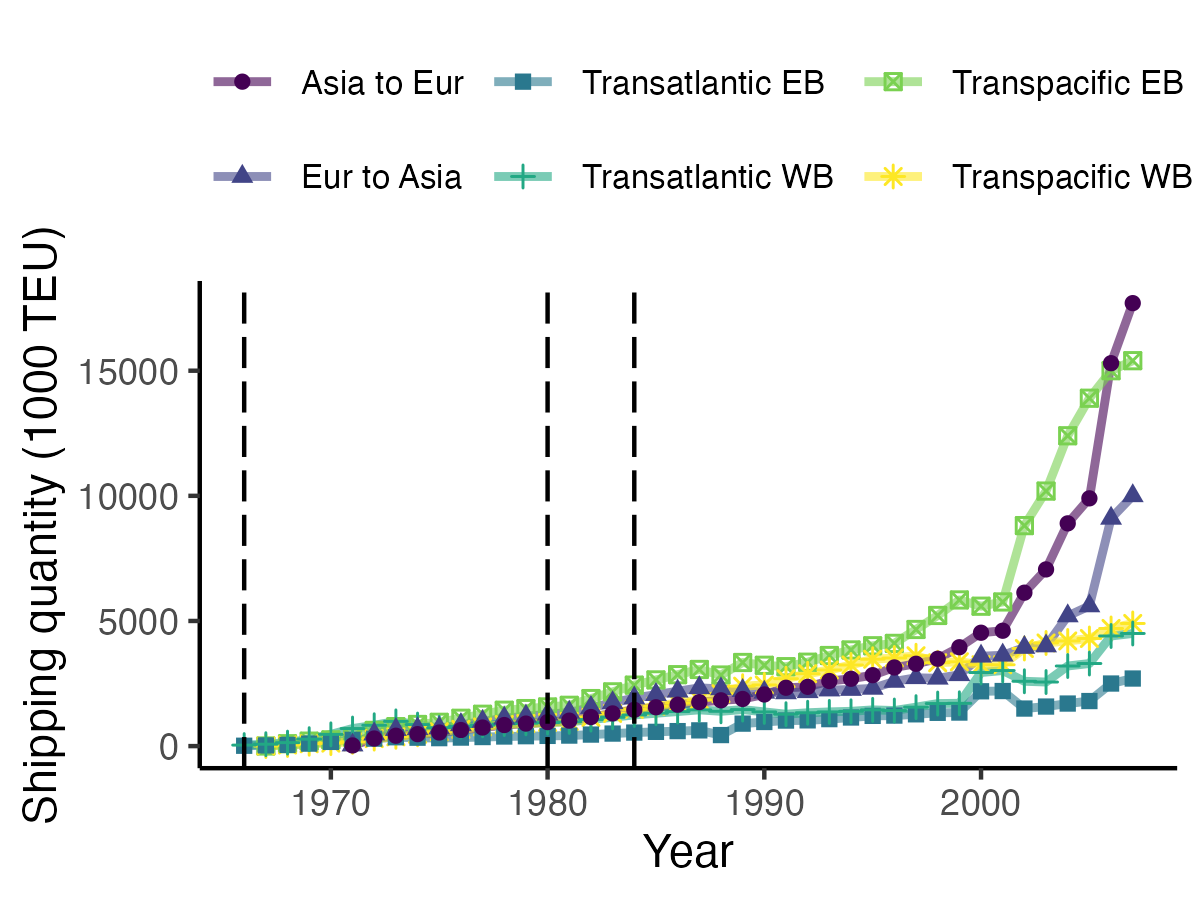}}
  \caption{Container freight rates and shipping quantities by route}
  \label{fg:container_freight_rate_and_shipping_quantity_each_route}
  \end{center}
\footnotesize
   Note: Freight rates are conference rates through 1992 and unified conference and non-conference rates from 1993 onward. Quantities are route-level container shipments. Vertical dashed lines mark the start of international container service in 1966, \textcolor{black}{the common empirical regime boundary in 1980, coinciding with Sea-Land's withdrawals}, and the U.S. Shipping Act of 1984.
\end{figure}

The route-year data contain freight rates and shipping quantities for both directions of the transpacific, transatlantic, and Asia--Europe markets.
Because no single source covers the early container era, \cite{matsuda2022unified} combine several historical sources to construct these six series for 1966--2009.
Figure \ref{fg:container_freight_rate_and_shipping_quantity_each_route} shows the full series; freight rates generally fell, particularly in the early 1980s, while shipping quantities expanded substantially.

The estimation sample covers 1973--1990.
It begins with the 1973 oil crisis, a major change in shipping costs, and ends before the Soviet collapse in 1991.
The sample also predates later regulatory changes, including the Ocean Shipping Reform Act of 1998.

\paragraph{Firm-level capacity and market composition}

The ship-level data come from successive editions of the \textit{Containerisation International Yearbook}.
For each ship, I observe its build year, operator, assigned route, and capacity in \textcolor{black}{TEU}.
I assign each ship to one of the three markets based on its reported main route and aggregate the two directions within each market.
Firms may operate in multiple markets, but ships rarely served multiple main markets during the sample period, \textcolor{black}{so each ship is assigned to a single market in the data}.
I use these records and conference status to construct firm-market-year capacity, entry, and exit and to proxy shipment shares with tonnage shares.

\begin{figure}[!t]
  \begin{center}
  \subfloat[Transpacific]{\includegraphics[width = 0.32\textwidth]
  {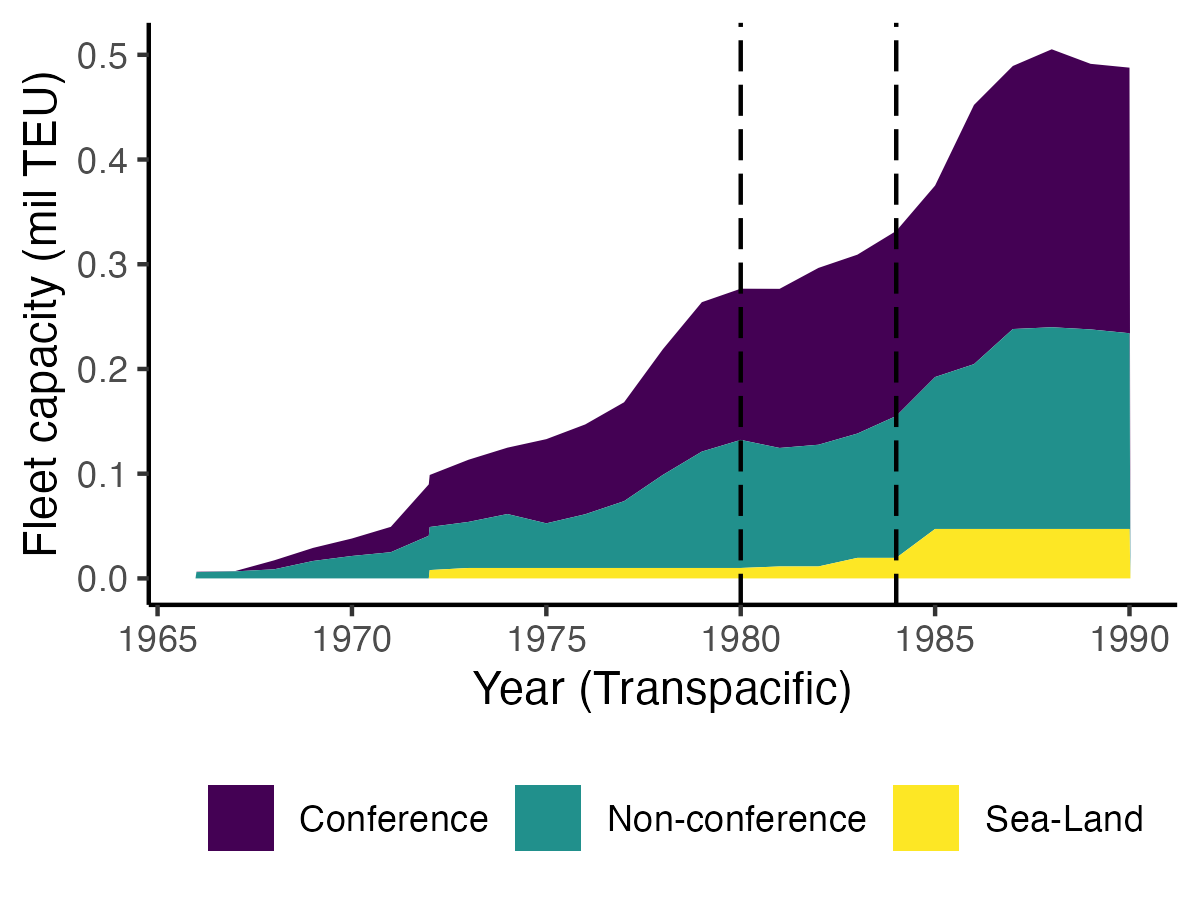}}
  \subfloat[Transatlantic]{\includegraphics[width = 0.32\textwidth]
  {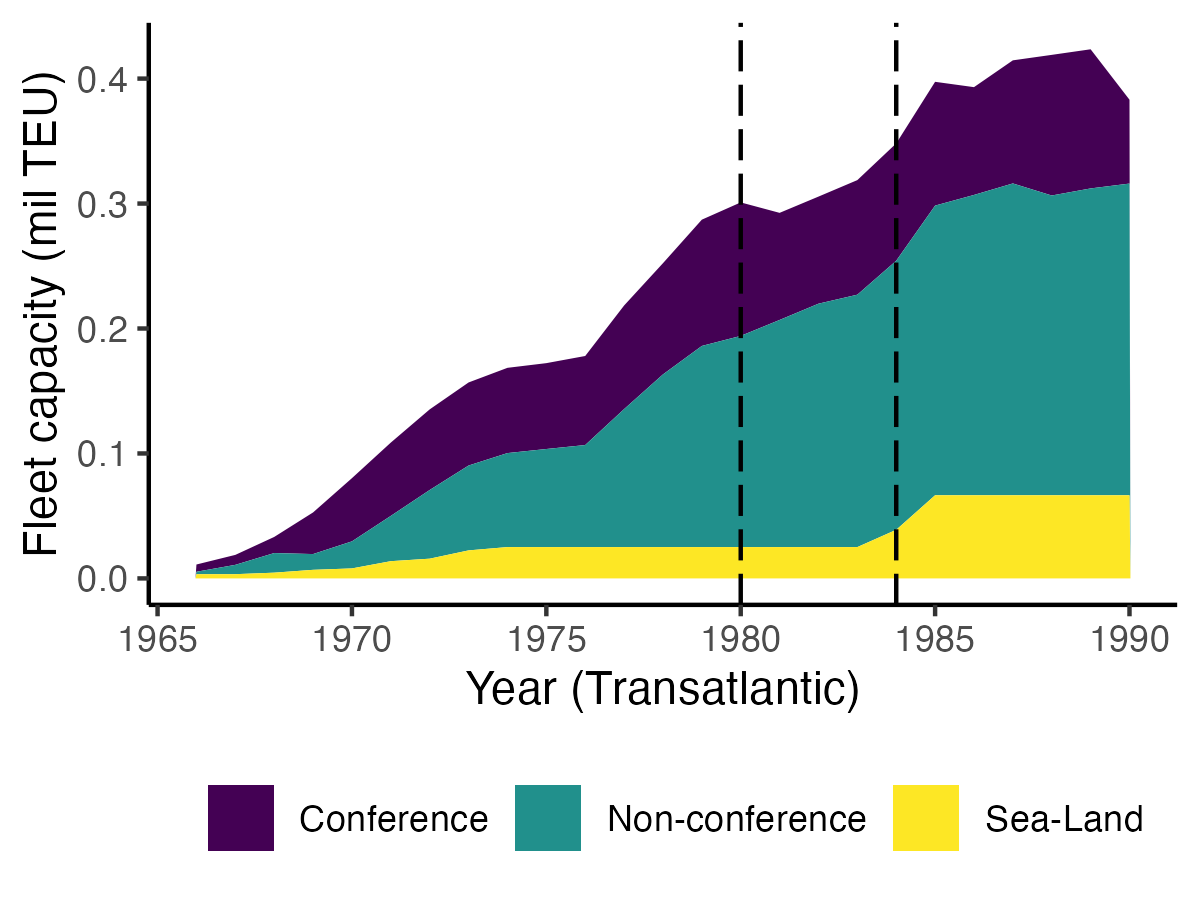}}
  \subfloat[Asia--Europe]{\includegraphics[width = 0.32\textwidth]
  {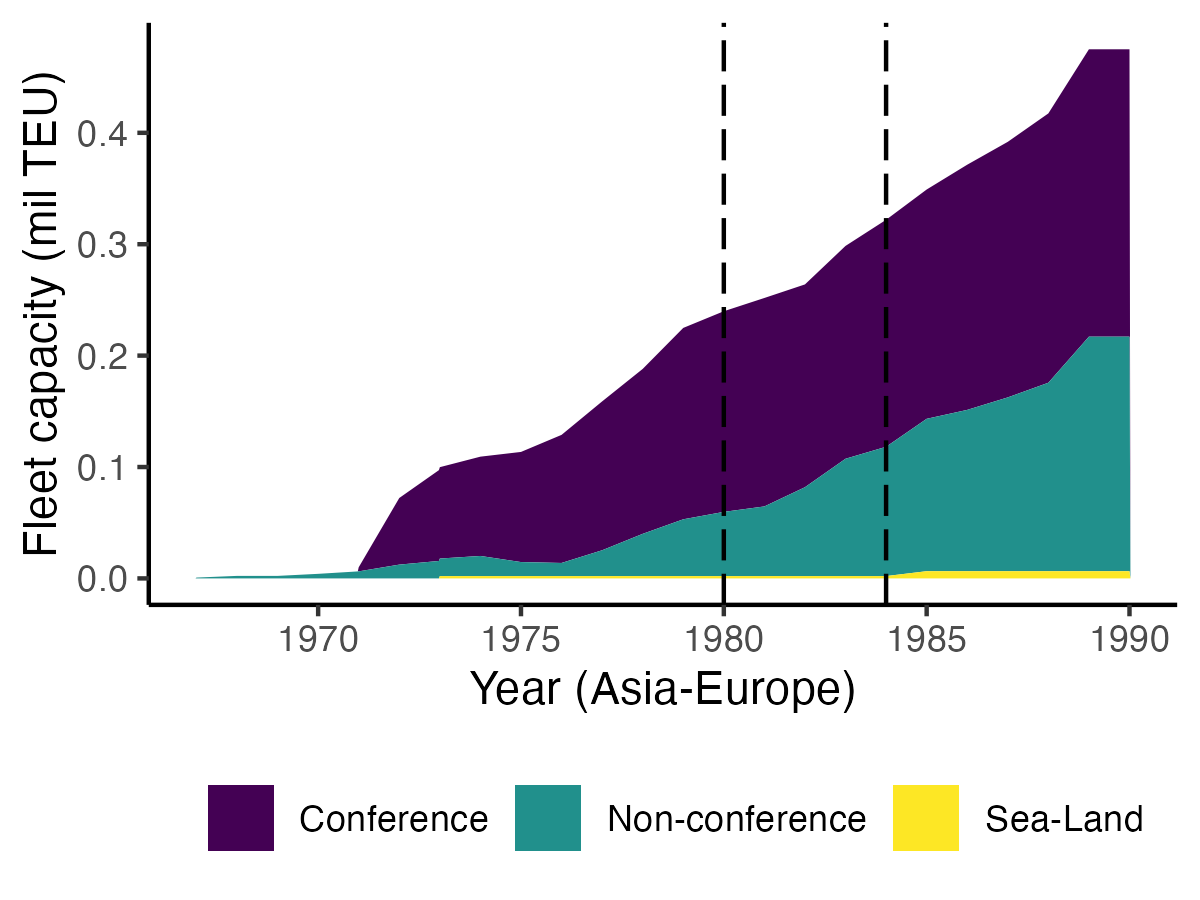}}
  \caption{Fleet capacity by conference status and market}
  \label{fg:cartel_level_share_transition_asia_and_europe}
  \end{center}
  \footnotesize
   Note: The stacked areas show the total TEU capacity of ships assigned to both directions of each market. Sea-Land is shown separately from the other conference and non-conference carriers.
\end{figure}

Figure \ref{fg:cartel_level_share_transition_asia_and_europe} shows market capacity by conference status, with Sea-Land reported separately.
Non-conference capacity was largest relative to conference capacity in the transatlantic market, conference capacity dominated Asia--Europe before 1980, and the transpacific market lay between these cases.
\textcolor{black}{Sea-Land supplied substantial capacity in the transpacific and transatlantic markets; its documented 1980 conference withdrawals occurred in the U.S.--Far East trade.}

\paragraph{Investment within conference markets}

\begin{figure}[!t]
  \begin{center}
  \subfloat[Transpacific]{\includegraphics[width = 0.32\textwidth]
  {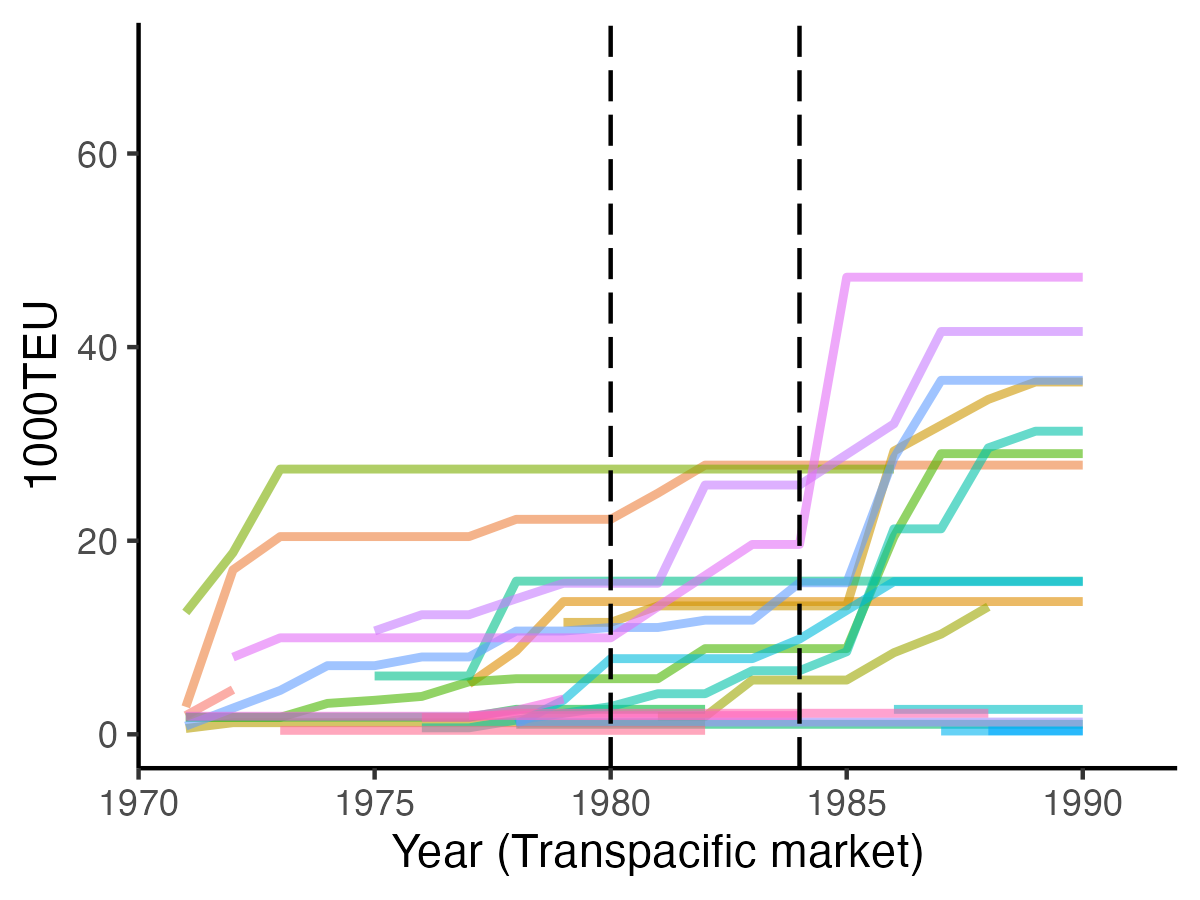}}
  \subfloat[Transatlantic]{\includegraphics[width = 0.32\textwidth]
  {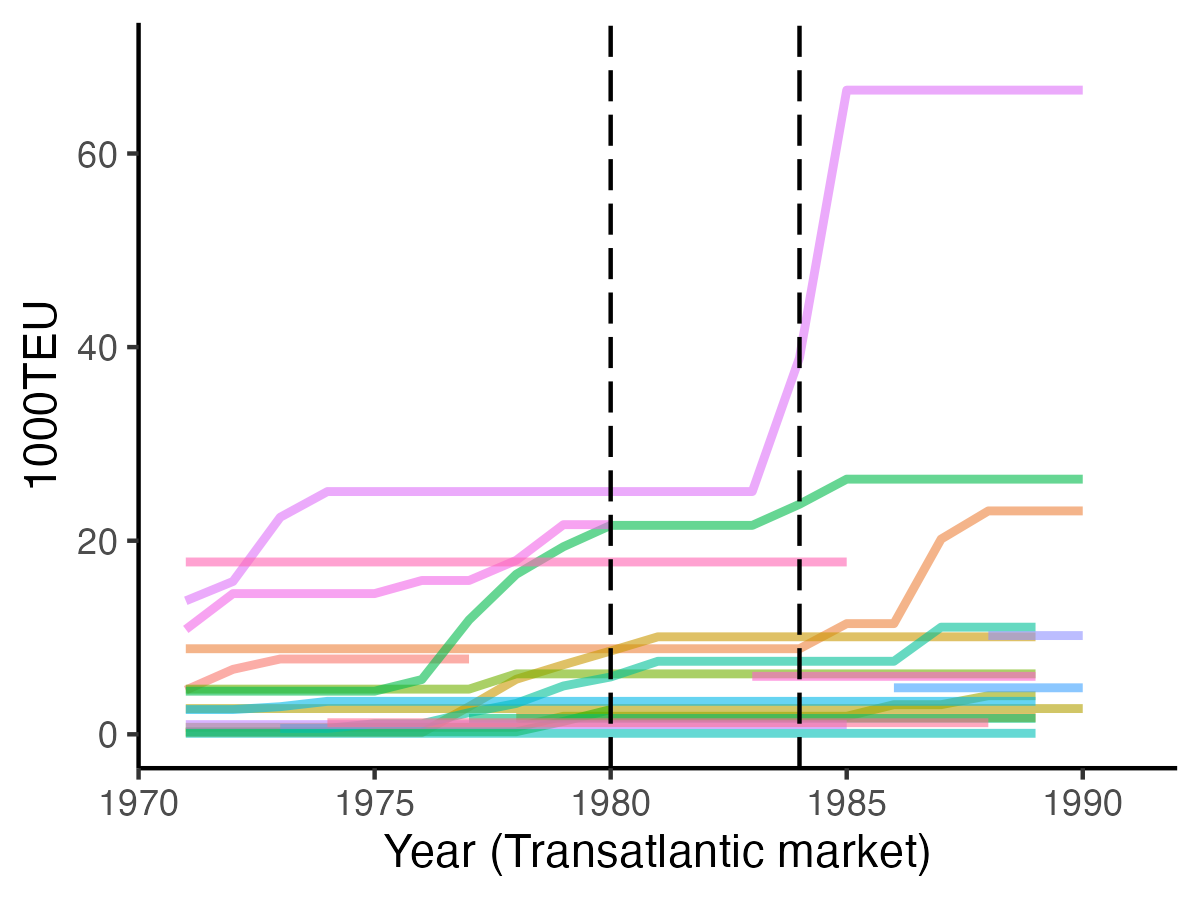}}
  \subfloat[Asia--Europe]{\includegraphics[width = 0.32\textwidth]
  {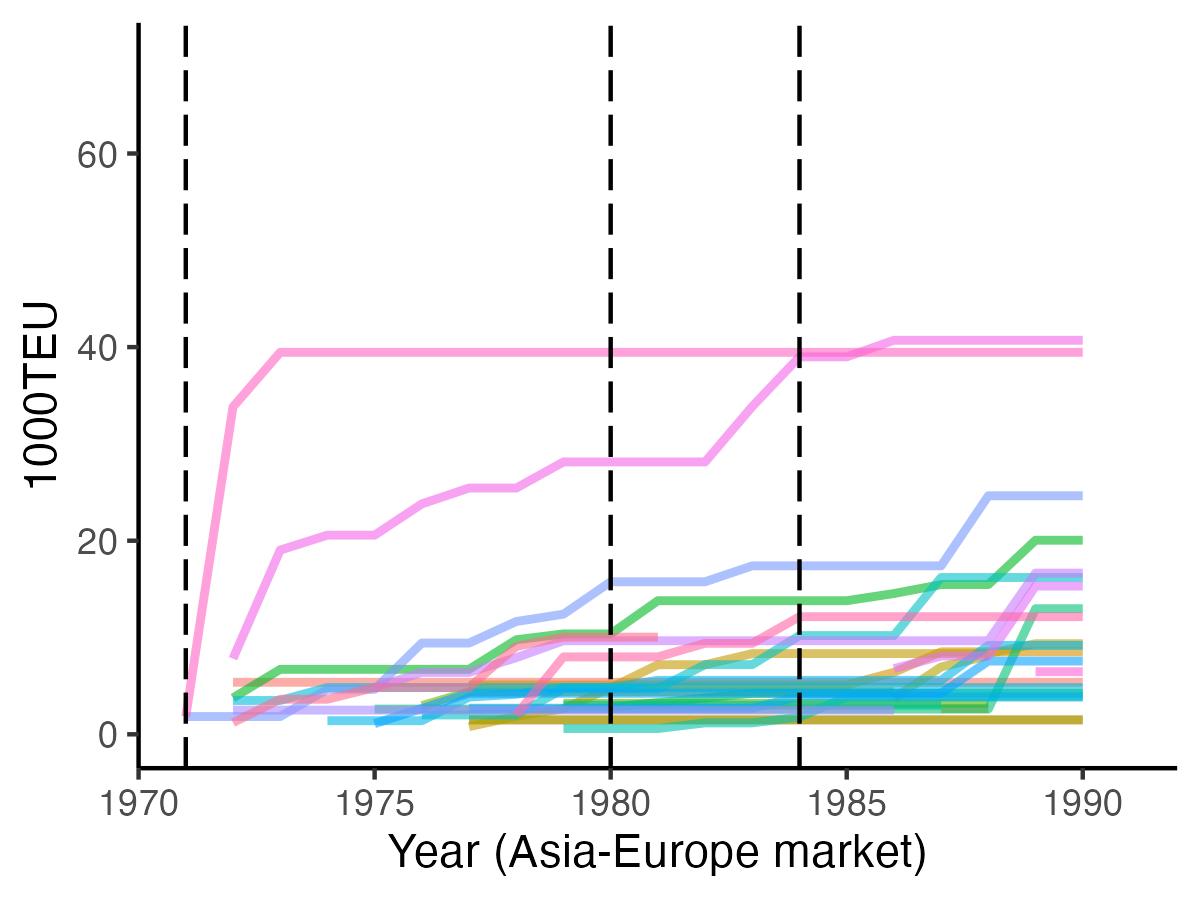}}
  \caption{Firm-market-year capacity paths}
  \label{fg:firm_level_tonnage_transition_asia_and_europe}
  \end{center}
  \footnotesize
   Note: Each line represents a firm's total TEU capacity in the corresponding conference market. Sea-Land is excluded.
\end{figure}

Figure \ref{fg:firm_level_tonnage_transition_asia_and_europe} plots each conference carrier's capacity path, excluding Sea-Land.
Capacity is persistent, but many expansions occur in discrete steps.
These patterns motivate the discrete capacity state in the dynamic model.

Tables \ref{tb:summary_statistics_of_route_year_level} and \ref{tb:summary_statistics_of_firm_market_year_level} summarize the route-, market-, and firm-level data.
There are more non-conference than conference firms in all three markets, especially in the transatlantic market, but individual non-conference firms are smaller on average.
Conference firms operated 5.57--\textcolor{black}{9.63} ships on average\textcolor{black}{, and across firm-market-year observations their mean share of whole-market capacity was 4\%, compared with 1--2\% for non-conference firms (Table \ref{tb:summary_statistics_of_firm_market_year_level})}.

\begin{table}[!t]
  \begin{center}
      \caption{Summary statistics: route- and market-year data}
      \label{tb:summary_statistics_of_route_year_level}
      \subfloat[Route-year variables]{\begin{tabular}{llllll}
\hline
& N & mean & sd & min & max \\ \hline
Conference quantity (million TEU): $Q_{rt}$ & 138 & 0.61 & 0.47 & 0.00 & 2.05 \\
Non-conference quantity (million TEU): $Q_{rt}^{non}$ & 138 & 0.42 & 0.33 & 0.00 & 1.30 \\
Conference freight rate (1,000 USD per TEU): $P_{rt}$ & 138 & 2.79 & 1.23 & 0.89 & 6.65 \\
Log destination GDP: $\log(X_{rt})$ & 138 & 28.26 & 0.73 & 26.33 & 29.42 \\
\hline
\end{tabular}
}\\
      \subfloat[Market-year variables]{\begin{tabular}{lllllll}
\hline
Market & Type &   & mean & sd & min & max \\ \hline
Asia-Europe & Conference & Number of firms & 18.45 & 6.94 & 2.00 & 25.00 \\
&  & Total capacity (million TEU) & 0.16 & 0.07 & 0.00 & 0.26 \\
& Non-conference & Number of firms & 20.67 & 12.68 & 2.00 & 37.00 \\
&  & Total capacity (million TEU) & 0.07 & 0.07 & 0.00 & 0.21 \\
Transatlantic & Conference & Number of firms & 14.08 & 3.46 & 6.00 & 17.00 \\
&  & Total capacity (million TEU) & 0.10 & 0.05 & 0.01 & 0.18 \\
& Non-conference & Number of firms & 61.04 & 36.56 & 3.00 & 102.00 \\
&  & Total capacity (million TEU) & 0.13 & 0.09 & 0.00 & 0.25 \\
Transpacific & Conference & Number of firms & 12.36 & 5.54 & 1.00 & 19.00 \\
&  & Total capacity (million TEU) & 0.14 & 0.11 & 0.00 & 0.31 \\
& Non-conference & Number of firms & 25.52 & 11.53 & 4.00 & 38.00 \\
&  & Total capacity (million TEU) & 0.09 & 0.06 & 0.01 & 0.19 \\
\hline
\end{tabular}
}
  \end{center}
  {\footnotesize
  Note: Panel (a) covers six directional routes through 1990: transatlantic from 1966, transpacific from 1967, and Asia--Europe from 1971. Panel (b) aggregates the two directions into three markets.
  \textcolor{black}{In panel (a), $P_{rt}$, $Q_{rt}$, $Q_{rt}^{non}$, and $X_{rt}$ denote the conference freight rate, conference quantity, non-conference quantity, and destination-region GDP, respectively.}
  }
\end{table}

\begin{table}[!t]
  \begin{center}
      \caption{Summary statistics: firm-market-year data}
      \label{tb:summary_statistics_of_firm_market_year_level}
      \subfloat[Conference firms]{\begin{tabular}{llllll}
\hline
Market &   & mean & sd & min & max \\ \hline
Asia-Europe & Number of ships & 5.57 & 4.59 & 1.00 & 20.00 \\
& Total capacity (1,000 TEU) & 8.87 & 10.04 & 0.60 & 40.70 \\
& Capacity share (conference): $\tilde{s}_{imt}/S_{mt}$ & 0.05 & 0.08 & 0.00 & 0.57 \\
& Capacity share (whole market): $\tilde{s}_{imt}/S_{mt}^{whole}$ & 0.04 & 0.06 & 0.00 & 0.47 \\
Transatlantic & Number of ships & 7.59 & 7.46 & 1.00 & 36.00 \\
& Total capacity (1,000 TEU) & 7.35 & 10.62 & 0.10 & 66.55 \\
& Capacity share (conference): $\tilde{s}_{imt}/S_{mt}$ & 0.07 & 0.09 & 0.00 & 0.50 \\
& Capacity share (whole market): $\tilde{s}_{imt}/S_{mt}^{whole}$ & 0.04 & 0.05 & 0.00 & 0.32 \\
Transpacific & Number of ships & 9.63 & 8.28 & 1.00 & 28.00 \\
& Total capacity (1,000 TEU) & 11.12 & 11.76 & 0.31 & 47.22 \\
& Capacity share (conference): $\tilde{s}_{imt}/S_{mt}$ & 0.08 & 0.11 & 0.00 & 1.00 \\
& Capacity share (whole market): $\tilde{s}_{imt}/S_{mt}^{whole}$ & 0.04 & 0.05 & 0.00 & 0.27 \\
\hline
\end{tabular}
}\\
      \subfloat[Non-conference firms]{\begin{tabular}{llllll}
\hline
Market &   & mean & sd & min & max \\ \hline
Asia-Europe & Number of ships & 3.99 & 4.92 & 1.00 & 30.00 \\
& Total capacity (1,000 TEU) & 3.42 & 5.05 & 0.11 & 28.01 \\
& Capacity share (non-conference): $\tilde{s}_{imt}^{non}/S_{mt}^{non}$ & 0.05 & 0.09 & 0.00 & 0.88 \\
& Capacity share (whole market): $\tilde{s}_{imt}^{non}/S_{mt}^{whole}$ & 0.02 & 0.07 & 0.00 & 0.88 \\
Transatlantic & Number of ships & 3.17 & 3.18 & 1.00 & 26.00 \\
& Total capacity (1,000 TEU) & 2.15 & 3.15 & 0.10 & 29.81 \\
& Capacity share (non-conference): $\tilde{s}_{imt}^{non}/S_{mt}^{non}$ & 0.02 & 0.04 & 0.00 & 0.54 \\
& Capacity share (whole market): $\tilde{s}_{imt}^{non}/S_{mt}^{whole}$ & 0.01 & 0.01 & 0.00 & 0.19 \\
Transpacific & Number of ships & 4.64 & 4.51 & 1.00 & 24.00 \\
& Total capacity (1,000 TEU) & 3.56 & 5.21 & 0.13 & 32.37 \\
& Capacity share (non-conference): $\tilde{s}_{imt}^{non}/S_{mt}^{non}$ & 0.04 & 0.06 & 0.00 & 0.53 \\
& Capacity share (whole market): $\tilde{s}_{imt}^{non}/S_{mt}^{whole}$ & 0.02 & 0.04 & 0.00 & 0.51 \\
\hline
\end{tabular}
}
  \end{center}
  {\footnotesize
  Note: The sample covers 1966--1990. Firm shipment shares are proxied by tonnage shares.
  \textcolor{black}{Here, $\tilde{s}_{imt}$ and $\tilde{s}_{imt}^{non}$ denote carrier $i$'s observed conference and non-conference market capacities, while $S_{mt}$, $S_{mt}^{non}$, and $S_{mt}^{whole}$ denote total conference, non-conference, and whole-market capacity, respectively.}
  }
\end{table}

\subsection{Institutional Background and Regime Changes}\label{subsec:industry_background}

Shipping conferences were route-level agreements among liner carriers that coordinated both prices and service.
Members jointly published tariffs and coordinated vessel deployment, schedules, and cargo allocation.
A carrier's cargo allocation was commonly linked to its tonnage, while non-conference carriers remained free to serve the same route.
Conferences supported their common tariffs with loyalty arrangements that made it costly for shippers to switch to outside carriers.
The 1974 United Nations Convention on a Code of Conduct for Liner Conferences also codified principles governing conference participation and cargo shares \citep{UNLinerCode1974}.

This system evolved alongside containerization, which reduced cargo-handling costs but raised the capital required for ships, containers, and terminals.
National and state-owned carriers from developing countries entered the world shipping market during the 1970s, often initially outside the conferences.
Entry thus expanded even as conference members continued to coordinate.

Two institutional features map directly into the model.
Collective tariff setting appears as a regime-specific wedge between conference prices and competitive route marginal costs.
\textcolor{black}{In the model, capacity-proportional cargo allocation} maps a carrier's capacity into its service quantity and profit, linking conference rules to entry, exit, and shipbuilding incentives.

Conference control weakened in two stages.
\textcolor{black}{Sea-Land, a major transpacific and transatlantic carrier, withdrew from a transpacific market in early 1980 to set rates independently \citep{FMC1980annual}.}
Conference discipline weakened, although conferences continued to publish common tariffs.
The Shipping Act of 1984 then required conferences to permit independent rate action, authorized service contracts, and removed antitrust protection from conference loyalty contracts \citep{ShippingAct1984,FMC1985annual}.
These changes made common tariffs harder to enforce on U.S.-related routes.

Historical accounts describe the sharp decline in conference freight rates during the early 1980s as the ``Container Crisis'' and associate it with the breakdown of shipping conferences \citep{broeze2002globalisation}.
The breakdown refers to the loss of effective conference control over common tariffs, not necessarily the formal dissolution of every conference.
Appendix \ref{sec:institutional_details} provides the underlying institutional details.

For the empirical analysis, I classify 1973--1979 as a strong conference regime, 1980--1983 as a weakened conference regime, and 1984 onward as a competitive regime.
\textcolor{black}{The common 1980 boundary is a maintained pooled classification for all six routes rather than an estimated causal effect of Sea-Land's withdrawal.}
The Act directly governed only U.S.-related routes.
I use the same boundary for Asia--Europe because its rates also remained low after 1984; this classification does not attribute their decline to U.S. legislation.

\subsection{Descriptive Evidence on the Container Crisis}\label{subsec:descriptive_evidence}

To document the magnitude of the two breaks, I estimate \textcolor{black}{ordinary least squares (OLS)} specifications with \textcolor{black}{the indicators $1(t\ge 1980)$ and $1(t\ge 1984)$, controlling for the post-oil-crisis indicator $1(t\ge 1974)$ used in the static model}.
I \textcolor{black}{also} control for destination \textcolor{black}{GDP} and the crude oil price:
\begin{align*}
    \log P_{rt}=&\textcolor{black}{\textcolor{black}{\delta_1} \mathbbm{1}\{t\geq 1974\}+\textcolor{black}{\delta_2} \mathbbm{1}\{t\geq 1980\}+\textcolor{black}{\delta_3} \mathbbm{1}\{t\geq 1984\}+\textcolor{black}{\delta_4}\log X_{rt}+\textcolor{black}{\delta_5}\log W_t}+\tau_{r}+\xi_{rt},
\end{align*}
where \textcolor{black}{$\delta_1,\ldots,\delta_5$ are regression coefficients}, $P_{rt}$ is the conference freight rate, $X_{rt}$ is destination GDP, $W_t$ is the crude oil price, $\tau_r$ is a route fixed effect, and $\xi_{rt}$ is an error term.
Specifications without route fixed effects omit $\tau_r$.

Table \ref{tb:ols_time_variation_of_price} reports specifications without route fixed effects in columns 1--3 and with route fixed effects in columns 4--6.\footnote{Appendix \ref{subsec:route_specific_freight_rate_trends} reports the route-specific year-coefficient paths underlying this summary.}
The signs of the two break coefficients are stable, and the route fixed effects absorb persistent differences in rate levels across routes.
In column 6, the coefficient \textcolor{black}{on $1(t\ge 1980)$ is $-0.385$, corresponding to rates 32\% lower during 1980--1983 than during 1974--1979, conditional on the controls.
The additional coefficient on $1(t\ge 1984)$ is $-0.283$, implying a further 25\% decline; the combined post-1984 difference is 49\% ($\exp(-0.385-0.283)-1=-0.49$).}

\begin{table}[!htbp]
  \begin{center}
      \caption{Changes in log conference freight rates}
      \begin{tabular}{lllllll}
\hline
& (1) & (2) & (3) & (4) & (5) & (6) \\ \hline
$1(t\ge 1974)$ & -0.439*** & 0.109 & -0.192* & -0.267** & 0.177 & -0.099 \\
& (0.151) & (0.074) & (0.103) & (0.096) & (0.105) & (0.099) \\
$1(t\ge 1980)$ & -0.605*** &  & -0.467*** & -0.469*** &  & -0.385*** \\
& (0.190) &  & (0.150) & (0.086) &  & (0.057) \\
$1(t\ge 1984)$ &  & -0.594*** & -0.338** &  & -0.422** & -0.283** \\
&  & (0.200) & (0.133) &  & (0.149) & (0.109) \\
Log destination GDP: $\log(X_{rt})$ & 0.059 & -0.081 & 0.137 & -0.132 & -0.291* & -0.024 \\
& (0.194) & (0.150) & (0.214) & (0.121) & (0.120) & (0.126) \\
Log crude oil price: $\log(W_t)$ & 0.318*** & -0.173 & 0.088 & 0.242** & -0.131 & 0.067 \\
& (0.059) & (0.110) & (0.066) & (0.063) & (0.093) & (0.081) \\
Route fixed effects &  &  &  & X & X & X \\
Observations & 108 & 108 & 108 & 108 & 108 & 108 \\
$R^2$ & 0.460 & 0.411 & 0.491 & 0.812 & 0.787 & 0.833 \\
Adjusted $R^2$ & 0.439 & 0.388 & 0.466 & 0.795 & 0.767 & 0.815 \\
RMSE & 0.33 & 0.35 & 0.32 & 0.20 & 0.21 & 0.19 \\
\hline
\end{tabular}

      \label{tb:ols_time_variation_of_price}
  \end{center}
  \footnotesize
   Note: The dependent variable is the log conference freight rate in U.S. dollars per TEU. The sample contains six routes over 1973--1990. Standard errors clustered by route are in parentheses. Significance levels are denoted by $^{*}p<0.1$, $^{**}p<0.05$, and $^{***}p<0.01$.
\end{table}

These estimates measure conditional price breaks, not the causal effects of the two events.
Changes in marginal costs, demand, capacity, or conduct could all contribute to the decline.
The structural model in the next section separates these channels and quantifies the associated changes in market structure and welfare.

\section{Model}\label{sec:model}

Cartels in capital-intensive industries affect welfare not only through prices but also by changing entry and long-lived investment: cartel rents may expand capacity by inducing additional entry and shipbuilding, but the associated resource costs reduce net social welfare.
This dynamic channel is central in liner shipping because \textcolor{black}{the model's conference price wedges and capacity-proportional allocation link} firms' profits to installed capacity.
The observed freight-rate decline therefore cannot reveal whether conferences merely redistributed surplus or also distorted entry and fleet composition, because it combines changes in conduct, demand, marginal cost, and capacity.
I use a structural model to recover the competitive cost curve and cartel price wedge, estimate firms' dynamic costs, and simulate market structure and welfare with and without the cartel effect.

The model has two parts.
The static part estimates demand and supply for conference routes.
It treats 1984 onward as the competitive regime and 1973--1983 as the conference regime, with separate cartel effects for 1973--1979 and 1980--1983.
This part maps demand and tonnage capacity into current profits.
The dynamic part uses these profits to estimate firms' entry, exit, and shipbuilding decisions and then simulates how market structure and welfare change under alternative regimes.

I focus on conference markets.
Because non-conference price data are unavailable, I do not model substitution or strategic interaction between conference and non-conference services.
Non-conference quantities are therefore treated as exogenous and excluded from the demand and supply system below.
\textcolor{black}{Accordingly, the counterfactual welfare calculations are partial-equilibrium measures for the conference segment: they hold the non-conference segment fixed and do not include substitution toward non-conference services, strategic responses by non-conference carriers, or welfare changes in the non-conference segment.}

I consider a finite-horizon, nonstationary dynamic game.
For each market-year, incumbents and potential entrants observe all firms' tonnage states and exogenous demand.
Current incumbents first earn static profits determined by these states and the cartel effect.
They then sequentially choose whether to exit, stay without investment, or invest in shipbuilding, and potential entrants choose whether to enter.
These choices update the next year's market structure.\footnote{The dynamic game follows the equilibrium concept and computational approach of \textcolor{black}{\cite{igami2017estimating,igami2018industry}}, while the static model differs by incorporating cartel effects rather than a static oligopoly game. The constant-elasticity route demand and utilization-based marginal cost parallel \cite{jeon2022learning}, whose model also includes demand learning, order books, scrapping, chartering, and capacity deployment in the modern industry. The present finite-horizon state is tailored to entry, exit, shipbuilding, and observed conference regime changes.}
The finite horizon and sequential move order permit backward induction.




\subsection{Demand and supply}
\paragraph{Demand}
Let $Q_{rt}$ denote conference shipping quantity on directed route $r$ in year $t$.\footnote{Appendix \ref{sec:data_details} describes its data construction.}
Let $P_{rt}(s_{rt},D_{rt})$ denote the conference freight rate, where \textcolor{black}{$s_{rt}$ collects firms' route-assigned TEU capacities} and \textcolor{black}{$D_{rt}$ is an exogenous demand state constructed from observed covariates and estimated coefficients}.

\textcolor{black}{Demand for conference shipping services} on route $r$ in year $t$ has constant elasticity:
\begin{align}
    Q_{rt}(s_{rt},D_{rt}) = \exp(D_{rt}) \left[P_{rt}(s_{rt},D_{rt})\right]^{\alpha_1},\label{eq:route_demand_model}
\end{align}

where $\alpha_1$ is the price elasticity of demand.

\paragraph{Supply}
Suppose $N_{rt}$ firms operate on route $r$ in year $t$\textcolor{black}{, and let $\mathcal N_{rt}$ denote their set}.
\textcolor{black}{Each firm observes $(s_{rt},D_{rt})$, where $s_{rt}=(s_{irt},s_{-irt})$ is the vector of representative TEU capacities, $s_{irt}$ is firm $i$'s own capacity, and $s_{-irt}$ collects its rivals' capacities.}
\textcolor{black}{The dynamic part and Appendix \ref{sec:appendix} describe the mapping from discrete capacity levels to these representative capacities.}

\paragraph{Individual supply equation}
Firm $i$'s marginal cost of providing service quantity $q_{irt}$ is
\begin{align}
    mc_{irt}(q_{irt},s_{irt})&= c_{rt}+\gamma_1 \frac{q_{irt}}{s_{irt}},\label{eq:individual_marginal_cost}
\end{align}
where $\gamma_1$ is the utilization-cost parameter and $c_{rt}$ is an additive route-year cost component common to firms on route $r$ in year $t$.
\textcolor{black}{Marginal cost rises with utilization, and greater capacity lowers the cost of serving a given quantity, as in the interior marginal-cost specification of \cite{jeon2022learning}.}
I assume no additional firm-level marginal-cost heterogeneity.
Tonnage $s_{irt}$ does not fully measure annual service capacity because service frequency is unobserved.

I define route marginal cost $MC_{rt}$ as the capacity-share-weighted average of the individual marginal costs in \eqref{eq:individual_marginal_cost}:
\begin{align*}
    MC_{rt}&\equiv \sum_{i=1}^{N_{rt}} \frac{s_{irt}}{S_{rt}} mc_{irt}(q_{irt},s_{irt})
    =c_{rt}+\gamma_1 \frac{Q_{rt}}{S_{rt}},
\end{align*}
where $S_{rt}=\sum_{i=1}^{N_{rt}}s_{irt}$.
In a competitive market, $P_{rt}=MC_{rt}$.
This specification is similar to \cite{porter1983study}.

To derive individual supply in the competitive regime, suppose firm $i$ chooses $q_{irt}$ to maximize variable profit
\begin{align*}
    \textcolor{black}{\pi_{irt}(q_{irt};s_{irt},s_{rt},D_{rt})}=P_{rt}(s_{rt},D_{rt})q_{irt}-\int_{0}^{q_{irt}}mc_{irt}(q,s_{irt})dq,
\end{align*}
given demand state $D_{rt}$, tonnage capacity $s_{irt}$ included in $s_{rt}$, and equilibrium price $P_{rt}(s_{rt},D_{rt})$.
The first-order condition is $P_{rt}(s_{rt},D_{rt}) = c_{rt}+\gamma_1 q_{irt}/s_{irt}$.
Solving for $q_{irt}$ gives firm $i$'s individual supply equation:
\begin{align}
    q_{irt}(P_{rt}(s_{rt},D_{rt}),s_{irt})=\frac{P_{rt}(s_{rt},D_{rt}) - c_{rt}}{\gamma_1}s_{irt}. \label{eq:individual_supply_curve}
\end{align}
Aggregating the individual supply equations gives $Q_{rt}=\sum_{i=1}^{N_{rt}}q_{irt}=\frac{P_{rt}-c_{rt}}{\gamma_1}S_{rt}$ and therefore the competitive route supply equation $P_{rt}=c_{rt}+\gamma_1 Q_{rt}/S_{rt}$.
In the competitive regime, each firm serves the quantity at which its marginal cost equals the route price.
Through 1983, \textcolor{black}{the capacity-proportional benchmark allocation} instead determines each firm's quantity; from 1984 onward, \eqref{eq:individual_supply_curve} determines it.

\paragraph{Route supply equation with and without cartel behavior}
The route price equals competitive route marginal cost plus the route cartel effect:
\begin{align}
    P_{rt}&=c_{rt}+\gamma_1 \frac{Q_{rt}}{S_{rt}}+C_t,\label{eq:supply_model}\\
    C_t&=\tilde{\gamma}_1\mathbbm{1}(t \le 1979)+\tilde{\gamma}_2\mathbbm{1}(1980 \le t \le 1983).\nonumber
\end{align}
The term $C_t$ is the route cartel effect, a regime-specific price wedge over competitive route marginal cost.
\textcolor{black}{Here, $\tilde{\gamma}_1$ and $\tilde{\gamma}_2$ are the wedges for 1973--1979 and 1980--1983, respectively.}
This direct approach is related to \cite{igami2015market}; Appendix \ref{sec:model_details} compares it with \cite{porter1983study} and discusses the modeling choice.\footnote{The alternative conduct-parameter approach requires specifying a static quantity-setting game \citep{matsumura2023revisiting,matsumura2023mpec}. The observed competitive regime identifies the cost curve without that restriction, while fixed schedules, tonnage quotas, and evolving capacity and entry make a stable conduct index a restrictive summary of conference behavior.}

\subsection{Equilibrium price, quantity, and profit in spot markets}\label{sec:equilbrium_price_and_quantity}

In the model, the industry regime changes exogenously from the collusive regime through 1983 to the competitive regime from 1984 onward.

\paragraph{Equilibrium prices and quantities}
Substituting \eqref{eq:route_demand_model} into \eqref{eq:supply_model}, I obtain the equilibrium price $P_{rt}^{*}(s_{rt},D_{rt})$ as the numerical solution to
\begin{align*}
    P_{rt}=c_{rt}+\gamma_1 \frac{\exp(D_{rt})(P_{rt})^{\alpha_1}}{S_{rt}}+C_t.
\end{align*}
The same equation covers both regimes: $C_t$ equals the relevant cartel effect through 1983 and zero from 1984 onward.
The fixed point is unique under the standard conditions $\alpha_1<0$ and $\gamma_1>0$.\footnote{Define $\Delta(P_{rt})=P_{rt}-c_{rt}-C_t-\gamma_1\exp(D_{rt})(P_{rt})^{\alpha_1}/S_{rt}$. Under $\alpha_1<0$ and $\gamma_1>0$, $\Delta'(P_{rt})>0$, $\lim_{P_{rt}\to 0}\Delta(P_{rt})=-\infty$, and $\lim_{P_{rt}\to\infty}\Delta(P_{rt})=\infty$, so the equation has a unique positive solution. The estimates in Section \ref{sec:estimation_results} satisfy these conditions.}
I obtain equilibrium quantity $Q_{rt}^{*}(s_{rt},D_{rt})$ by substituting $P_{rt}^{*}(s_{rt},D_{rt})$ into \eqref{eq:route_demand_model}.

\paragraph{Static profits in market $m$ and route $r$}

For the dynamic part, I construct market-year variables by aggregating the corresponding route-year variables. \textcolor{black}{Let $s_{imt}$ denote firm $i$'s discrete capacity level and $s_{mt}$ the vector of firm counts by capacity level, as formally defined in Section \ref{sec:dynamics}. Let $\bar{s}_{imt}$ denote the market-specific representative TEU capacity assigned to $s_{imt}$. For each $r\in\mathcal R_m$, I set $s_{irt}=\bar{s}_{imt}$ and let $s_{rt}$ collect these route-level capacities.} 
Firm $i$ in market $m$ and year $t$ has static profit
\begin{align}
        &\pi_{imt}(s_{imt}, s_{mt},D_{mt})=\sum_{r \in \textcolor{black}{\mathcal R_m}} \pi_{irt}(s_{irt},s_{rt},D_{rt}),\label{eq:market_level_profit}\\
        &\pi_{irt}(s_{irt},s_{rt},D_{rt})=P_{rt}^{*}(s_{rt},D_{rt})q_{irt}-\int_{0}^{q_{irt}}mc_{irt}(q,s_{irt})dq,\nonumber\\
        &\text{s.t.}\quad  q_{irt}=\begin{cases}
        Q_{rt}^{*}(s_{rt},D_{rt})\omega_{irt} \quad\quad  \text{ if } t \le 1983\\
        \frac{P_{rt}^{*}(s_{rt},D_{rt}) - c_{rt}}{\gamma_1}s_{irt}\quad  \text{ otherwise },
        \end{cases}\quad\omega_{irt}\in[0,1],\forall i, \sum_{i=1}^{N_{rt}} \omega_{irt}=1,\nonumber
\end{align}
where $\omega_{irt}$ is the service quota within the cartel and \textcolor{black}{$D_{mt}=(D_{rt})_{r \in \mathcal R_m}$} contains the demand states for both directions in market $m$ and year $t$.
\textcolor{black}{Because carrier-level liftings and confidential pool formulas are unavailable, I approximate each member's unobserved service share by its share of route-assigned carrying capacity, $\omega_{irt}=s_{irt}/\sum_{i=1}^{N_{rt}}s_{irt}$. This approximation is equivalent to assuming similar utilization across conference members within a route-year. Historical evidence indicates that negotiated pool shares were related to members' carrying capacity \citep[p.~30]{bennathan1989deregulation}.\footnote{\textcolor{black}{Contemporaneous sources support capacity as an observable proxy for service contribution: a Canadian government study discusses capacity, output, and minimum sailing and cargo obligations in pool bargaining \citep[pp.~119--124]{canada1978containerization}; a Federal Maritime Commission notice records joint revisions to participation percentages and minimum sailings \citep[p.~27322]{FMC1973GulfJapanPool}; and UNCTAD reports describe negotiated, often confidential shares and traditional pools based on actual liftings and freight earnings \citep[paras.~133--136]{UNCTAD1972Regulation} and \citep[paras.~160--162]{UNCTAD1986Guidelines}. They do not establish a universal tonnage-proportional rule.}} I therefore use capacity proportionality as the preferred, institutionally grounded modeling benchmark.}
I use the same firm capacities on both directional routes in a market because each ship serves both directions of a round trip. \textcolor{black}{Section \ref{sec:counterfactual} varies the allocation rule $\omega_{irt}$ for the counterfactual simulations.}

Firms do not strategically choose utilization in the spot market. Conditional on the state and quota, quantities and profits are determined without a Cournot quantity-setting game.
Appendix \ref{sec:qualitative_analysis_static_dynamic_link} analyzes how capacity choices affect static profits and reports profit estimates by capacity level.

\subsection{Dynamics}\label{sec:dynamics}

Ships are durable and costly to build, so a pricing regime can affect welfare beyond its immediate effect on freight rates by changing capacity and market structure. I capture this intertemporal margin with a finite-horizon game in annual periods. The model uses firms' decisions from 1973 through 1983 and treats 1984 as the terminal year. This cutoff focuses the estimation on the institutionally coherent pre-breakdown period and avoids imposing unverifiable assumptions on firms' beliefs during the uncertain transition after the cartel breakdown. It covers the three conference markets---transpacific, transatlantic, and Asia--Europe---and treats them as independent round-trip markets.

\paragraph{Firms, actions, and states}
The players are \textcolor{black}{strategic} incumbent carriers $i\in\mathcal{N}_{mt}$ and a fixed pool of potential entrants $i\in\mathcal{N}_{mt}^{pe}$. Potential entrants choose whether to stay out or enter, $\mathcal{A}^{pe}=\{x,e\}$; incumbents choose whether to exit, continue without investment, or build enough ships to move to a higher capacity level, $\mathcal{A}^{inc}=\{x,k,b\}$. Each incumbent's tonnage is represented by a capacity level $s_{imt}\in\{1,2,3,4\}$, ordered from smallest to largest, while potential entrants have state zero. \textcolor{black}{For $l\in\{1,2,3,4\}$, let $\mathcal N_{mt}^{l}=\{i\in\mathcal N_{mt}:s_{imt}=l\}$ denote the set of level-$l$ incumbents.} The endogenous market state is $s_{mt}=(N_{mt}^{1},N_{mt}^{2},N_{mt}^{3},N_{mt}^{4})$, where $N_{mt}^{l}$ is the number of incumbents in level $l$. \textcolor{black}{These counts exclude the fixed background carriers used in the static capacity calculations; Appendix \ref{sec:algorithm} describes that computational device.} Firms also observe the demand state $D_{mt}$ from the static model and independently distributed, action-specific type-I extreme-value payoff shocks, \textcolor{black}{whose component for action $a$ is denoted by $\varepsilon_{imt}^{a}$}.

\paragraph{Transitions and payoffs}
Exit removes an incumbent, continued operation leaves its capacity level unchanged, investment raises it by one level up to level 4, and entry places a firm at level 1 in the next year. These actions therefore determine the next market state; demand follows its estimated path, and private shocks are independent across firms and periods. For each feasible action $a$, let $\pi_{imt}^{a}(s_{imt},s_{mt},D_{mt},\varepsilon_{imt}^{a})$ denote the resulting action-specific payoff. Following the timing in \cite{igami2017estimating}, every current incumbent earns the static profit from Section \ref{sec:equilbrium_price_and_quantity} before making its dynamic choice. It then pays the exit cost $\psi$ if it exits, the operating cost $\phi$ if it remains active, and the \textcolor{black}{additional investment cost $I(s_{imt};\iota_1,\iota_2)$, defined below}, if it builds. A potential entrant pays the entry cost $\kappa^e$.

\paragraph{Timing}
Each year begins with static market outcomes for the current incumbents. Incumbents then move by capacity level from level 4 to level 1; each group observes the actions of earlier groups, draws its private payoff shocks, and moves simultaneously. Potential entrants move last. The state is then updated. \textcolor{black}{In the terminal period $T$,} the terminal ex-ante value capitalizes the period profit that firms expect under their maintained belief that the 1980--1983 regime continues, $\mathcal V_{imT}\textcolor{black}{(s_{imT},s_{mT},D_{mT})}=\pi_{imT}\textcolor{black}{(s_{imT},s_{mT},D_{mT})}/(1-\beta)$, \textcolor{black}{where $\beta$ is the annual discount factor}.
\textcolor{black}{This terminal condition is a boundary (franchise-value) approximation rather than a continuation of the full dynamic game: it holds terminal static profit fixed and does not model post-terminal operating costs or exit, entry, and build options. It is applied identically across all counterfactual scenarios.} Appendix \ref{sec:dynamic_setup_details} reports the capacity thresholds, exact transition and payoff equations, potential-entrant counts, and full within-year timing.

\subsubsection{Dynamic optimization problem}\label{sec:dynamic_optimization}

\paragraph{Beliefs}
Firms have rational expectations about rivals' strategies. To isolate the conference mechanism, I condition their dynamic decisions on the estimated paths of demand and marginal costs that determine static profits; equivalently, firms have perfect foresight over these exogenous profit components.\footnote{\cite{jeon2022learning} identifies demand learning using quarterly demand histories, direct shipbuilding and demolition prices, and firm-level investment and scrapping decisions. The annual historical data here do not contain comparable information with which to separately identify belief formation from dynamic costs.} They anticipate the weakening of the conference regime in 1980 but not its breakdown in 1984. \textcolor{black}{The common 1980 boundary is a maintained belief restriction, so the dynamic-cost estimates are conditional on this imposed profit path.} Thus, the terminal value is evaluated under the belief that the 1980--1983 regime continues. Appendix \ref{sec:pf_robustness} shows that allowing firms to anticipate the 1984 breakdown produces similar estimates.

\paragraph{Choice-specific value functions}
The available actions depend only on the firm's capacity type:
\begin{align*}
\mathcal{A}(s_{imt})=
\begin{cases}
\{x,e\} & \text{if }s_{imt}=0,\\
\{x,k,b\} & \text{if }s_{imt}\in\{1,2,3\},\\
\{x,k\} & \text{if }s_{imt}=4.
\end{cases}
\end{align*}
\textcolor{black}{Let $a_{imt}\in\mathcal A(s_{imt})$ denote firm $i$'s action and let $\varepsilon_{imt}=(\varepsilon_{imt}^{a})_{a\in\mathcal A(s_{imt})}$ denote its private payoff-shock vector.}
Let $\bar V_{imt}^{a}$ denote the shock-inclusive choice-specific value, and let $\mathcal{V}_{imt}$ denote the ex-ante value before the firm observes its private payoff-shock vector. Because static market outcomes precede the dynamic choices, $\pi_{imt}$ is common to all incumbent choice-specific values. For example, the value of continuing without investment is
\begin{align}
\bar V_{imt}^{k}(s_{imt},s_{mt},D_{mt},\varepsilon_{imt}^{k})
&=\pi_{imt}(s_{imt},s_{mt},D_{mt})-\phi+\varepsilon_{imt}^{k}\nonumber\\
&\quad+\beta E\!\left[\mathcal{V}_{imt+1}(s_{imt+1},s_{mt+1},D_{mt+1})
\mid s_{imt},s_{mt},a_{imt}=k\right]. \label{eq:representative_csvf}
\end{align}
The expectation integrates over rivals' choices and the resulting market-state transition. The exit, build, and entry values replace the current payoff and own-state transition according to Section \ref{sec:dynamics}. After observing $\varepsilon_{imt}$, the ex-post value is
\begin{align*}
V_{imt}(s_{imt},s_{mt},D_{mt},\varepsilon_{imt})
=\max_{a\in\mathcal{A}(s_{imt})}\bar V_{imt}^{a}(s_{imt},s_{mt},D_{mt},\varepsilon_{imt}^{a}).
\end{align*}

\paragraph{Ex-ante value and choice probabilities}
The ex-ante value entering continuation payoffs is
\begin{align*}
\mathcal{V}_{imt}(s_{imt},s_{mt},D_{mt})
&\equiv E_{\varepsilon_{imt}}\!\left[
V_{imt}(s_{imt},s_{mt},D_{mt},\varepsilon_{imt})
\mid s_{imt},s_{mt},D_{mt}\right].
\end{align*}
Define the deterministic component of a choice-specific value as $\widetilde V_{imt}^{a}=\bar V_{imt}^{a}-\varepsilon_{imt}^{a}$. The type-I extreme-value assumption yields
\begin{align}
\mathcal{V}_{imt}(s_{imt},s_{mt},D_{mt})
=\sigma\left[\textcolor{black}{\gamma_{\mathrm E}}+\log\sum_{a\in\mathcal{A}(s_{imt})}
\exp\left(\frac{\widetilde V_{imt}^{a}(s_{imt},s_{mt},D_{mt})}{\sigma}\right)\right], \label{eq:integrated_value_generic}
\end{align}
where \textcolor{black}{$\gamma_{\mathrm E}$ is Euler's constant} and $\sigma$ is the logit scale, normalized to one when profits are measured in billions of dollars. The corresponding conditional choice probability (CCP) is
\begin{align}
\Pr(a_{imt}=a\mid s_{imt},s_{mt},D_{mt})
=\frac{\exp\left(\widetilde V_{imt}^{a}/\sigma\right)}
{\sum_{a'\in\mathcal{A}(s_{imt})}\exp\left(\widetilde V_{imt}^{a'}/\sigma\right)}. \label{eq:ccp_generic}
\end{align}
Conditioning on the realized actions of earlier movers is suppressed. Type symmetry implies that firms at the same capacity level use the same CCP. These CCPs induce the transition distribution of the market state, and the finite-horizon problem is solved by backward induction from the terminal value. The dynamic parameter vector is $\theta_{\pi}=(\kappa^{e},\psi,\phi,\iota_1,\iota_2)$. Appendix \ref{sec:dynamic_optimization_details} gives the type-specific value functions, expands the expectation in \eqref{eq:representative_csvf}, and derives the market-state transition from the CCPs.

\subsubsection{Equilibrium concept}\label{sec:dynamic_equilibrium}

I use a finite-horizon perfect Bayesian equilibrium in type-symmetric strategies, following \textcolor{black}{\cite{igami2017estimating,igami2018industry,igami2020mergers}}. A firm's type is its capacity level. Conditional on its private payoff shocks, each firm chooses a pure action; integrating the optimal-choice indicator over the shock distribution yields the CCP in \eqref{eq:ccp_generic}. At each turn, a firm maximizes expected discounted profit given earlier movers' actions and correct beliefs about the strategies of firms that move later.

Because private shocks affect rivals only through observed choices and capacity types move sequentially, each turn reduces to the problem of a representative firm of the relevant capacity type. The finite horizon then permits backward induction, as in \cite{igami2017estimating}. Economically, the model transmits the conference regime into entry, exit, and investment through its effect on current and expected future profits. Appendix \ref{sec:dynamic_optimization_details} provides the full equilibrium construction.
\paragraph{Technical specifications}

I impose three additional specifications.
First, the cost for an incumbent at capacity level $s_{imt}$ to move up one level is
\begin{align*}
I(s_{imt};\iota_1,\iota_2)=
\begin{cases}
\iota_1 & \text{if }s_{imt}\in\{1,2\},\\
\iota_2 & \text{if }s_{imt}=3.
\end{cases}
\end{align*}
Second, I set the discount factor to the standard annual value $\beta=0.9$.
\textcolor{black}{Third, I exclude the build action for level-4 firms, consistent with the observed upper bound of the capacity state.}

\section{Identification and Estimation}

\subsection{Demand parameters}
Following \cite{kalouptsidi2014aer} and \cite{jeon2022learning}, I specify constant-elasticity route demand and use route-level fleet characteristics as price instruments.
The historical sources consistently report prices, quantities, and fleet composition at an annual rather than quarterly frequency, so the estimation uses route-year observations.
The instrument set is adapted to the young fleet: the tonnage share of ships over 20 years old provides little variation before 1986, so the preferred specification uses average ship age and \textcolor{black}{the tonnage share of ships over 15 years old}.\textcolor{black}{\footnote{Adding the log crude oil price as an additional cost shifter leaves the parameters of interest robust: the demand elasticity and the cost-side slope and cartel wedges retain their signs and significance. However, the oil-price terms are not significant at the 5\% level, and the first-stage $F$ statistics fall in both equations, so the preferred demand and cost specifications omit it. With route fixed effects, a mileage-weighted oil price is numerically equivalent to the aggregate oil price because the mileage weight is route-constant. The oil price therefore contributes no cross-route variation by construction, and its aggregate time-series variation is largely absorbed by the post-1973 and regime indicators, leaving little identifying variation in either equation.}}

Taking logs of the demand equation \eqref{eq:route_demand_model}, I estimate $\alpha=(\alpha_1,\alpha_2,\alpha_3,\alpha_4,\alpha_5,\alpha_r)$ from
\begin{align}
    \log(Q_{rt}) ={}&
    \alpha_1 \log P_{rt}
    + \alpha_2 \textcolor{black}{\log(X_{rt})}
    + \alpha_3 1(t\ge 1974) \notag\\
    &+ \alpha_4 1(t\le 1979)
    + \alpha_5 1(1980\le t\le 1983)
    + \alpha_r + \zeta_{rt}. \label{eq:log_demand_model}
\end{align}
The demand state is $D_{rt}=\alpha_2 \textcolor{black}{\log(X_{rt})} + \alpha_3 1(t\ge 1974) + \alpha_4 1(t\le 1979) + \alpha_5 1(1980\le t\le 1983) +\alpha_r$.
Here, $X_{rt}$ denotes destination-region GDP in levels and $\log(X_{rt})$ its logarithm, \textcolor{black}{$\alpha_r$ is a route fixed effect}, and the residual $\zeta_{rt}$ is an econometric demand error that is not part of $D_{rt}$ or the state used in the dynamic model.
The parameter of primary interest is the demand elasticity $\alpha_1$.
Because route-year price $P_{rt}$ is correlated with $\zeta_{rt}$, I estimate \eqref{eq:log_demand_model} using instrumental variables (IVs).
The price is instrumented with the average age of ships deployed on each route and the tonnage share of ships over 15 years old.\footnote{Appendix \ref{sec:iv_specification_search} reports estimates using lagged and alternative fleet-composition instruments.}
Both instruments are functions of the current fleet state, which is determined by past entry and shipbuilding decisions and is therefore predetermined relative to the current \textcolor{black}{independent and identically distributed (i.i.d.)} route-year demand shock.
The exclusion restriction is that, conditional on log GDP, route fixed effects, and regime dummies, the fleet-age measures affect conference quantity only through price; in particular, it rules out a direct effect through service quality.
I add route fixed effects and regime dummies to control for route- and regime-level unobserved heterogeneity.
The regime dummies capture broad shifts over time rather than demand-side effects of the cartel.
I cluster the standard errors at the route level.

The demand parameters are identified from time-series and cross-sectional variation across the six main routes under the constant-elasticity functional form.
Following \cite{jeon2022learning}, identification also uses the fact that opposite-direction routes share vessel supply while facing different demand shocks.

\subsection{Cost parameters}
Based on the supply equation \eqref{eq:supply_model}, I estimate $\gamma=(\gamma_0,\gamma_1,\gamma_2,\tilde{\gamma}_1,\tilde{\gamma}_2,\gamma_{r})$ using IVs:
\begin{align}
    P_{rt}={}&\gamma_0 1(t\ge 1974)+\gamma_1 \frac{Q_{rt}}{S_{rt}}+\gamma_2'Y_{rt}\nonumber\\
    &+\tilde{\gamma}_1 1(t\le 1979)+\tilde{\gamma}_2 1(1980\le t\le 1983)+\gamma_{r}+\eta_{rt},\label{eq:industry_cost_model}
\end{align}
where $\gamma_{0}$ is the coefficient on the post-1973 time control, $\gamma_{1}$ is the supply slope, $\gamma_{2}$ is the vector of coefficients on the cost shifters, $\tilde{\gamma}_1$ and $\tilde{\gamma}_2$ are the regime-specific price wedges, $\gamma_{r}$ is a route fixed effect, and $\eta_{rt}$ is an i.i.d. error term. Like the demand error $\zeta_{rt}$, the supply error $\eta_{rt}$ is an econometric residual and is not included in the state used in the dynamic model. Route marginal cost is
\begin{align*}
MC_{rt}=\gamma_0 1(t\ge 1974)+\gamma_1 \frac{Q_{rt}}{S_{rt}}+\gamma_2'Y_{rt}+\gamma_r.
\end{align*}
The term $\gamma_0 1(t\ge 1974)+\gamma_2'Y_{rt}+\gamma_r$ is the additive route-year cost component $c_{rt}$ in Section \ref{sec:model}.
The cost shifters $Y_{rt}$ are average ship size and the tonnage share of ships over 20 years old; both describe fleet technology in the route-year cost component.
I instrument $Q_{rt}/S_{rt}$ with log GDP for the destination region. The exclusion restriction is that, conditional on the cost shifters, the post-1973 time control, route fixed effects, and regime dummies, destination GDP affects freight rates only through shipping quantity, not through marginal cost.

The cost parameters are identified by time-series and cross-sectional variation across the six main routes, given the functional form and \textcolor{black}{the maintained common regime classification} described in Section \ref{sec:data_and_institiutional_background}. \textcolor{black}{Accordingly, the 1980--1983 wedge is a pooled regime parameter, not the causal effect of Sea-Land's withdrawal.}





\subsection{Dynamic parameters}\label{subsec:dynamic_parameter_estimation}
I estimate $\theta_{\pi}=(\kappa^e,\psi,\phi,\iota_1,\iota_2)$ separately for each market by maximum likelihood.
For each candidate parameter vector, the inner loop solves the finite-horizon game by backward induction and computes the CCPs. The outer loop maximizes the likelihood of the observed entry, exit, continuation, and shipbuilding choices in 1973--1983; the 1984 terminal year enters only through continuation values.
The parameters are identified by how these choices vary across capacity and market states, given the estimated static profits. For example, a higher investment cost lowers the build probability relative to continuation.
I implement this nested fixed-point estimator following \cite{rust1987optimal} and \cite{igami2017estimating}. Appendix \ref{sec:dynamic_estimation_details} provides the likelihood, computation, and inference details.

\section{Estimation Results}\label{sec:estimation_results}

\subsection{Demand for container shipping}

Table \ref{tb:demand_estimate_results_after_1973_before_1990} reports the demand estimates.\footnote{Appendix \ref{sec:iv_specification_search} and Table \ref{tb:demand_iv_specification_comparison} compare alternative instrument sets; Table \ref{tb:demand_estimate_results_after_1973_before_1990} reports the preferred specification.}
\textcolor{black}{The two fleet-age instruments enter the first stage with opposite signs: negative for average ship age and positive for the share of ships over 15 years old. The negative coefficient on average ship age is consistent with lower rates on routes served by older fleets, possibly compensating shippers for lower sailing speed or service frequency. The positive coefficient on the share of ships over 15 years old is consistent with higher operating costs when older vessels make up a larger share of route capacity.}
The IV estimate of the price elasticity is $-1.111$: a 1\% increase in the freight rate reduces conference quantity by about 1.11\%.\footnote{\cite{jeon2022learning} estimates an elasticity of $-3.89$ using quarterly aggregate route data for 2001:Q2--2014:Q4. The present estimate pertains to annual conference prices and quantities in 1973--1990.}

\begin{table}[!htbp]
  \begin{center}
  \caption{Demand estimates}
  \label{tb:demand_estimate_results_after_1973_before_1990}
  \begin{tabular}{lll}
\hline
& 1st
($\log(P_{rt})$) & 2nd
($\log(Q_{rt})$) \\ \hline
$\log(P_{rt})$ &  & -1.111*** \\
&  & (0.354) \\
log GDP: $\log(X_{rt})$ & 0.256 & 0.410* \\
& (0.249) & (0.217) \\
$1(t\ge 1974)$ & 0.104 & 0.041 \\
& (0.084) & (0.131) \\
$1(t\le 1979)$ & 0.419** & 0.552* \\
& (0.162) & (0.304) \\
$1(1980\le t\le 1983)$ & 0.175 & 0.168 \\
& (0.113) & (0.111) \\
Avg ship age & -0.133*** &  \\
& (0.024) &  \\
Share (15+ y.o. ships) & 1.102 &  \\
& (1.013) &  \\
Route FE & X & X \\
Num.Obs. & 108 & 108 \\
First-stage F & 19.0 &  \\
R2 Adj. & 0.841 & 0.804 \\
\hline
\end{tabular}

  \end{center}
  \footnotesize
  \parbox{0.95\textwidth}{Note: The sample covers six routes in 1973--1990. Column 1 reports the first stage for $\log(P_{rt})$, and Column 2 reports the IV demand equation. The excluded instruments are average ship age and the tonnage share of ships over 15 years old. Both columns include log destination GDP, $1(t\ge 1974)$, the two regime indicators, and route fixed effects. Standard errors in parentheses are clustered by route. The first-stage F statistic is a route-clustered joint Wald statistic. Significance levels are denoted by $^{*}p<0.1$, $^{**}p<0.05$, and $^{***}p<0.01$.}
\end{table}

\subsection{Cost parameters and cartel effects}

Table \ref{tb:cost_estimate_results_after_1973_before_1990} reports the cost estimates.\footnote{Appendix \ref{sec:iv_specification_search} and Table \ref{tb:cost_iv_specification_comparison} compare alternative cost specifications; Table \ref{tb:cost_estimate_results_after_1973_before_1990} reports the preferred specification.}
The estimated utilization-cost parameter is $\hat{\gamma}_1=596.922$, so a one-unit increase in $Q_{rt}/S_{rt}$ raises route marginal cost by about \$597 per TEU, holding the other cost controls fixed.
This positive slope is consistent with the utilization-cost pattern estimated for the modern industry by \cite{jeon2022learning}.
\textcolor{black}{The negative first-stage coefficient on log GDP indicates that, at annual frequency, destination GDP shifts utilization net of longer-run capacity adjustment; it should not be read as a pure short-run demand shift.}
The estimated cartel wedges are \$1,807 per TEU in 1973--1979 and \textcolor{black}{\$998 per TEU} in 1980--1983, relative to the competitive regime from 1984 onward.
The first wedge equals 33\% of the mean 1973--1979 freight rate on the Asia-to-Europe route and 70\% on the westbound transatlantic route.
Thus, the estimates imply a large initial cartel wedge that \textcolor{black}{was substantially smaller in 1980--1983}.

\begin{table}[!htbp]
  \begin{center}
  \caption{Cost estimates and regime-specific price wedges}
  \label{tb:cost_estimate_results_after_1973_before_1990}
  \begin{tabular}{lll}
\hline
& 1st
($\frac{Q_{rt}}{S_{rt}}$) & 2nd
($P_{rt}$) \\ \hline
$Q_{rt}/S_{rt}$ &  & 596.922* \\
&  & (327.028) \\
Avg ship size (1,000 TEU) & 0.789 & 1817.274 \\
& (1.160) & (1827.054) \\
Share (20+ y.o. ships) & 4.213 & -1491.488 \\
& (3.639) & (2927.073) \\
$1(t\ge 1974)$ & -0.047 & 113.941 \\
& (0.427) & (320.233) \\
$1(t\le 1979)$ & -0.162 & 1807.343*** \\
& (0.216) & (411.167) \\
$1(1980\le t\le 1983)$ & -0.326* & 997.543*** \\
& (0.158) & (368.733) \\
log GDP: $\log(X_{rt})$ & -0.698** &  \\
& (0.239) &  \\
Route FE & X & X \\
Num.Obs. & 108 & 108 \\
First-stage F & 8.5 &  \\
R2 Adj. & 0.913 & 0.714 \\
\hline
\end{tabular}

  \end{center}
  \footnotesize
  \parbox{0.95\textwidth}{Note: The sample covers six routes in 1973--1990. Column 1 reports the first stage for $Q_{rt}/S_{rt}$, and Column 2 reports the IV cost equation. The excluded instrument is log destination GDP. Both columns include average ship size, the tonnage share of ships over 20 years old, $1(t\ge 1974)$, the two regime indicators, and route fixed effects. Standard errors in parentheses are clustered by route. The first-stage F statistic is a route-clustered Wald statistic. Significance levels are denoted by $^{*}p<0.1$, $^{**}p<0.05$, and $^{***}p<0.01$.}
\end{table}

\begin{figure}[!htbp]
  \begin{center}
  \subfloat[Demand state $D_{rt}$]{\includegraphics[width = 0.45\textwidth]
  {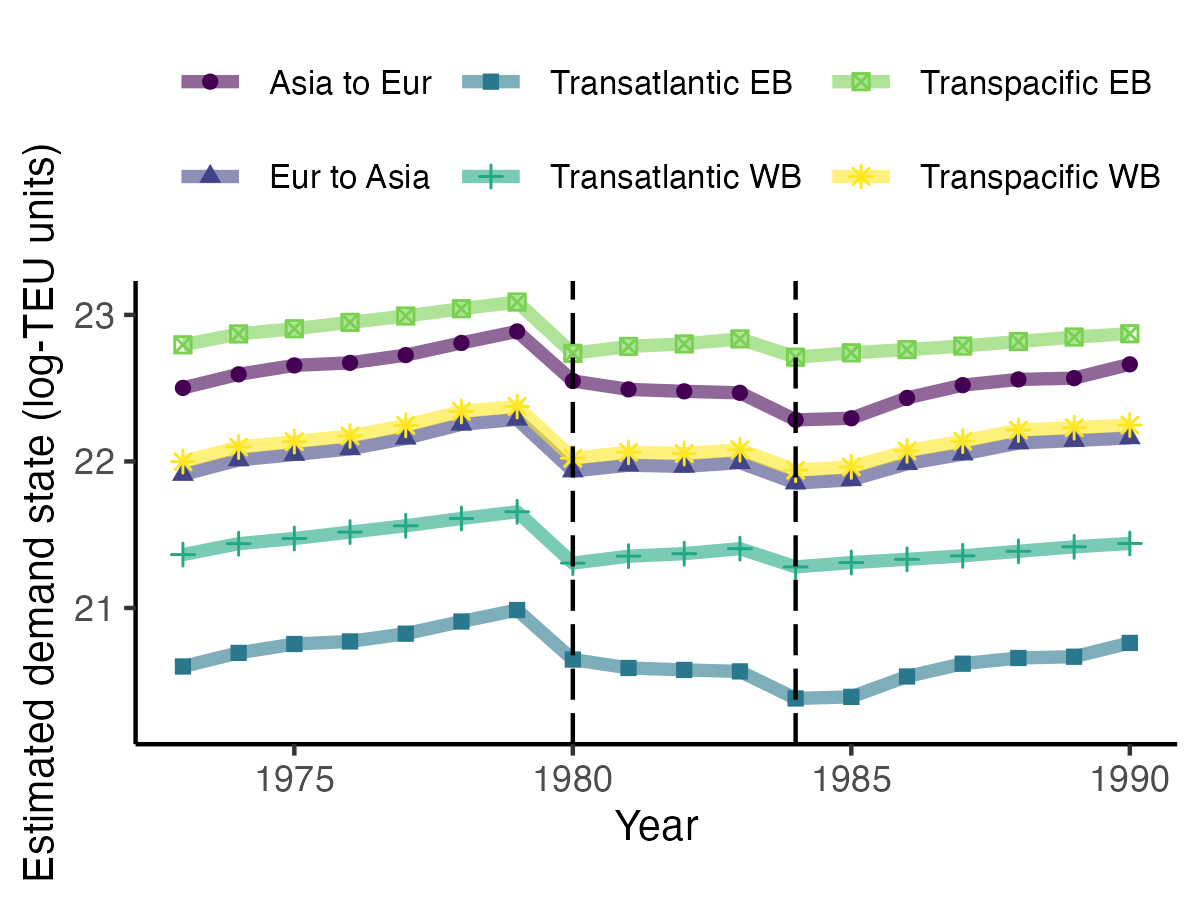}}
  \subfloat[Route marginal cost $MC_{rt}$]{\includegraphics[width = 0.45\textwidth]
  {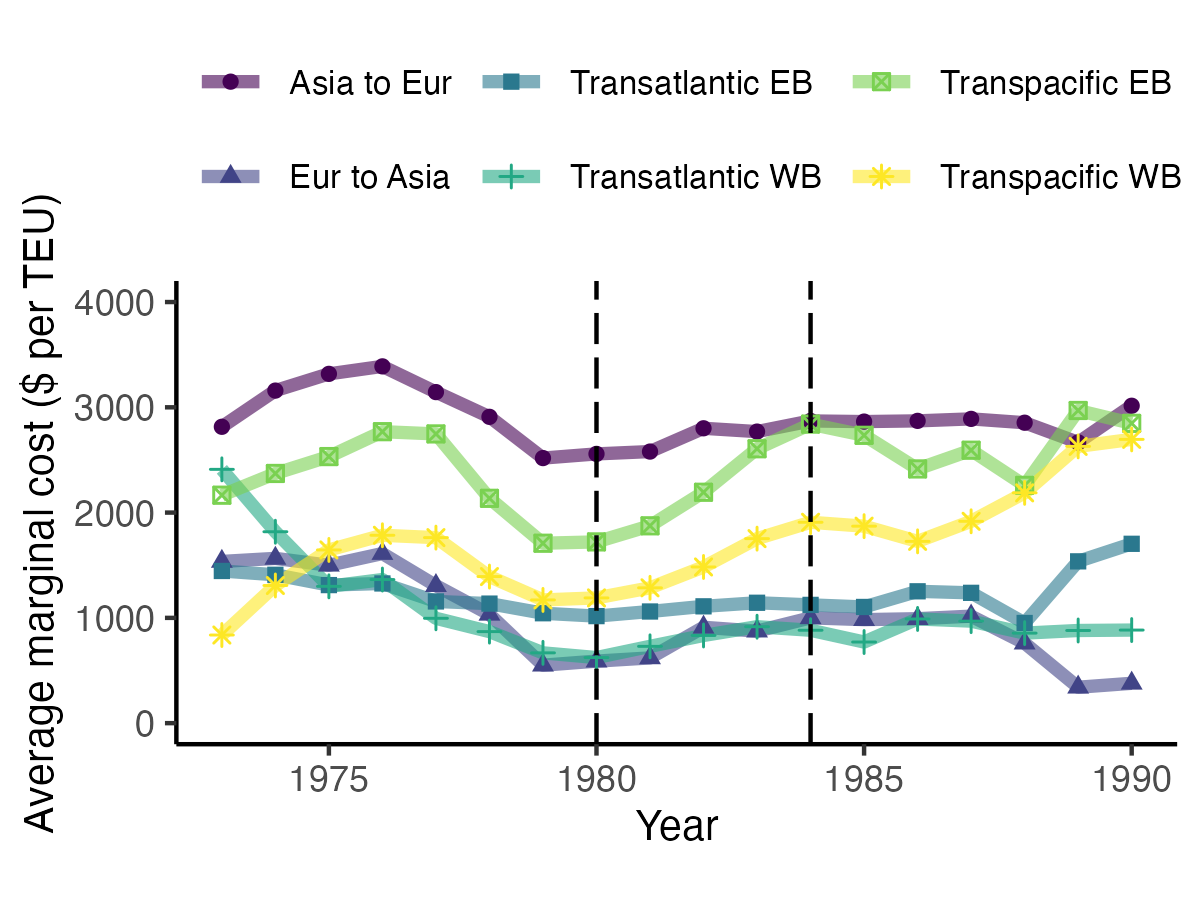}}
  \caption{Estimated demand state and route marginal cost}
  \label{fg:demand_state_results}
  \end{center}
\end{figure}

Figure \ref{fg:demand_state_results}(a) shows that the estimated demand state rises through 1979, declines at the 1980 and 1984 regime boundaries, and gradually recovers after 1984.
Figure \ref{fg:demand_state_results}(b) plots the estimated route marginal cost $c_{rt}+\hat{\gamma}_1Q_{rt}/S_{rt}$.
\textcolor{black}{Over the sample period, marginal cost rises on both transpacific routes, declines on the Europe-to-Asia and westbound transatlantic routes, is nonmonotonic on the Asia-to-Europe route, and remains broadly flat on the eastbound transatlantic route.}

\begin{figure}[!htbp]
  \begin{center}
  \subfloat[Transpacific \textcolor{black}{westbound}]{\includegraphics[width = 0.32\textwidth]
  {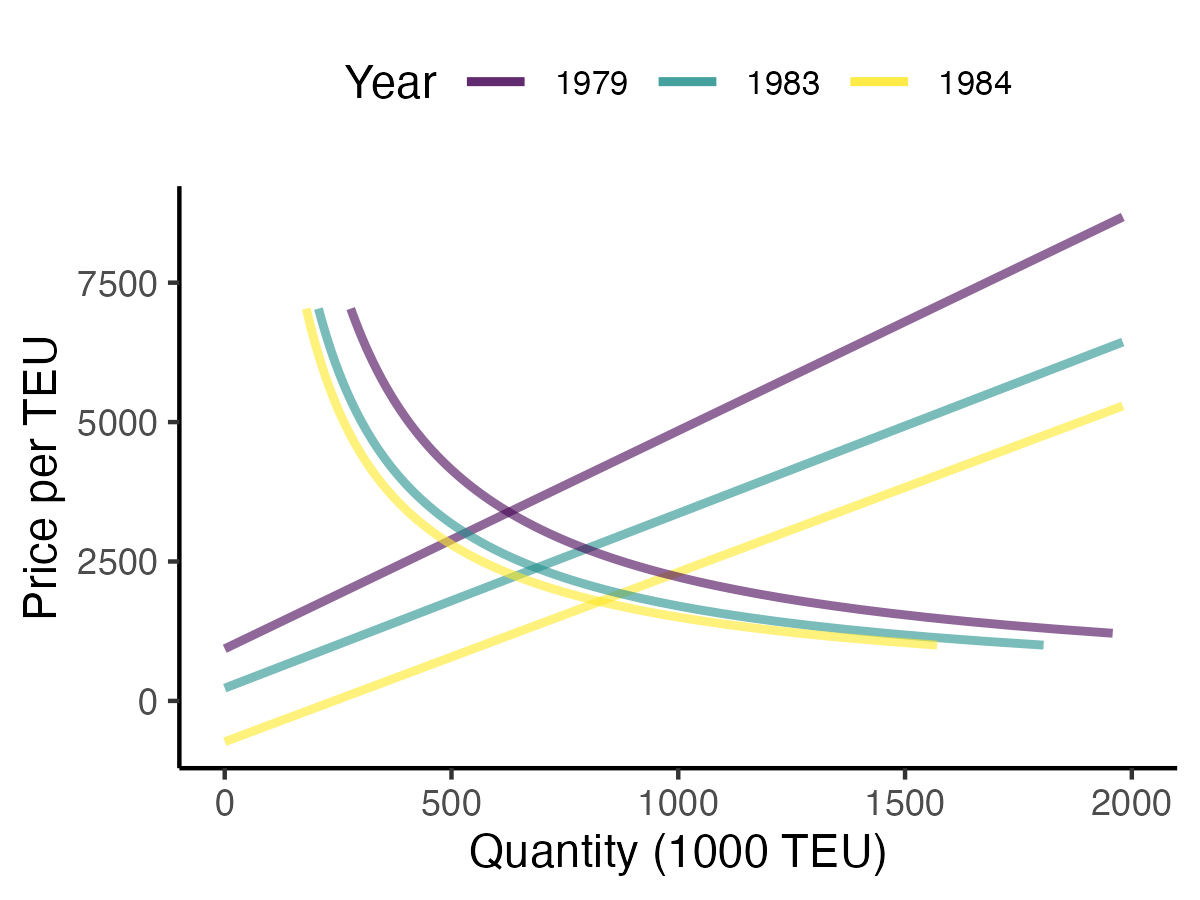}}
  \subfloat[Transatlantic \textcolor{black}{westbound}]{\includegraphics[width = 0.32\textwidth]
  {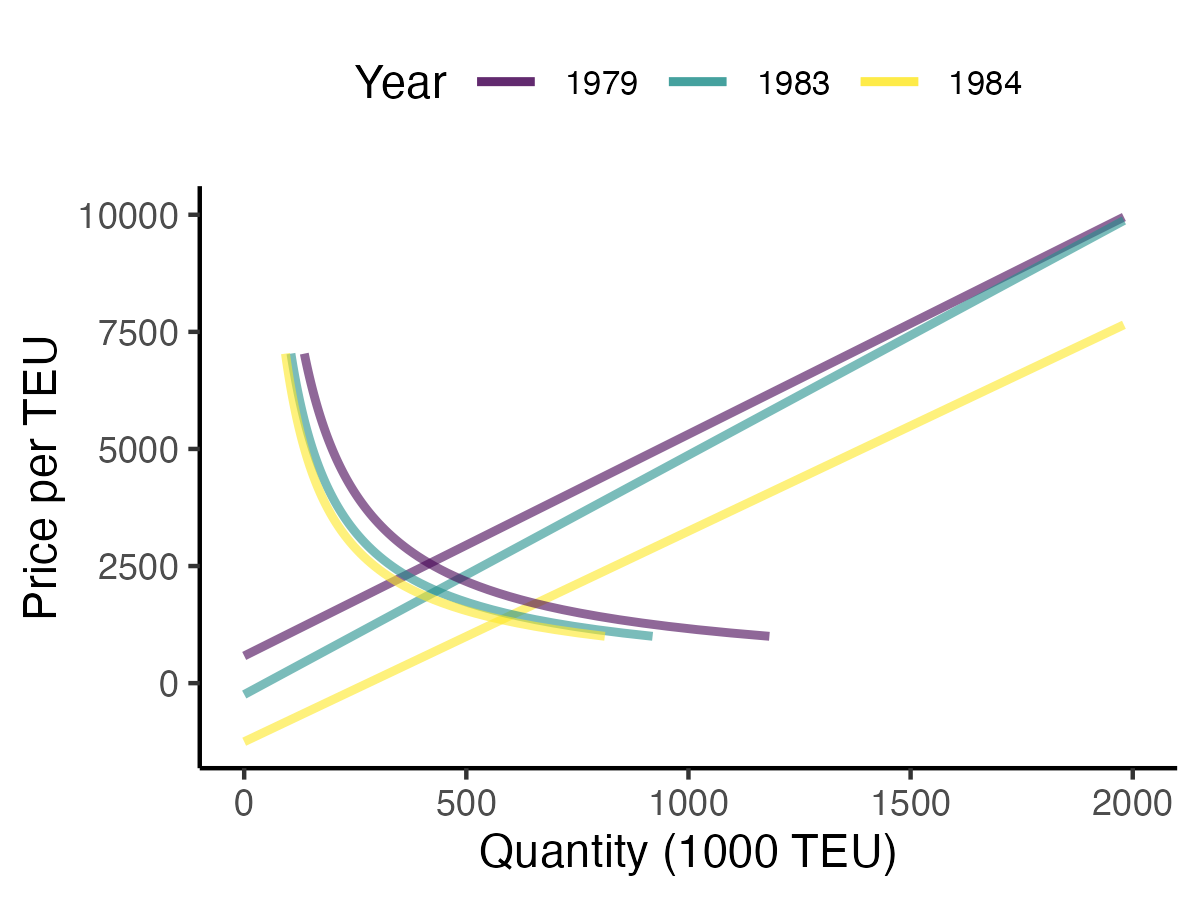}}
  \subfloat[Europe to Asia]{\includegraphics[width = 0.32\textwidth]
  {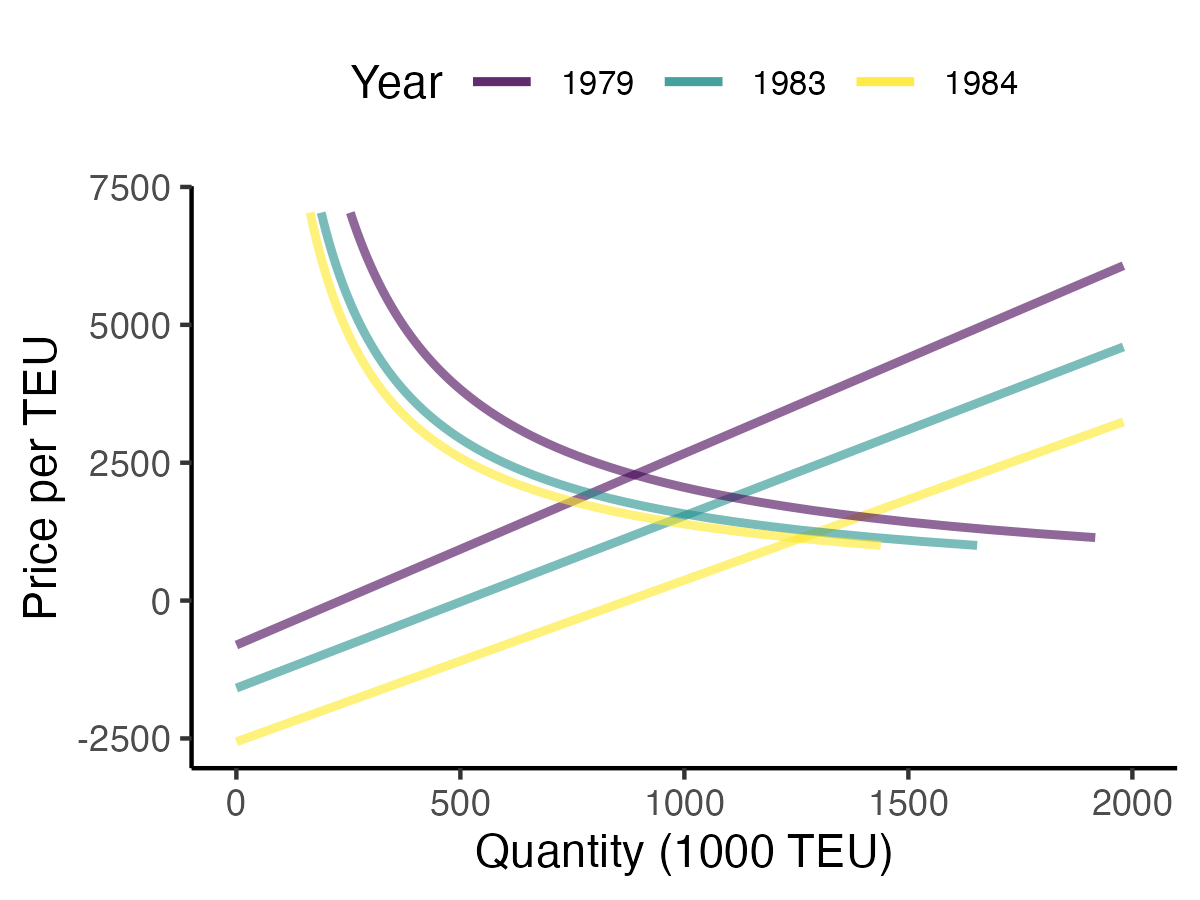}}\\
  \subfloat[Transpacific \textcolor{black}{eastbound}]{\includegraphics[width = 0.32\textwidth]
  {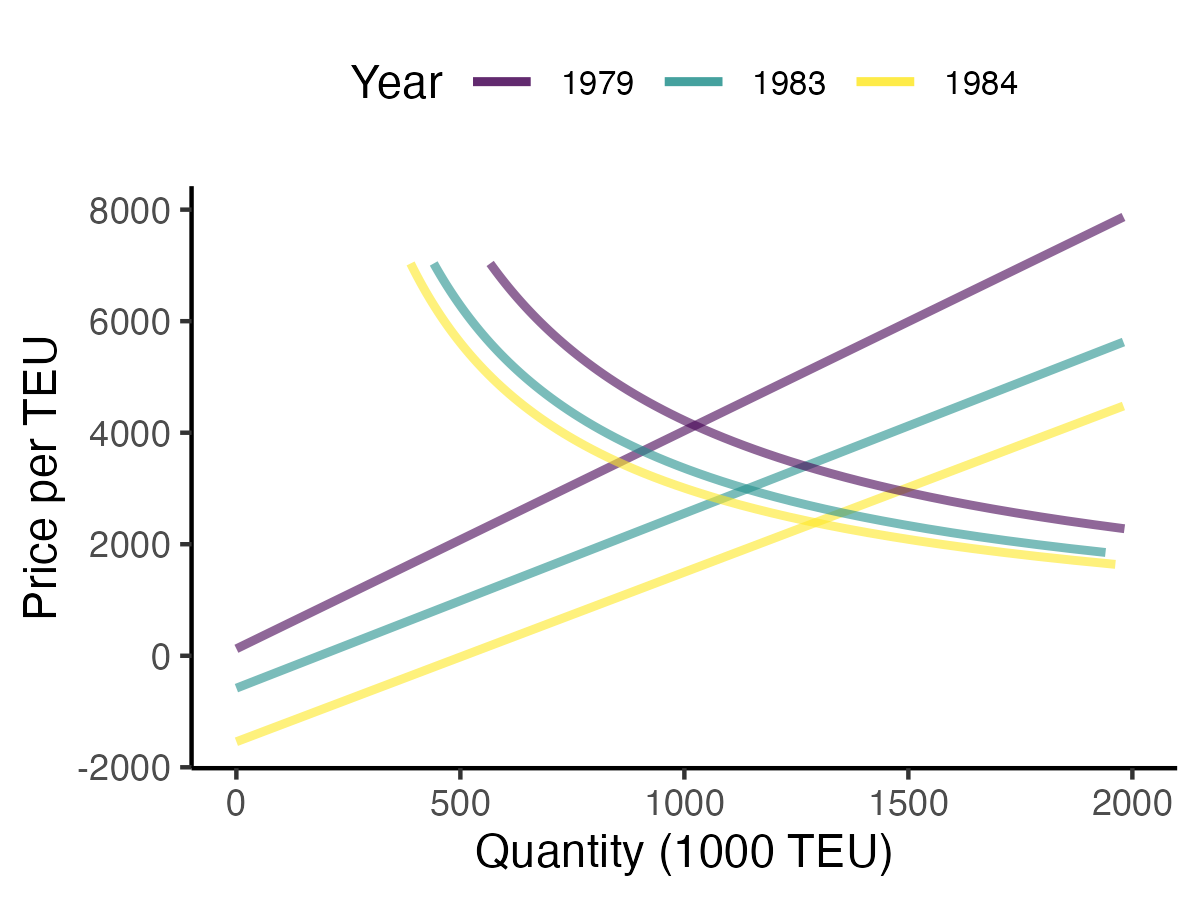}}
  \subfloat[Transatlantic \textcolor{black}{eastbound}]{\includegraphics[width = 0.32\textwidth]
  {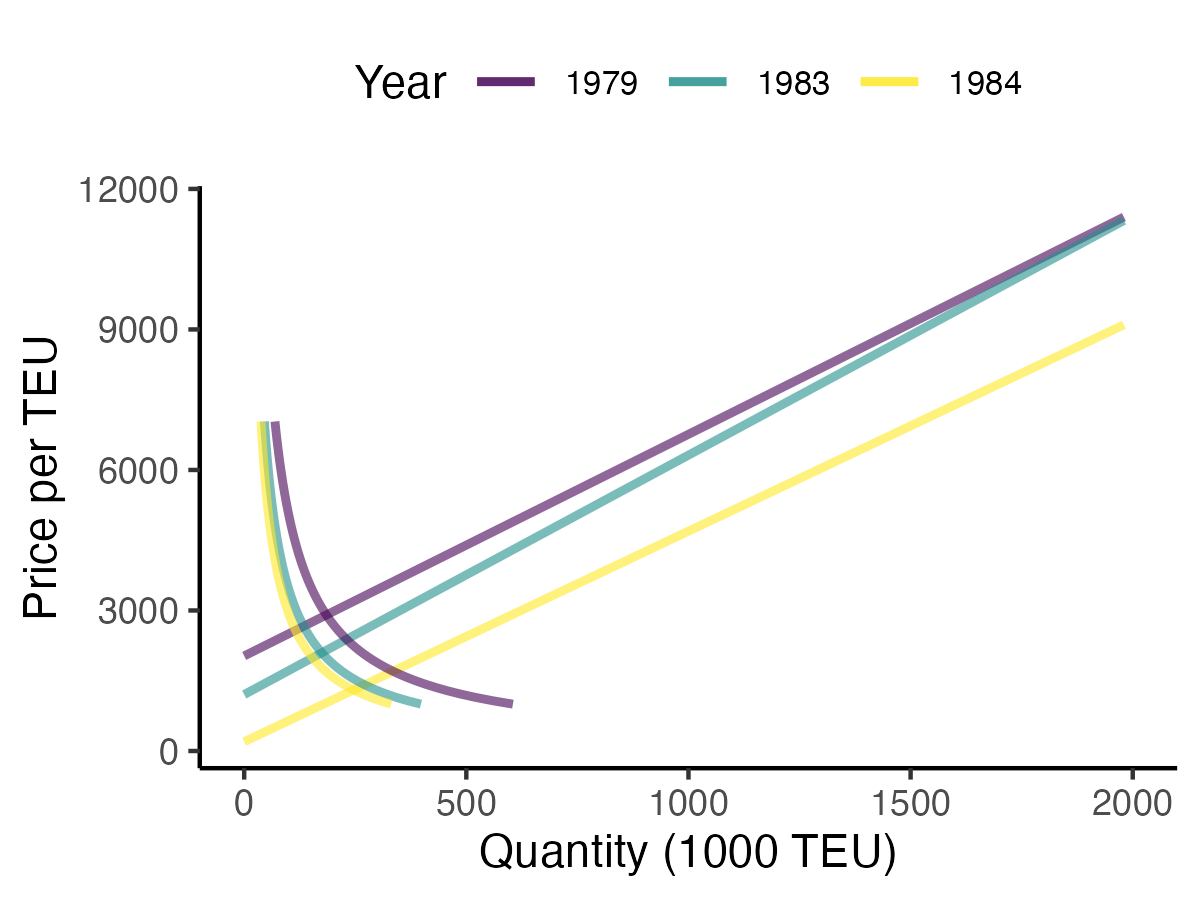}}
  \subfloat[Asia to Europe]{\includegraphics[width = 0.32\textwidth]
  {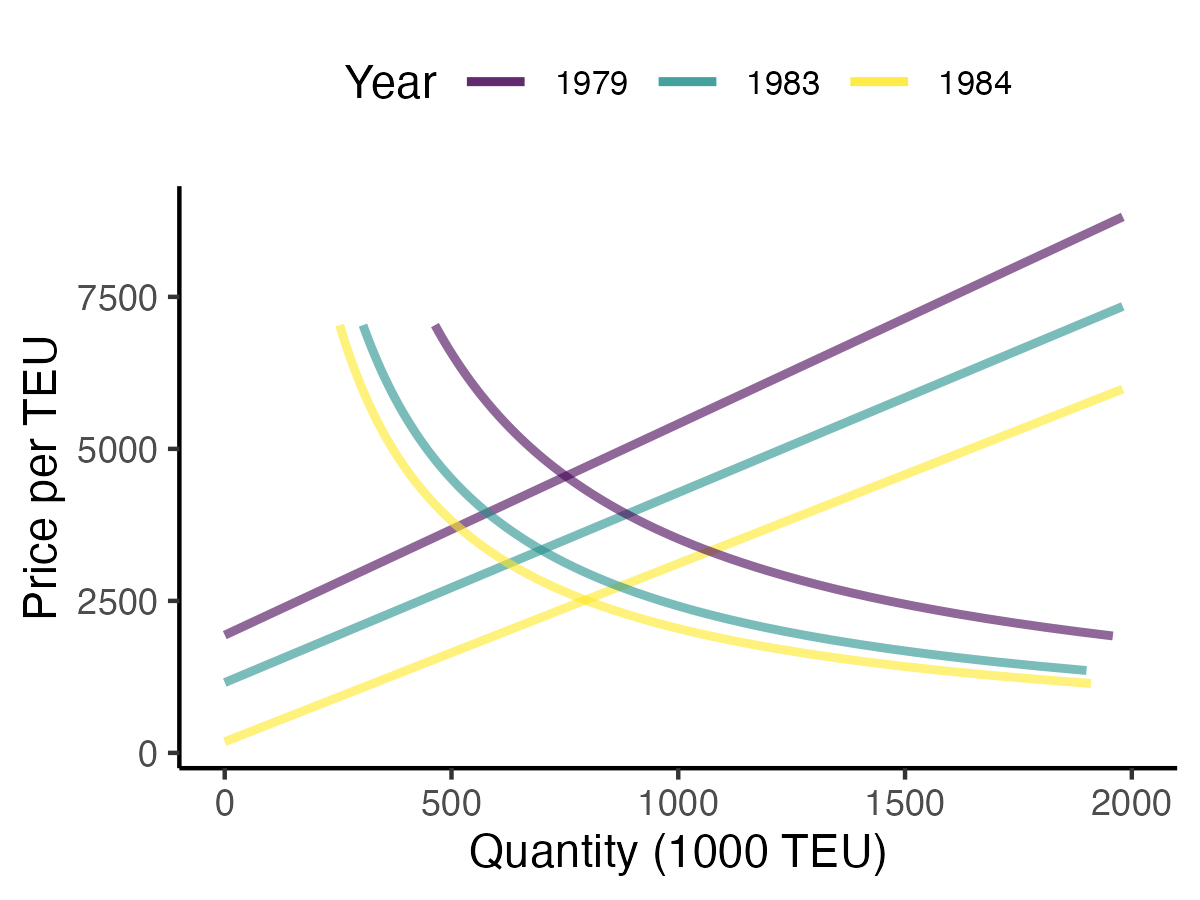}}
  \caption{Estimated demand and supply equations}
  \label{fg:demand_and_supply_plot_eur_to_asia}
  \end{center}
  \footnotesize
   Note: Each panel shows the estimated demand and supply equations in 1979, 1983, and 1984. Their intersections determine equilibrium freight rates and quantities.
\end{figure}

Figure \ref{fg:demand_and_supply_plot_eur_to_asia} compares the static equilibria in 1979 under the strong conference regime, in 1983 under the weakened conference regime, and in 1984 under competition.
Across all six routes, the supply equation shifts downward as the cartel wedge declines and then disappears.
Equilibrium freight rates are highest in 1979, lower in 1983, and lowest in 1984, while equilibrium quantities generally increase despite modest inward shifts in demand.
Route-specific demand, marginal cost, and capacity account for the remaining differences across routes.
These route-year equilibria determine the static profits used in the dynamic model.

\subsection{Dynamic parameters}

Table \ref{tb:dynamic_estimate} reports the dynamic parameter estimates for each market.\footnote{With profits expressed in billions of dollars, I impose $\sigma=1$, which fixes the scale of the i.i.d. payoff shocks at one billion dollars. The reported parameters are payoff-equivalent reduced-form costs under this maintained scale; their dollar interpretation and the implied counterfactual responsiveness are conditional on this normalization.}
Entry costs range from \$2.9 billion to \$4.2 billion across markets, and exit costs range from \$0.7 billion to \$3.4 billion.
Annual operating costs range from \$0.04 billion in Asia--Europe to \$0.50 billion in the transatlantic market.
The investment parameters are reduced-form sunk costs of crossing a capacity threshold, not construction prices for individual ships.
They can absorb the construction or acquisition of multiple vessels, complementary investments in containers and terminal facilities, and one-time fleet-deployment and organizational adjustment costs.
They should therefore be interpreted as firm-market-level capacity-expansion costs.
The cost of moving from capacity level 1 or 2 to the next level, $\iota_1$, ranges from \$3.3 billion to \$3.8 billion.
For level-3 firms, the identified cost of reaching level 4, $\iota_2$, is \$4.1 billion in Asia--Europe and \$9.8 billion in the transatlantic market; it is not finitely identified in the transpacific market.

\begin{table}[!t]
  \begin{center}
      \caption{Dynamic parameters}
      \begin{tabular}[t]{lccc}
\toprule
Parameter & Transpacific & Transatlantic & Asia--Europe\\
\midrule
Entry cost: $\kappa^{e}$ & 4.153 & 2.937 & 3.867\\
 & {}[3.695,4.666] & {}[2.404,3.587] & {}[3.441,4.346]\\
Exit cost: $\psi$ & 0.668 & 3.369 & 2.762\\
 & {}[0.055,1.536] & {}[2.758,4.115] & {}[1.516,5.033]\\
Operation cost: $\phi$ & 0.206 & 0.5 & 0.044\\
 & {}[0.09,0.313] & {}[0.409,0.61] & {}[0.004,0.154]\\
Investment cost: $\iota_1$ & 3.344 & 3.649 & 3.833\\
 & {}[2.807,3.984] & {}[2.987,4.457] & {}[3.411,4.307]\\
Investment cost: $\iota_2$ & Not identified & 9.766 & 4.135\\
 &  & {}[8.691,11.634] & {}[2.772,6.168]\\
Logit scale: $\sigma$ (normalized) & 1.000 & 1.000 & 1.000\\
 &  &  & \\
Log Likelihood & -67.691 & -53.687 & -67.857\\
\bottomrule
\end{tabular}

      \label{tb:dynamic_estimate}
  \end{center}
  \footnotesize
  Note: Each market aggregates its eastbound and westbound routes. Cost parameters are measured in billions of U.S. dollars, and the logit scale $\sigma$ is normalized to one. The investment parameters are reduced-form sunk costs of moving between capacity levels, not prices of individual ships. \textcolor{black}{Brackets report conditional likelihood-slice ranges based on a 90\% chi-squared cutoff, holding the other parameters at their maximum likelihood values (Appendix \ref{sec:dynamic_estimation_details}). Reported log-likelihood values are kernels that omit data-only combinatorial constants.} In the transpacific market, no level-3-to-4 investment is observed during the 1973--1983 estimation sample, so $\iota_2$ is not finitely identified; I treat the corresponding investment action as unavailable.
\end{table}

\begin{figure}[!t]
  \begin{center}
  \subfloat[Transpacific]{\includegraphics[width = 0.33\textwidth]
  {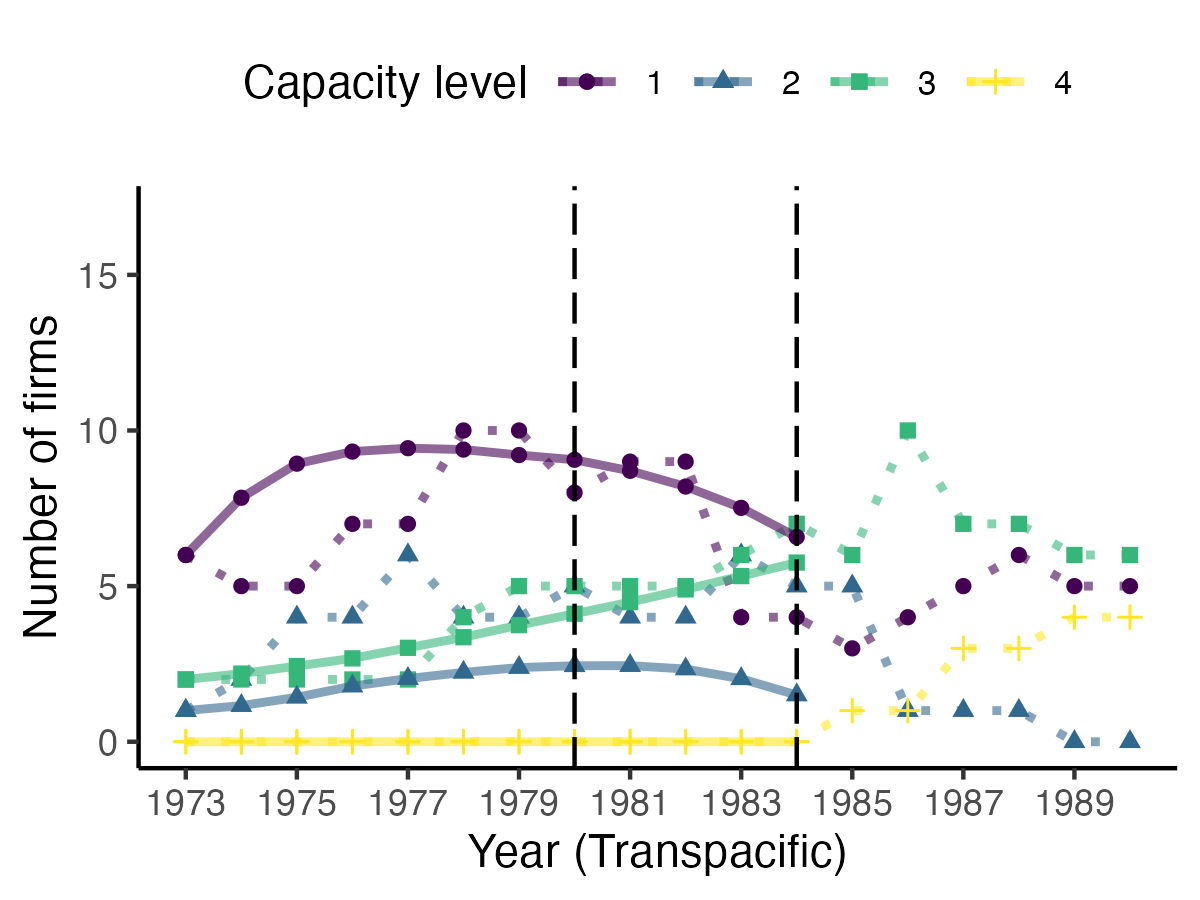}}
  \subfloat[Transatlantic]{\includegraphics[width = 0.33\textwidth]
  {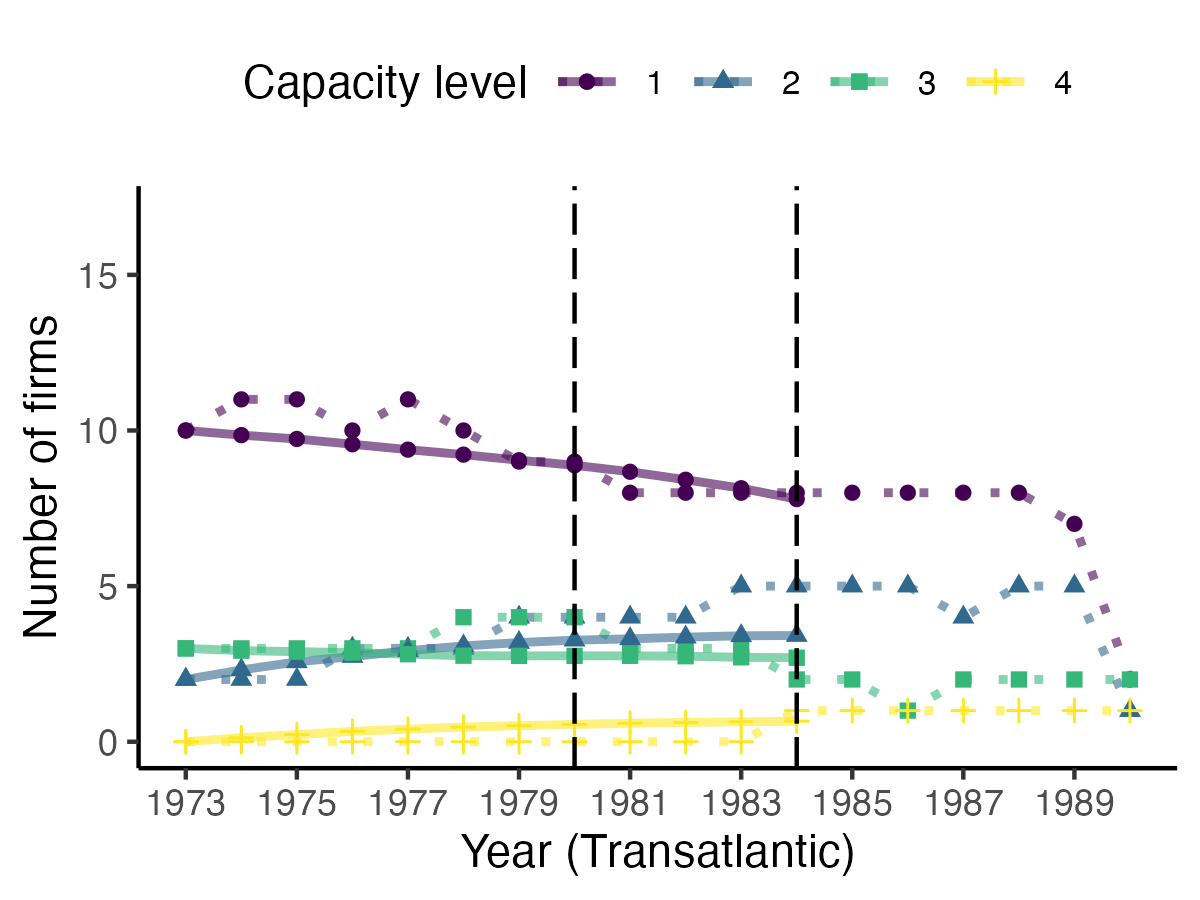}}
  \subfloat[Asia--Europe]{\includegraphics[width = 0.33\textwidth]
  {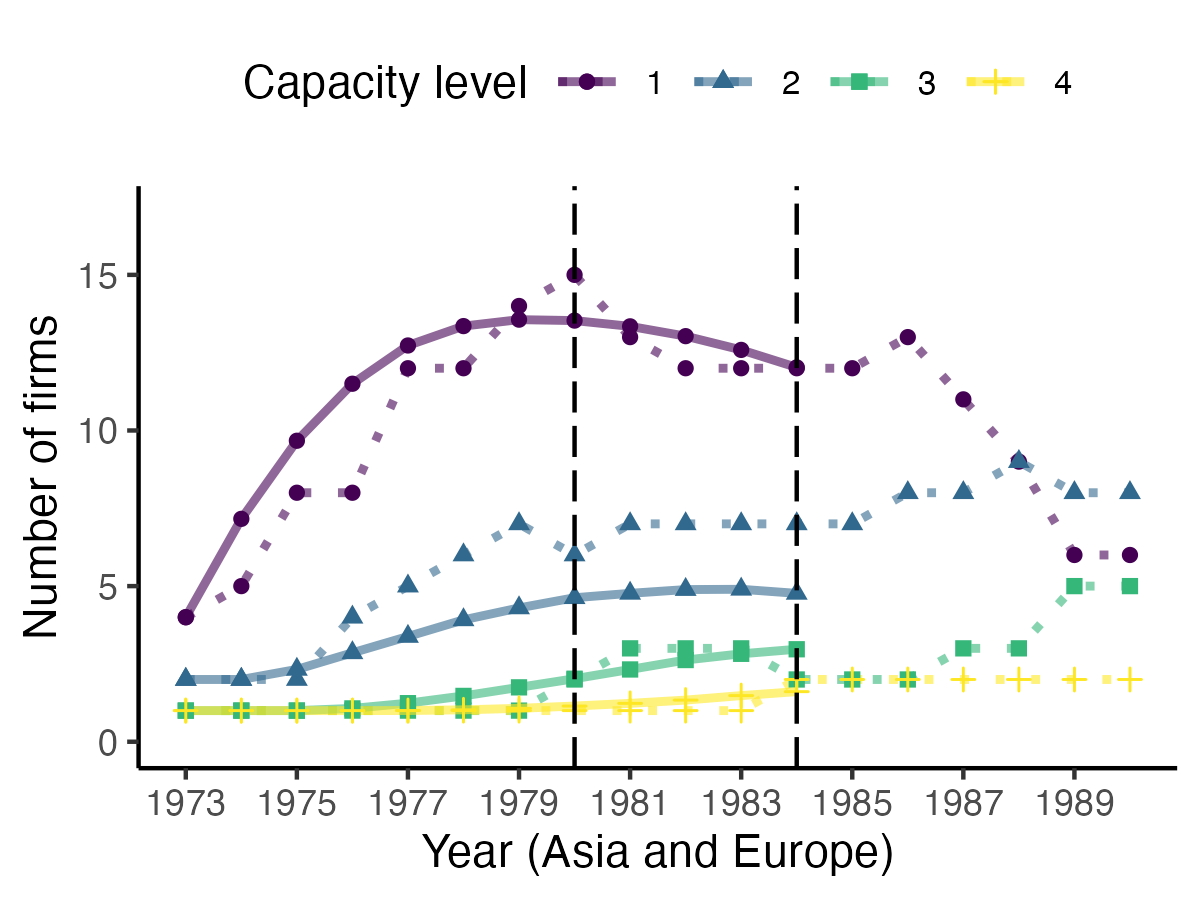}}
  \caption{Observed and simulated market structure}
  \label{fg:state_transition_asia_and_eur}
  \end{center}
  \footnotesize
  Note: Solid lines report the mean of 1,000 simulated paths, and dotted lines report the data. Simulated paths end in 1984; the data continue through 1990.
\end{figure}

Figure \ref{fg:state_transition_asia_and_eur} compares the observed market states with the mean of 1,000 simulated equilibrium paths through the 1984 terminal year.
The model reproduces the broad pre-1984 changes in the number of firms by capacity level, including the expansion of small firms in the transpacific and Asia--Europe markets and the decline of small firms in the transatlantic market.
The simulated paths smooth some year-to-year movements in the data but provide the benchmark market-structure paths for the counterfactuals in Section \ref{sec:counterfactual}.

\subsection{Welfare Along the Benchmark Path}

Table \ref{tb:welfare_benchmark} reports discounted consumer surplus, producer surplus, social welfare, and net social welfare along the benchmark equilibrium path.
Consumer surplus accounts for most of gross surplus in every market and regime, while operating, exit, entry, and shipbuilding costs create a substantial gap between gross surplus and net social welfare.
The transpacific market is largest in both consumer and producer surplus. The resource-cost adjustment is proportionally most important in the transatlantic market, where benchmark net social welfare is less than one-tenth of gross surplus.
Because the regime totals combine different years, market states, and discount factors, they describe the benchmark path rather than the causal effect of the cartel.
\textcolor{black}{Section \ref{sec:counterfactual} quantifies the model-implied effect by comparing the benchmark with alternative conference scenarios.}

\begin{table}[!htbp]
  \begin{center}
  \caption{Welfare along the benchmark path}
  
\begin{tabular}[t]{llrrrr}
\toprule
Market & Regime & CS & PS & SW & Net SW\\
\midrule
Transpacific & 1973-1979 & 244.71 & 22.29 & 267.00 & 205.14\\
 & 1980-1983 & 69.46 & 6.25 & 75.71 & 62.93\\
Transatlantic & 1973-1979 & 64.20 & 7.56 & 71.76 & 7.83\\
 & 1980-1983 & 18.16 & 2.23 & 20.39 & 1.37\\
Asia--Europe & 1973-1979 & 205.76 & 20.18 & 225.94 & 164.06\\
 & 1980-1983 & 58.18 & 6.48 & 64.66 & 52.20\\
\bottomrule
\end{tabular}

  \label{tb:welfare_benchmark}
  \end{center}
  \footnotesize
  Note: Entries are discounted to 1973 and measured in billions of U.S. dollars. Each row sums annual flows over the indicated years. CS denotes consumer surplus, PS producer surplus, $SW=CS+PS$, and Net SW subtracts operating, exit, entry, and shipbuilding costs from SW. Appendix \ref{sec:welfare_measurement} defines these measures.
\end{table}
\FloatBarrier

\section{Counterfactuals}\label{sec:counterfactual}
Shipping conferences affected the industry through two distinct mechanisms: collective control of route prices and the internal division of conference cargo. The first determined the rents available to conference carriers; the second determined how those rents were distributed across carriers of different sizes. Because both mechanisms changed current profits, they could alter entry, exit, shipbuilding, and hence future shipping capacity. Their welfare effects therefore cannot be inferred from contemporaneous freight rates alone. The central welfare question is whether conference rents financed valuable capacity expansion or instead induced entry and shipbuilding whose resource costs exceeded their service gains.

I use the model to answer two counterfactual questions. What market structure would have emerged without conference price coordination? How would quota rules favoring smaller or larger carriers have changed investment and welfare? The first exercise removes the estimated price wedges and, as an upper bound, imposes hypothetical full collusion; the second changes the internal quota rule. For each exercise, I recompute static profits, solve for equilibrium choice probabilities, and simulate 1,000 market paths through the 1984 terminal state. I measure welfare over the conference period 1973--1983, with all annual flows discounted to 1973. The 1984 state affects pre-1984 choices through the terminal value but is not included in the reported welfare totals.

\subsection{With and Without Shipping Cartels}
In the first counterfactual, I eliminate the cartel effect from the profit function and assume that individual service supply is determined by equation \eqref{eq:individual_supply_curve} throughout the sample. This non-cartel counterfactual parallels the removal of shipbuilding subsidies in \cite{kalouptsidi2017res}. I solve for a new equilibrium under $\tilde{\gamma}_1=\tilde{\gamma}_2=0$ and use the resulting equilibrium choice probabilities to simulate 1,000 paths in each market.

As a complementary upper bound, the model imposes hypothetical full collusion.\footnote{At $\hat{\alpha}_1=-1.111$, the condition below implies a price about ten times marginal cost. \textcolor{black}{For $\alpha_1\geq -1$, the static problem has no finite solution; the data do not reject the boundary case $\alpha_1=-1$.} This scenario is therefore an upper bound, not a policy counterfactual. The welfare results below are conditional on the preferred pooled additive-wedge and cost specification and on the selected admissible demand estimate ($\hat{\alpha}_1<-1$); Appendix \ref{sec:iv_specification_search} shows that both are sensitive, so the reported surplus levels should be interpreted conditionally.} This replaces the estimated cartel wedge with the joint-profit-maximizing markup while retaining the capacity-proportional benchmark allocation. \textcolor{black}{The comparison asks how closely the estimated conference regime approached monopoly behavior.} In each period, the conference price satisfies
\begin{align*}
P_{rt}\left(1+\frac{1}{\alpha_1}\right)=c_{rt}+\gamma_1\frac{Q_{rt}}{S_{rt}}.
\end{align*}

Entry, exit, and shipbuilding then respond to the resulting profits.

\begin{figure}[!t]
  \begin{center}
  \subfloat[Transpacific]{\includegraphics[width = 0.33\textwidth]  {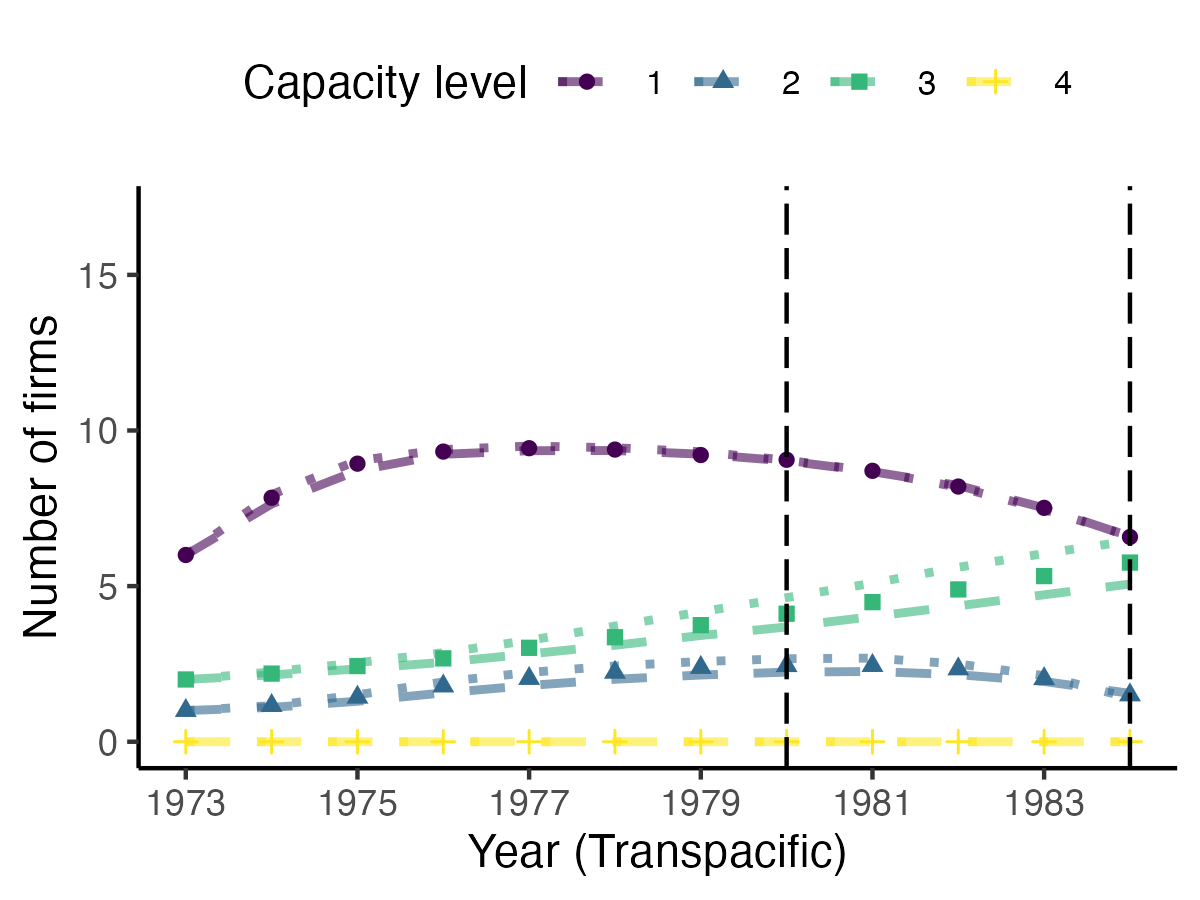}}
  \subfloat[Transatlantic]{\includegraphics[width = 0.33\textwidth]  {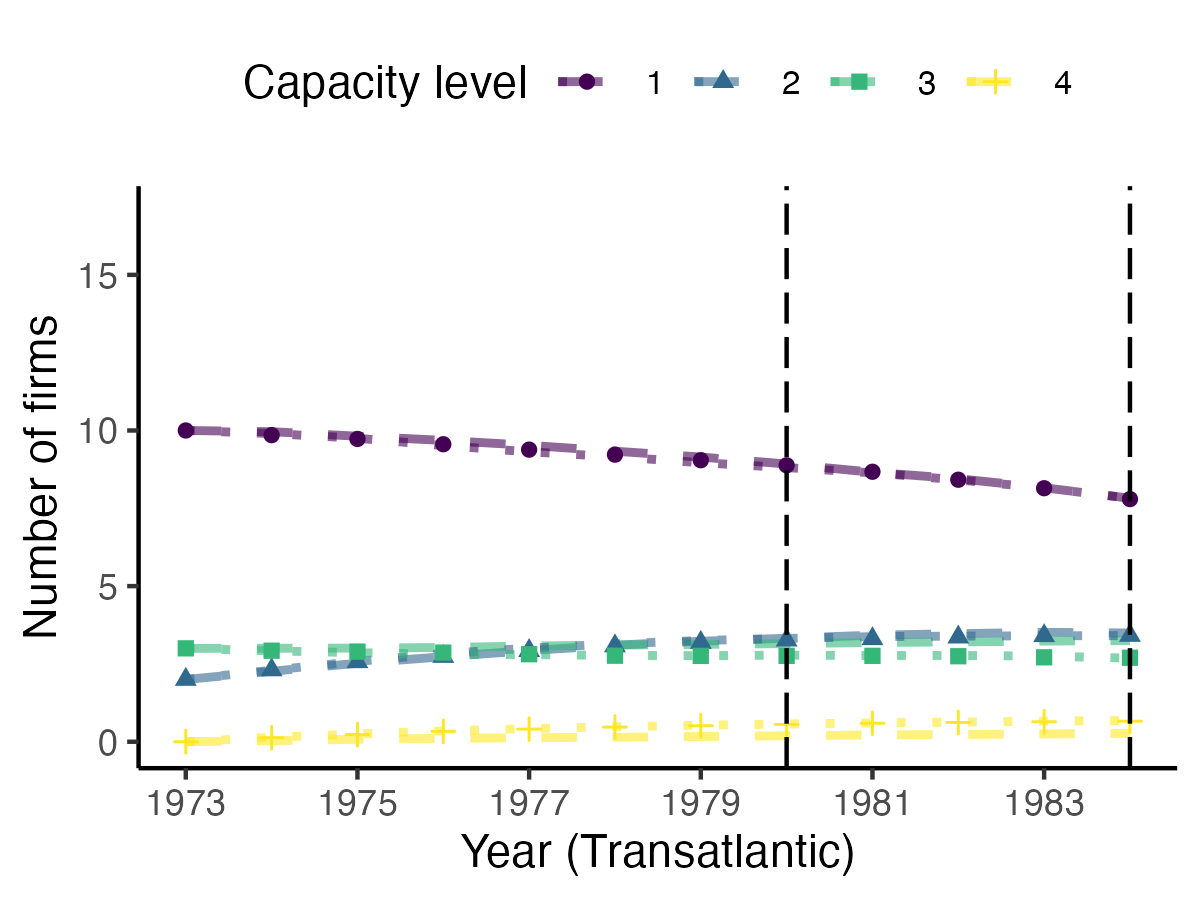}}
  \subfloat[Asia--Europe]{\includegraphics[width = 0.33\textwidth]  {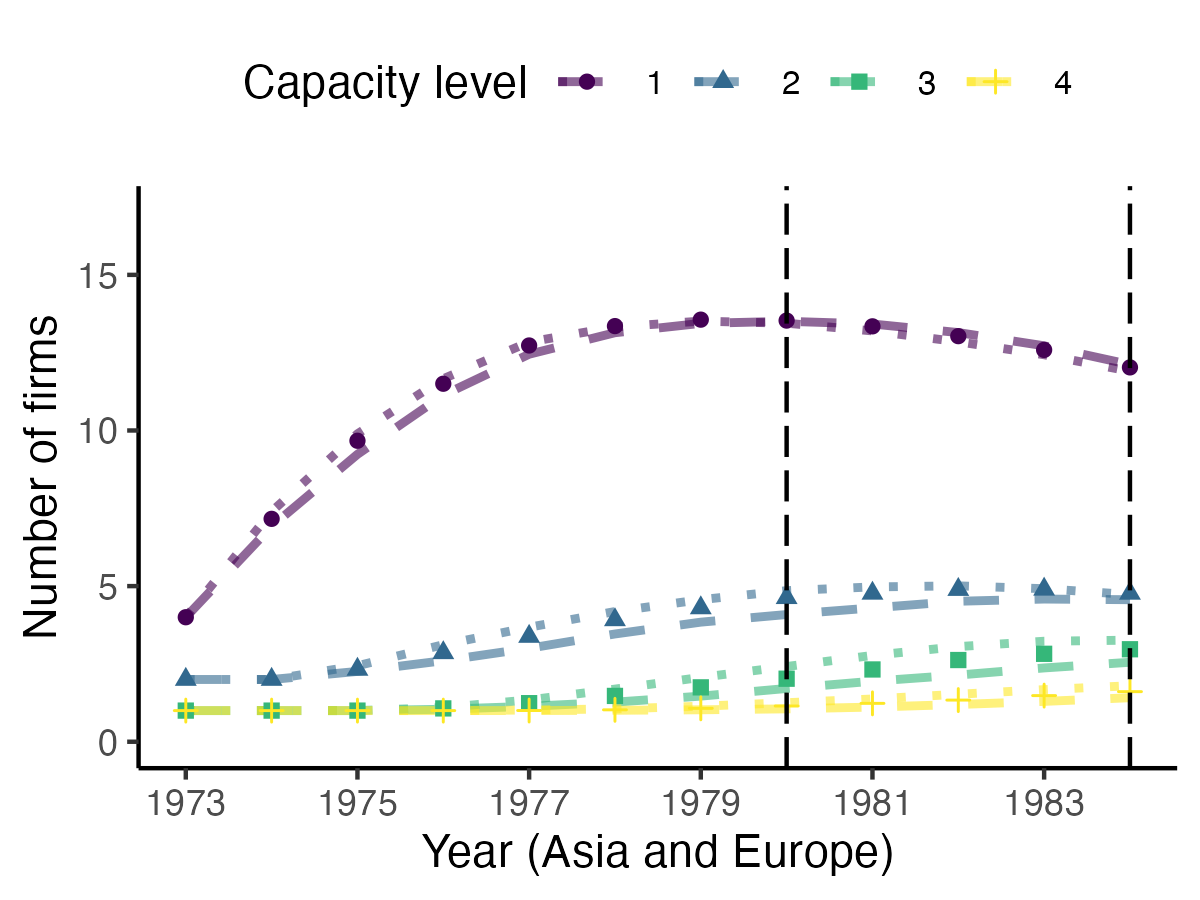}}
  \caption{Market structure under benchmark, non-cartel, and full-collusion scenarios}
  \label{fg:state_transition_counterfactual_without_cartel}
  \end{center}
  \footnotesize
  Note: Points show mean simulated firm counts under the benchmark. Dashed lines show the counterfactual without the conference regime, and dotted lines show the hypothetical full-collusion scenario. Colors and shapes identify capacity levels. Each path is based on 1,000 simulations.
\end{figure}

\textcolor{black}{Figure \ref{fg:state_transition_counterfactual_without_cartel} presents the state transitions under the benchmark, non-cartel, and full-collusion scenarios. Removing the cartel shifts the transatlantic market away from the largest capacity class: in 1984, the mean number of level-4 firms falls from 0.66 to 0.27, while the number of level-3 firms rises from 2.70 to 3.25. It also lowers the mean number of level-3 firms from 5.76 to 5.07 in the transpacific market and reduces level-3 and level-4 firms in Asia--Europe from 2.97 to 2.55 and from 1.61 to 1.41, respectively. Thus, cartel rents encouraged capacity expansion in all three markets, but the capacity class that responded differed across markets.}

\textcolor{black}{Under hypothetical full collusion, the higher rents produce a pronounced shift toward larger capacity classes in the transpacific and Asia--Europe markets but only a modest change in the transatlantic market. By 1984, the mean number of level-3 firms rises from 5.76 to 6.45 in the transpacific market. In Asia--Europe, level-3 and level-4 firms rise from 2.97 to 3.26 and from 1.61 to 1.80, respectively. In the transatlantic market, the level-4 count rises only from 0.66 to 0.68, while the level-3 count is nearly unchanged.}

Table \ref{tb:welfare_counterfactual_without_cartel} compares welfare under these scenarios with the benchmark. \textcolor{black}{Over 1973--1983, removing the cartel lowers producer surplus by about 17--27\% and raises consumer surplus by about 2--4\%.} The modest increase in consumer surplus reflects two opposing effects and the relatively low price elasticity of demand. Removing the cartel wedge directly lowers prices, but the resulting decline in profits discourages entry and shipbuilding, reduces total capacity, and raises marginal cost through the supply equation. The estimated elasticity, close to one in absolute value, further limits the quantity response to lower freight rates.

\begin{table}[!b]
  \begin{center}
  \setlength{\tabcolsep}{5.5pt}
  \caption{Welfare evaluation (non-cartel and full-collusion scenarios)}
  
\begin{tabular}[t]{llrrrrrrrr}
\toprule
\multicolumn{2}{c}{ } & \multicolumn{4}{c}{Present value} & \multicolumn{4}{c}{Change from benchmark} \\
\cmidrule(l{3pt}r{3pt}){3-6} \cmidrule(l{3pt}r{3pt}){7-10}
Market & Regime & CS & PS & SW & Net SW & CS & PS & SW & Net SW\\
\midrule

\multicolumn{10}{l}{\textit{Non-cartel}}\\
Transpacific & 1973-1979 & 248.93 & 18.41 & 267.34 & 211.70 & ( 0.02) & (-0.17) & ( 0.00) & (0.03)\\
 & 1980-1983 & 70.07 & 5.37 & 75.44 & 63.63 & ( 0.01) & (-0.14) & ( 0.00) & (0.01)\\
Transatlantic & 1973-1979 & 67.01 & 5.53 & 72.54 & 13.20 & ( 0.04) & (-0.27) & ( 0.01) & ( 0.69)\\
 & 1980-1983 & 18.49 & 1.66 & 20.15 & 1.91 & ( 0.02) & (-0.26) & (-0.01) & ( 0.39)\\
Asia--Europe & 1973-1979 & 210.96 & 16.12 & 227.08 & 172.43 & ( 0.03) & (-0.20) & ( 0.01) & ( 0.05)\\
 & 1980-1983 & 58.95 & 5.44 & 64.39 & 53.41 & ( 0.01) & (-0.16) & ( 0.00) & ( 0.02)\\

\multicolumn{10}{l}{\textit{Full collusion}}\\
Transpacific & 1973-1979 & 229.30 & 26.18 & 255.48 & 187.10 & (-0.06) & (0.17) & (-0.04) & (-0.09)\\
 & 1980-1983 & 65.24 & 7.60 & 72.84 & 58.38 & (-0.06) & (0.22) & (-0.04) & (-0.07)\\
Transatlantic & 1973-1979 & 62.31 & 7.80 & 70.11 & 5.42 & (-0.03) & (0.03) & (-0.02) & (-0.31)\\
 & 1980-1983 & 17.43 & 2.35 & 19.78 & 0.54 & (-0.04) & (0.05) & (-0.03) & (-0.61)\\
Asia--Europe & 1973-1979 & 191.51 & 23.29 & 214.80 & 147.03 & (-0.07) & (0.15) & (-0.05) & (-0.10)\\
 & 1980-1983 & 54.45 & 7.58 & 62.03 & 47.58 & (-0.06) & (0.17) & (-0.04) & (-0.09)\\
\bottomrule
\end{tabular}

  \label{tb:welfare_counterfactual_without_cartel}
  \end{center}
  \footnotesize
  Note: \textcolor{black}{Each row sums annual flows over the indicated subperiod (1973--1979 or 1980--1983); all entries are discounted to 1973 and measured in billions of U.S. dollars.} The non-cartel block removes the conference wedges and quota constraint; the full-collusion block imposes static joint-profit maximization under the capacity-proportional benchmark allocation. CS denotes consumer surplus, PS producer surplus, $SW=CS+PS$, and Net SW subtracts operating, exit, entry, and shipbuilding costs from SW. Parentheses report proportional changes from the \textcolor{black}{corresponding benchmark subperiod}. Appendix \ref{sec:welfare_measurement} defines the welfare measures.
\end{table}

Measured by the sum of consumer and producer surplus, removing conference price coordination changes welfare by less than 1\% in every market: cartel pricing mainly transfers surplus from consumers to producers. This measure, however, excludes \textcolor{black}{the resource costs associated with the resulting market structure}. \textcolor{black}{Entry and shipbuilding costs fall by 12--23\%, and net social welfare rises by 2.7\% in the transpacific market, 4.4\% in Asia--Europe, and 64.3\% in the transatlantic market.} The large transatlantic percentage reflects low benchmark net social welfare after the operating costs of maintaining many active firms.
\textcolor{black}{The transatlantic estimate is a conference-segment effect relative to a small benchmark denominator, not an industry-wide welfare effect.} Because the calculation ends with the conference period, it does not credit the accumulated large-ship capacity for returns after the sample period.

\textcolor{black}{Within the model and the 1973--1983 welfare horizon, these findings weaken an efficiency-based justification for centralized price and quantity controls during an industry's early stage.} In this setting, cartel pricing mainly redistributed surplus toward producers and accelerated large-ship investment; whether that investment improved welfare depends on the horizon over which its costs and returns are evaluated.

\textcolor{black}{Hypothetical full collusion produces larger welfare losses than the estimated conference regime. Over 1973--1983, producer surplus rises by about 4--18\%, consumer surplus falls by about 3--7\%, and their sum falls by about 2--5\% across markets. Entry and shipbuilding costs rise by about 4--14\%. Net social welfare falls by 8.4\% in the transpacific market, 10.0\% in Asia--Europe, and 35.2\% in the transatlantic market, but remains positive in all three. Using the 1973--1983 average model-implied price difference from the no-cartel scenario at observed market states, the estimated conference effect is 15.8\%, 14.7\%, and 31.4\% of the full-collusion effect in the transpacific, Asia--Europe, and transatlantic markets, respectively. Thus, full collusion remains substantially stronger than the estimated conference regime in every market, although the transatlantic market comes closest to that benchmark.}

\FloatBarrier

\subsection{Alternative Internal Allocation Rules}
Market-share division is central to cartel organization \citep{marshall2014economics}, but its dynamic effects are usually difficult to measure because internal allocation rules are rarely observed. \textcolor{black}{Shipping conferences provide an unusual setting in which route-assigned capacity data allow an explicit capacity-proportional benchmark to be defined and compared with alternative model allocation rules.} Holding a route's conference price and quantity fixed, changing this rule reallocates rents across capacity levels and thereby changes the incentives to enter or build larger ships. Economically, the rule acts as targeted support within the conference: favoring a capacity class raises its current revenue without changing total conference revenue in a given state.

\paragraph{Two discrete alternative allocation rules}
\textcolor{black}{The benchmark allocates} each route's conference quantity in proportion to members' tonnage. I first consider two discrete alternatives that shift conference cargo toward smaller or larger carriers. \textcolor{black}{These rules are motivated by policies that favor selected firm types through bid preferences or subsidies, thereby changing participation and investment incentives \citep{krasnokutskaya2011bid,fan2015competition}.} Let
\begin{align*}
    \textcolor{black}{\omega_{irt}^{h}=\frac{w_{irt}^{h}s_{irt}}{\sum_{j=1}^{N_{rt}}w_{jrt}^{h}s_{jrt}}, \qquad h\in\{S,L\}},
\end{align*}
where $w_{irt}^{S}=1.25$ for capacity levels 1 and 2 and $0.75$ otherwise; $w_{irt}^{L}$ reverses these weights. Thus, $\omega^{S}$ shifts quotas toward small carriers and $\omega^{L}$ toward large carriers. Conditional on a market state, these rules leave the route price and total quantity unchanged but redistribute profits across capacity levels. The resulting profit differences alter entry, exit, and shipbuilding incentives.

\begin{figure}[!b]
  \begin{center}
  \subfloat[Transpacific]{\includegraphics[width = 0.33\textwidth]  {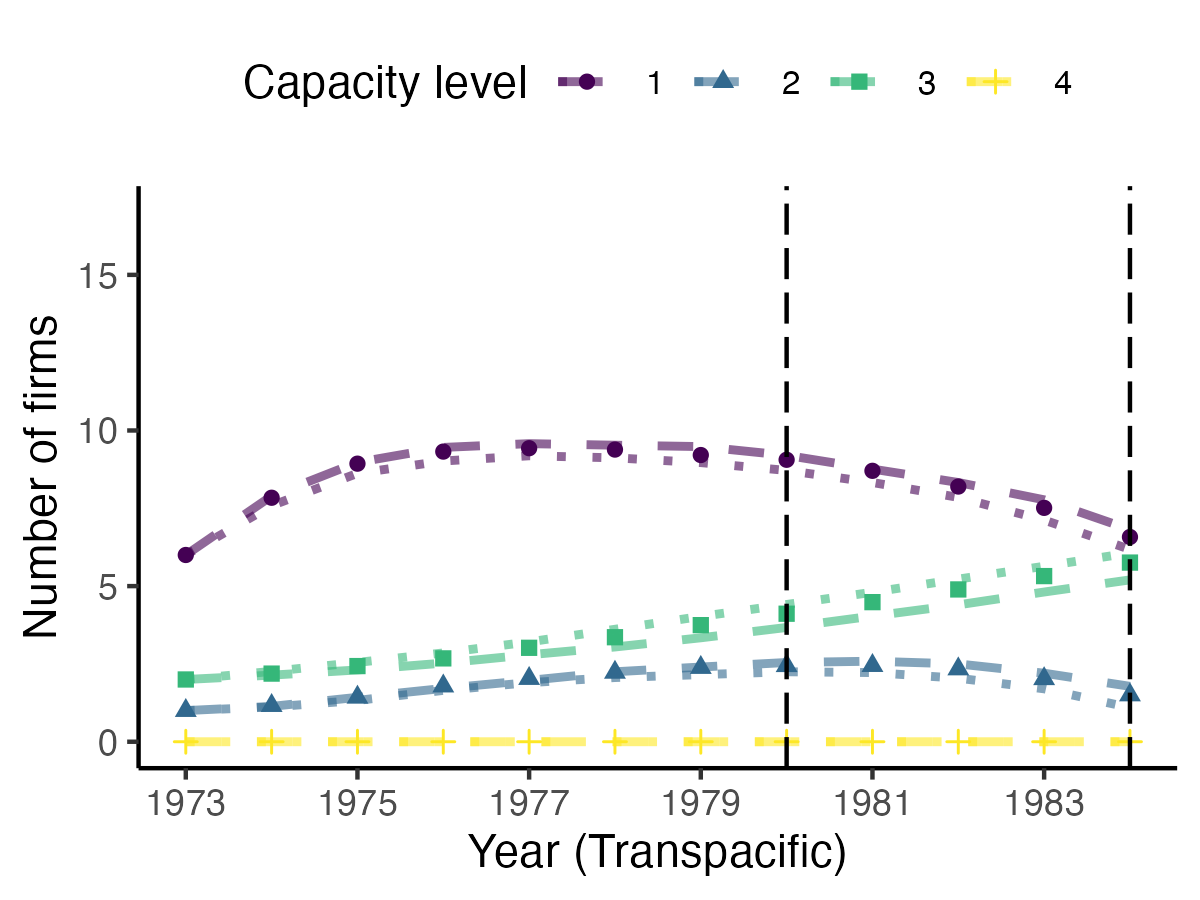}}
  \subfloat[Transatlantic]{\includegraphics[width = 0.33\textwidth]  {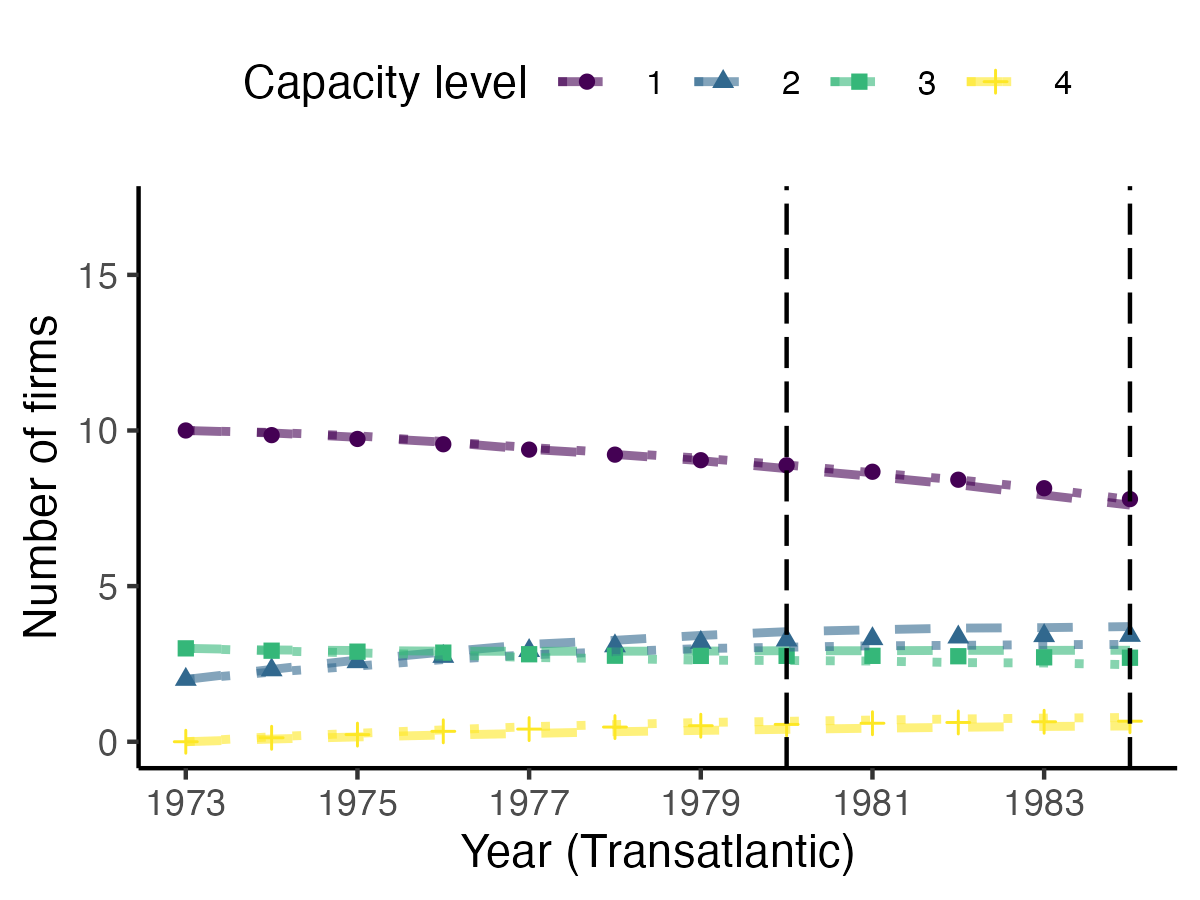}}
  \subfloat[Asia--Europe]{\includegraphics[width = 0.33\textwidth]  {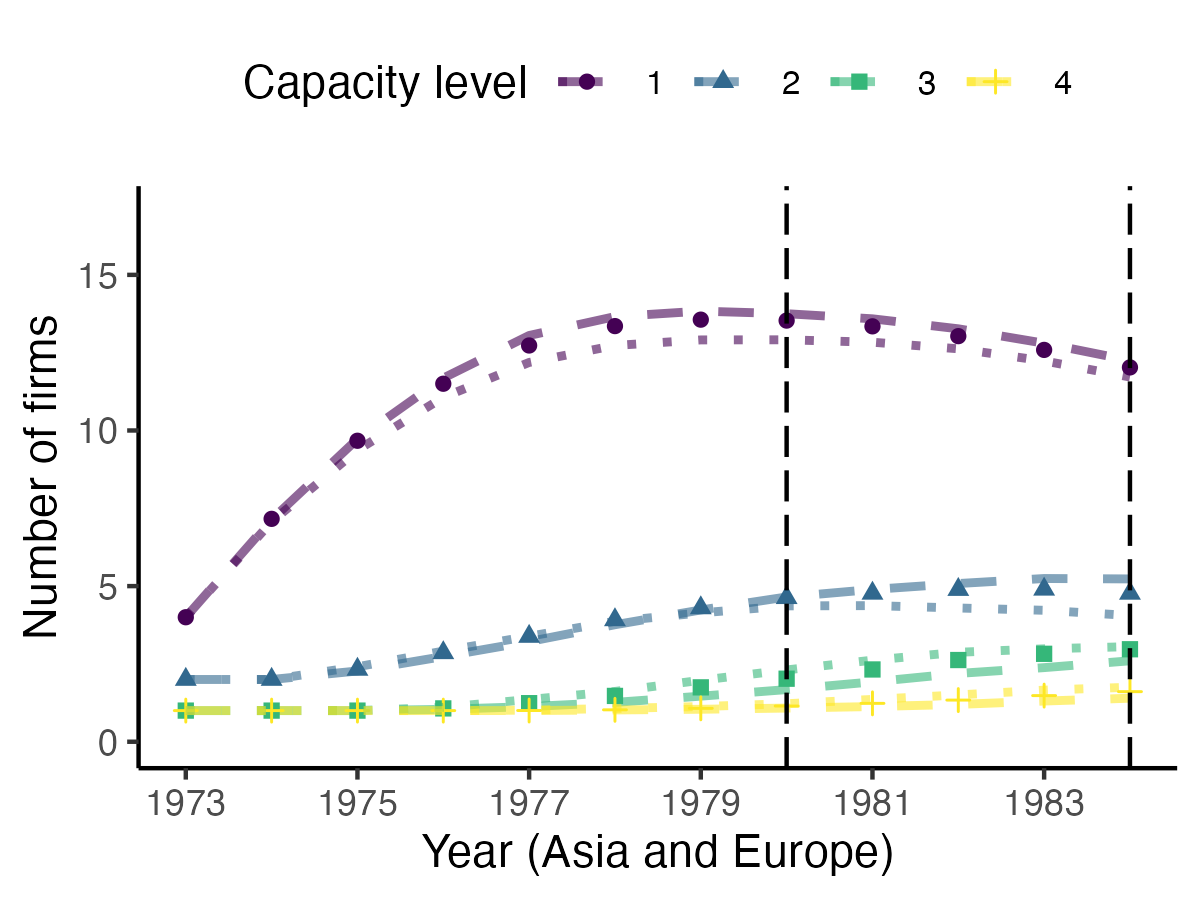}}
  \caption{Market structure under alternative internal allocation rules}
  \label{fg:state_transition_different_inner_allocation}
  \end{center}
  \footnotesize
  Note: Points show mean simulated firm counts under the benchmark. Dashed and dotted lines show $\omega^{S}$ and $\omega^{L}$, respectively. Colors and shapes identify capacity levels. Each path is based on 1,000 simulations.
\end{figure}

\textcolor{black}{Figure \ref{fg:state_transition_different_inner_allocation} shows these responses. In the transatlantic market, $\omega^{S}$ raises the mean numbers of level-2 and level-3 firms in 1984 and lowers the number of level-4 firms from 0.66 to 0.51. By contrast, $\omega^{L}$ reduces the two middle classes and raises the number of level-4 firms to 0.78. In the other markets, capacity similarly shifts toward the levels favored by each rule. In the model, these equilibrium responses reflect a participation--upgrading trade-off: $\omega^{S}$ raises the value of entry at level 1 but weakens incentives to move into and through the large-capacity states, leaving more firms at intermediate levels, whereas $\omega^{L}$ lowers the value of entry but strengthens incentives to upgrade into levels 3 and 4.}

\textcolor{black}{Table \ref{tb:welfare_counterfactual_inner_allocation_rule_1} shows that neither discrete rule has a uniform welfare ranking across markets. Relative to the benchmark, changes in net social welfare range from $-10.3\%$ to $13.1\%$. In the transatlantic market, $\omega^{S}$ raises net social welfare by 13.1\%, whereas $\omega^{L}$ lowers it by 10.3\%. The large-carrier rule raises net social welfare by 0.8\% in the transpacific market and by 0.1\% in Asia--Europe; the small-carrier rule raises it by 0.8\% in Asia--Europe and leaves it nearly unchanged in the transpacific market. No discrete alternative therefore dominates the capacity-proportional benchmark across markets.}

\begin{table}[!t]
  \begin{center}
  \setlength{\tabcolsep}{5.5pt}
  \caption{Welfare under alternative internal allocation rules}
  
\begin{tabular}[t]{llrrrrrrrr}
\toprule
\multicolumn{2}{c}{ } & \multicolumn{4}{c}{Present value} & \multicolumn{4}{c}{Change from benchmark} \\
\cmidrule(l{3pt}r{3pt}){3-6} \cmidrule(l{3pt}r{3pt}){7-10}
Market & Regime & CS & PS & SW & Net SW & CS & PS & SW & Net SW\\
\midrule

\multicolumn{10}{l}{\textit{Small-carrier rule ($\omega^{S}$)}}\\
Transpacific & 1973-1979 & 244.39 & 21.38 & 265.77 & 205.42 & (0.00) & (-0.04) & ( 0.00) & ( 0.00)\\
 & 1980-1983 & 69.26 & 5.90 & 75.16 & 62.55 & (0.00) & (-0.06) & (-0.01) & (-0.01)\\
Transatlantic & 1973-1979 & 64.02 & 7.29 & 71.31 & 9.02 & ( 0.00) & (-0.04) & (-0.01) & ( 0.15)\\
 & 1980-1983 & 18.07 & 2.11 & 20.18 & 1.40 & ( 0.00) & (-0.05) & (-0.01) & ( 0.02)\\
Asia--Europe & 1973-1979 & 205.62 & 19.51 & 225.13 & 165.07 & (0.00) & (-0.03) & ( 0.00) & ( 0.01)\\
 & 1980-1983 & 57.98 & 6.10 & 64.08 & 52.85 & (0.00) & (-0.06) & (-0.01) & ( 0.01)\\

\multicolumn{10}{l}{\textit{Large-carrier rule ($\omega^{L}$)}}\\
Transpacific & 1973-1979 & 244.82 & 21.59 & 266.41 & 206.94 & (0.00) & (-0.03) & (0.00) & (0.01)\\
 & 1980-1983 & 69.54 & 6.08 & 75.62 & 63.32 & (0.00) & (-0.03) & (0.00) & (0.01)\\
Transatlantic & 1973-1979 & 64.28 & 7.55 & 71.83 & 6.84 & (0.00) & (0.00) & (0.00) & (-0.13)\\
 & 1980-1983 & 18.22 & 2.25 & 20.47 & 1.41 & (0.00) & (0.01) & (0.00) & ( 0.03)\\
Asia--Europe & 1973-1979 & 205.81 & 19.69 & 225.50 & 164.70 & (0.00) & (-0.02) & (0.00) & ( 0.00)\\
 & 1980-1983 & 58.27 & 6.34 & 64.61 & 51.82 & (0.00) & (-0.02) & (0.00) & (-0.01)\\
\bottomrule
\end{tabular}

  \label{tb:welfare_counterfactual_inner_allocation_rule_1}
  \end{center}
  \footnotesize
  Note: \textcolor{black}{Each row sums annual flows over the indicated subperiod (1973--1979 or 1980--1983); all entries are discounted to 1973 and measured in billions of U.S. dollars.} The two blocks report the small-carrier rule $\omega^{S}$ and the large-carrier rule $\omega^{L}$. CS denotes consumer surplus, PS producer surplus, $SW=CS+PS$, and Net SW subtracts operating, exit, entry, and shipbuilding costs from SW. Parentheses report proportional changes from the \textcolor{black}{corresponding benchmark subperiod}. Appendix \ref{sec:welfare_measurement} defines the welfare measures.
\end{table}
\FloatBarrier

\paragraph{A one-parameter family of allocation rules}
The two discrete rules establish the direction of the investment response but provide only two benchmarks. To examine how \textcolor{black}{the capacity-proportional benchmark} trades off within-period shipping costs against dynamic investment incentives, I embed it in
\begin{align}
    \omega_{irt}(\lambda)=\frac{s_{irt}^{\lambda}}{\sum_{j=1}^{N_{rt}}s_{jrt}^{\lambda}}, \quad \lambda \ge 0, \label{eq:allocation_rule_lambda}
\end{align}
where $\lambda=1$ is \textcolor{black}{the capacity-proportional benchmark} and $\lambda=0$ assigns equal quotas. Relative to \textcolor{black}{the benchmark}, $0\leq\lambda<1$ favors smaller carriers, whereas $\lambda>1$ assigns cargo more than proportionally to tonnage. In particular, quota per unit of tonnage is proportional to $s_{irt}^{\lambda-1}$, so under $\lambda>1$ larger carriers operate at higher utilization rates than smaller carriers. This allocation is feasible in the model because there is no hard utilization constraint; large values of $\lambda$ should therefore be interpreted as design bounds rather than literal conference rules. For fixed $(P_{rt}^{*},Q_{rt}^{*})$ and a fixed market state, $\lambda=1$ equalizes marginal costs across conference members and minimizes aggregate variable shipping cost among feasible quota allocations. It need not maximize dynamic welfare because changing $\lambda$ also changes profit differences across capacity levels and hence the equilibrium capacity path.

\begin{figure}[!htbp]
  \begin{center}
  \subfloat[Transpacific]{\includegraphics[width = 0.33\textwidth]{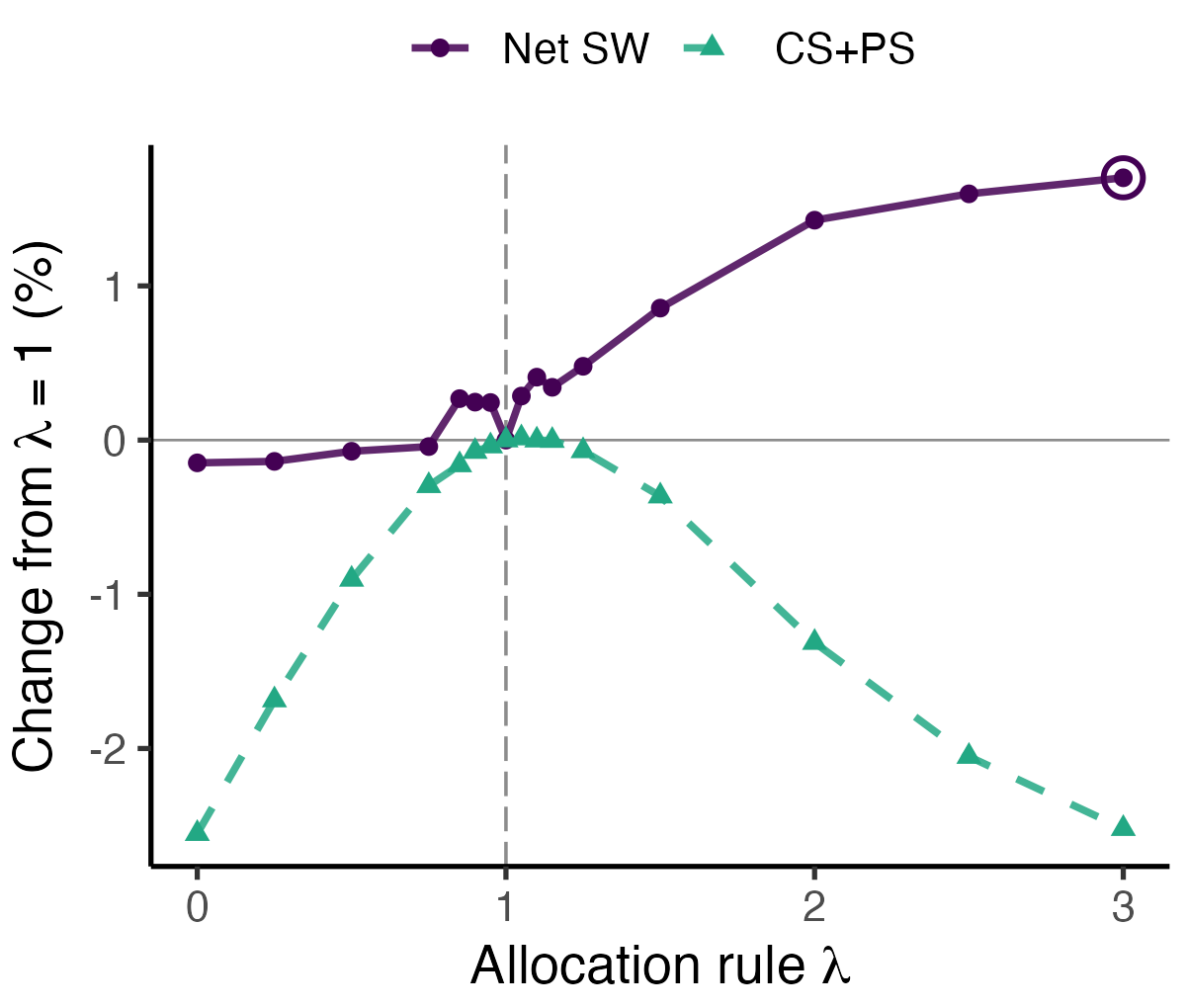}}
  \subfloat[Transatlantic]{\includegraphics[width = 0.33\textwidth]{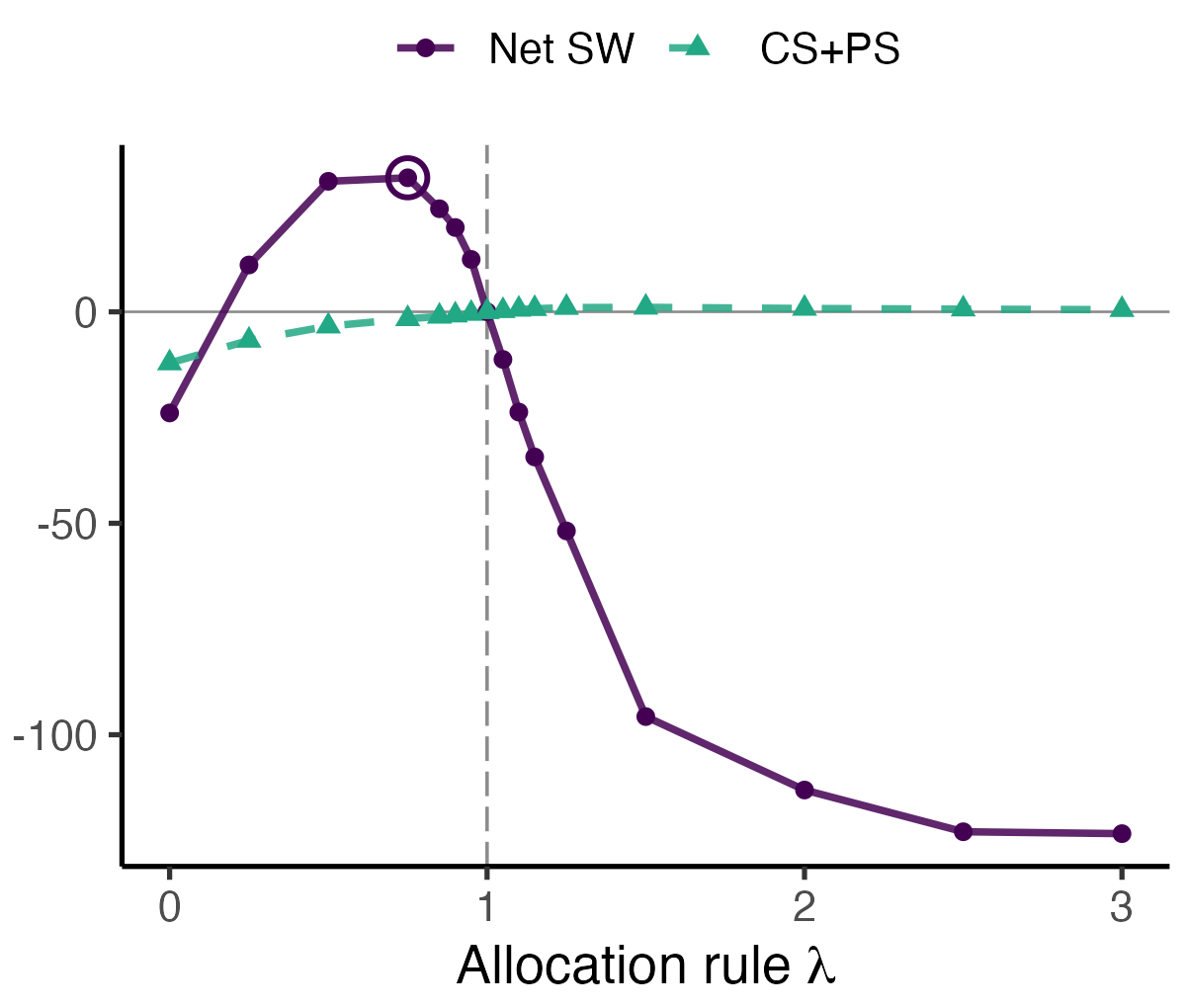}}
  \subfloat[Asia--Europe]{\includegraphics[width = 0.33\textwidth]{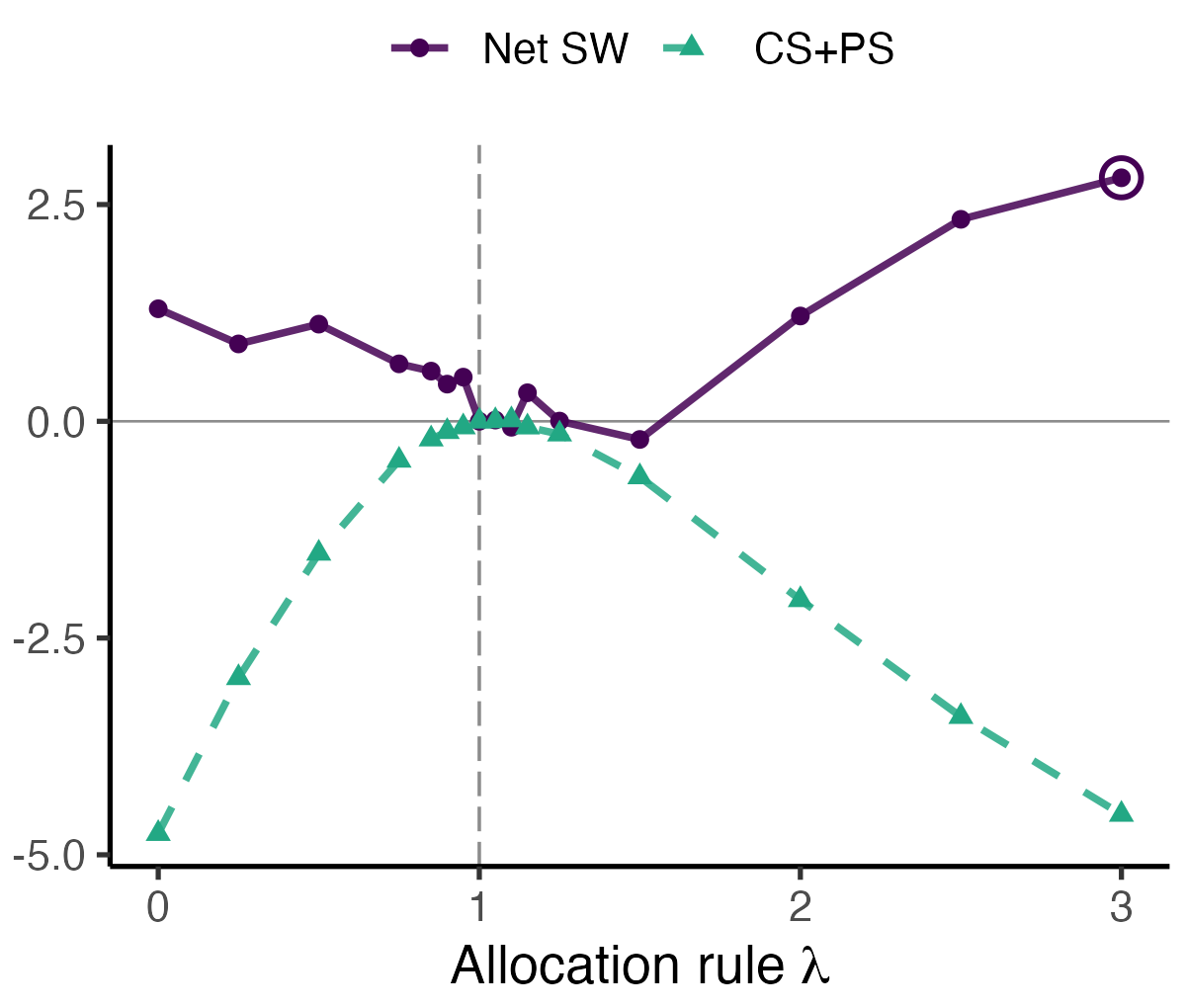}}
  \caption{Welfare under counterfactual allocation rules $\omega(\lambda)$, relative to \textcolor{black}{the capacity-proportional benchmark $\lambda=1$}}
  \label{fg:optimal_allocation_rule_lambda}
  \end{center}
  \footnotesize
  Note: Each panel reports the percentage change from \textcolor{black}{the capacity-proportional benchmark} ($\lambda=1$, vertical dashed line) in gross surplus ($CS+PS$) and net social welfare ($CS+PS-FC-XC-SC$), where CS denotes consumer surplus, PS producer surplus, FC operating costs, XC exit costs, and SC entry and shipbuilding costs. Vertical scales differ across panels to show within-market variation. Open circles mark the highest net social welfare among the evaluated grid points; a circle at an endpoint does not identify an interior optimum. Present values are discounted to 1973 over 1973--1983. Appendix \ref{sec:welfare_measurement} defines the welfare measures.
\end{figure}

\textcolor{black}{Figure \ref{fg:optimal_allocation_rule_lambda} reports the welfare ranking across allocation rules. Among the evaluated grid points, transatlantic net social welfare is highest at $\lambda=0.75$, 31.7\% above the benchmark. As $\lambda$ rises above one and quotas become increasingly concentrated on large carriers, net social welfare falls: it is 51.8\% lower at $\lambda=1.25$ and turns negative by $\lambda=2$. At $\lambda=3$, it is $-\$2.15$ billion, or 123.4\% below the benchmark. These large percentages partly reflect the low level of transatlantic benchmark net social welfare. The latter value of $\lambda$ is the upper boundary of the grid and should be read as an extreme illustration of a strong large-carrier tilt, not as an estimated optimum. At the same endpoint, net social welfare instead rises by 1.7\% in the transpacific market and by 2.8\% in Asia--Europe, despite declines in gross surplus of 2.5\% and 4.5\%, respectively. Thus, strengthening the same quota tilt can have opposite welfare effects across markets and need not preserve the ranking based on gross surplus.}

\begin{figure}[!htbp]
  \begin{center}
  \subfloat[Transpacific]{\includegraphics[width = 0.33\textwidth]{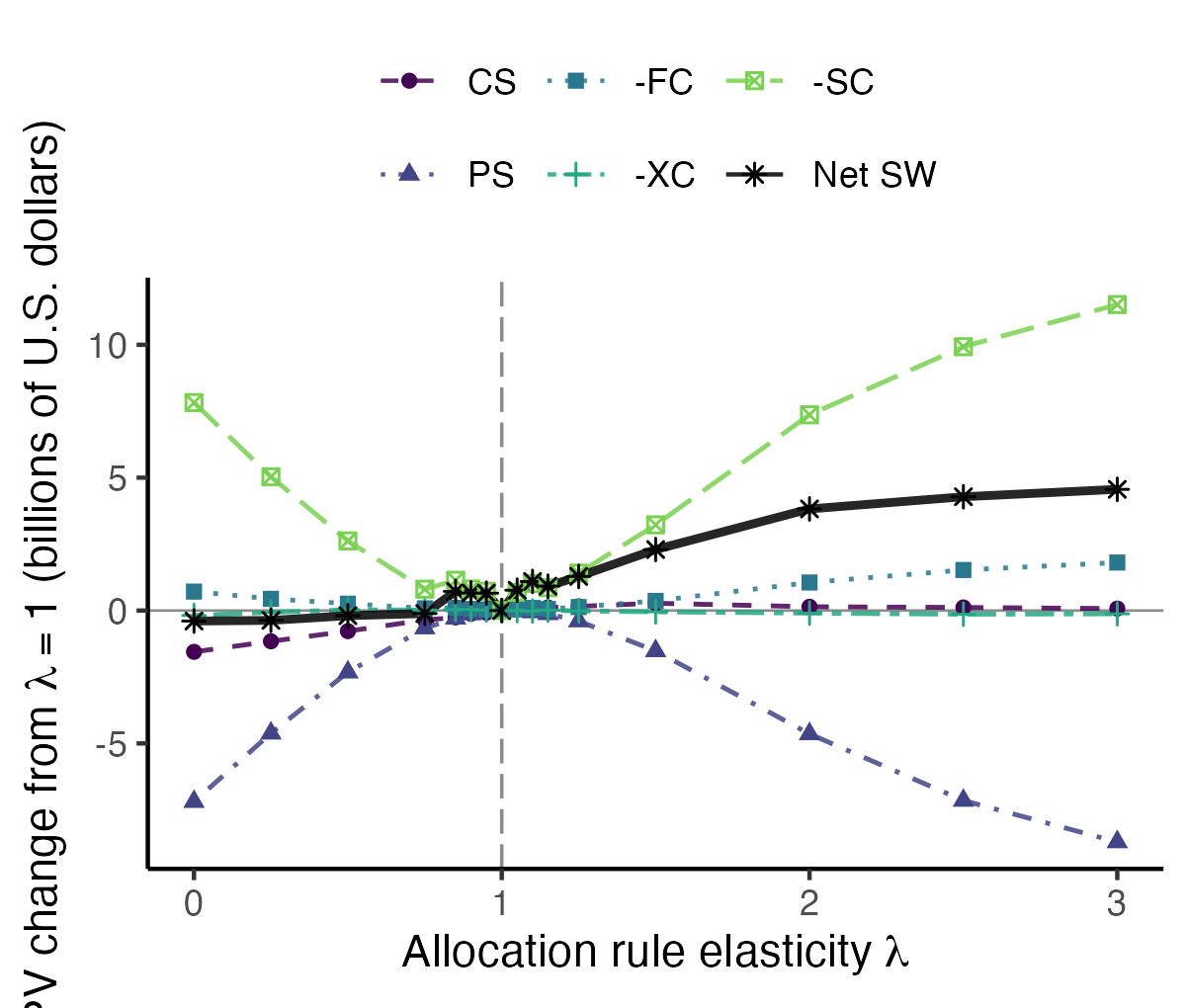}}
  \subfloat[Transatlantic]{\includegraphics[width = 0.33\textwidth]{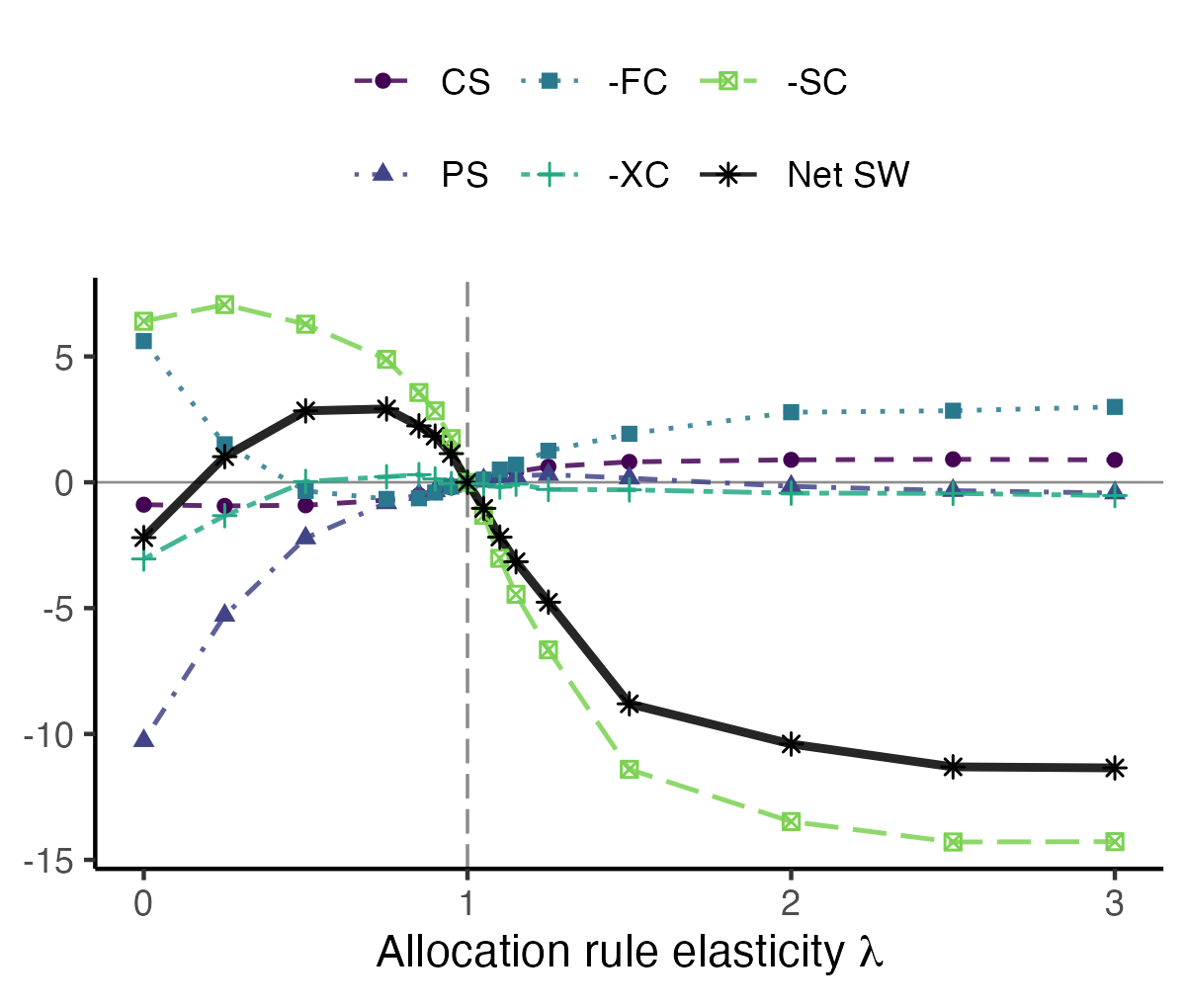}}
  \subfloat[Asia--Europe]{\includegraphics[width = 0.33\textwidth]{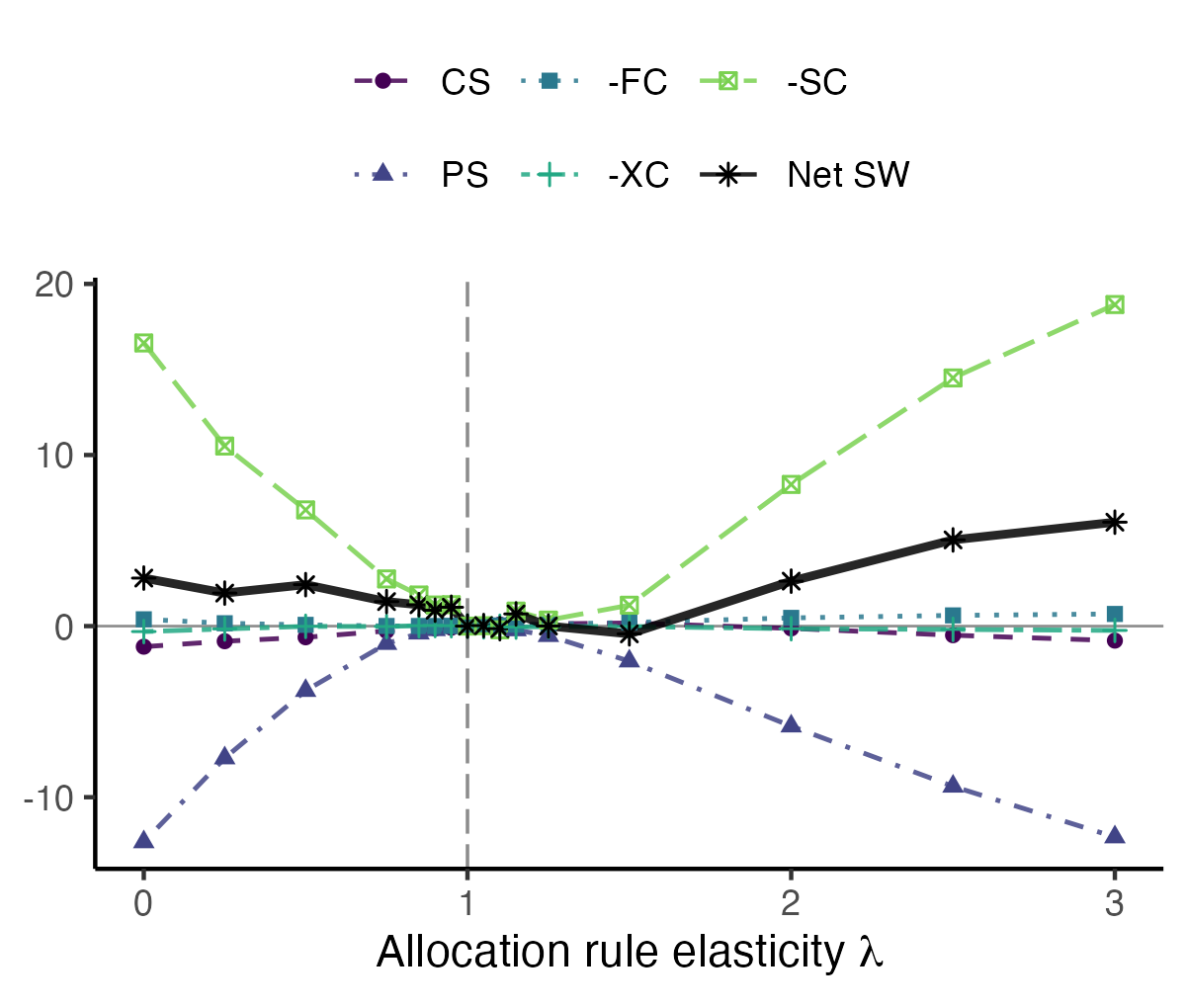}}
  \caption{Decomposition of the welfare effect of the allocation rule $\omega(\lambda)$}
  \label{fg:optimal_allocation_rule_lambda_components}
  \end{center}
  \footnotesize
  Note: Each line reports the change from $\lambda=1$ in the present value of consumer surplus (CS), producer surplus (PS), operating-cost savings ($-FC$), exit-cost savings ($-XC$), entry and shipbuilding-cost savings ($-SC$), and net social welfare. Values are discounted to 1973 over 1973--1983 and measured in billions of U.S. dollars. Appendix \ref{sec:welfare_measurement} defines the welfare measures.
\end{figure}

\textcolor{black}{Figure \ref{fg:optimal_allocation_rule_lambda_components} explains these rankings. The cross-market divergence is driven primarily by entry and shipbuilding costs. A large-carrier tilt raises the relative continuation value of higher capacity; because capacity choices are discrete, this change can either induce costly upgrades or discourage entry and lower-capacity operation. The upgrade margin dominates in the transatlantic market: at $\lambda=1.25$, the mean number of level-4 firms in 1984 rises from 0.66 to 0.94, and entry and shipbuilding costs rise by 26.0\%. At $\lambda=0.75$, by contrast, the level-4 count falls to 0.19 and these costs fall by 19.1\%. The contraction margin dominates in the other markets. In the transpacific market, where the level-3-to-4 build action is unavailable because its cost is not identified, no level-4 investment occurs anywhere on the grid; at the endpoint $\lambda=3$, a stronger large-carrier tilt shifts capacity toward level 3 and reduces operating costs by 9.8\% and entry and shipbuilding costs by 20.9\%, more than offsetting higher exit costs. In Asia--Europe, the same endpoint reduces the numbers of smaller carriers and lowers operating costs by 14.5\% and entry and shipbuilding costs by 27.5\%, again outweighing higher exit costs. Thus, the welfare sign depends on whether the quota tilt triggers costly capacity expansion or reduces the number of firms incurring entry and expansion costs.}

\FloatBarrier

\textcolor{black}{Within the model's conference segment and 1973--1983 horizon,} the counterfactuals show that conference design affected welfare through prices and rent allocation. Price coordination transferred surplus to carriers and encouraged costly capacity expansion, so removing it raised net social welfare despite modest consumer gains. At fixed route price and quantity, alternative quotas changed investment and welfare rankings across markets; minimizing within-period shipping costs did not ensure dynamic efficiency. Antitrust policy should therefore account for how cartel rents and quota rules shape entry and investment, not only current prices.

\section{Conclusion}\label{sec:conclusion}
This paper asks how cartels affect welfare when they shape not only prices but also entry and long-lived investment. In capital-intensive industries, cartel rents may support valuable capacity expansion or induce investment whose resource costs exceed its benefits. I combine historical route and fleet data from early containerization and estimate demand, marginal cost, regime-specific price wedges, and carriers' entry, exit, and shipbuilding decisions. The estimated conference wedge declines by about 45\% in 1980--1983 relative to 1973--1979, linking the weakening of conference control to firms' incentives to enter and expand capacity.

\textcolor{black}{Over 1973--1983 in the modeled conference segment, removing the conference regime raises consumer surplus by about 2--4\% and lowers producer surplus by about 17--27\%, while changing their sum by less than 1\%. After operating, exit, entry, and shipbuilding costs are included, removal raises net social welfare in all three markets, whereas hypothetical full collusion lowers it in all three, most sharply in the transatlantic market, but leaves it positive in every market.} For a fixed route quantity, \textcolor{black}{the capacity-proportional benchmark allocation} equalizes marginal costs across conference members, but alternative quotas change investment incentives and yield different dynamic rankings across markets. In capital-intensive industries, cartel welfare therefore depends not only on prices but also on how rents are allocated and converted into long-lived capacity.

\renewcommand{\refname}{References}
\bibliographystyle{ecca}
\bibliography{ship_cartel}

\newpage
\appendix
\setbibunitsuffix{@appendix}
\begin{bibunit}[ecca]
\noindent{\LARGE{\textbf{Appendix (For Online Publication)} }}
\onehalfspacing

This appendix is prepared for online publication and contains supplementary materials. Table \ref{tb:contents_of_appendix} lists the contents.
\begin{table}[!htbp]
    \centering
    \caption{Contents of the Appendix}
    \label{tb:contents_of_appendix}
    \begin{tabular}{c >{\raggedright\arraybackslash}p{0.26\textwidth} >{\raggedright\arraybackslash}p{0.55\textwidth}}
        Appendix & Contents & Coverage\\\midrule
        \ref{sec:institutional_details} & Institutional details & Conference rules, containerization, and regime changes\\
        \ref{sec:appendix} & Computation and estimation details & Dynamic estimation, algorithm, likelihood-slice ranges, computational time, discretization\\
        \ref{sec:data_details} & Data details & Data construction and merger issues\\
        \ref{sec:model_details} & Model details & Cartel conduct, dynamic setup, welfare measurement, and the capacity-profit link\\
        \ref{sec:data_fitting_and_additional_results}& Additional estimation results& Data fitting, specification and instrument selection, and robustness to anticipating the breakdown\\
    \end{tabular}
\end{table}

\section{Institutional Details}\label{sec:institutional_details}

This section provides the institutional details underlying the concise discussion in Section \ref{subsec:industry_background}.
It first describes how conferences coordinated their members and responded to non-conference carriers, then summarizes the spread of containerization and the two events used to define the empirical regimes.

\subsection{Conference Agreements and Enforcement}

Shipping conferences sought to stabilize liner markets through agreements over rates and service and through measures that limited competition from carriers outside the conference \citep{fox1992empirical,morton1997entry,clyde1998market,podolny1999social,fusillo2003excess,marin2003exclusive}.
Their arrangements covered three broad areas: limiting rate competition among members, retaining shippers within the conference, and responding to non-conference entry.
Rate agreements set and jointly updated tariffs for individual products.
Vessel allocation agreements coordinated tonnage, voyages, ports of call, operating schedules, and cargo allocation.
These arrangements made conferences explicit price- and quantity-coordinating cartels.

The United Nations Convention on a Code of Conduct for Liner Conferences, adopted in 1974 and entering into force in 1983 for its contracting states, set out members' loading and pool-participation rights and principles for dividing conference trade in trades to which it applied \citep{UNLinerCode1974}.
Under Article 2, national shipping lines of the two trading countries generally received equal participation rights, while third-country lines could receive a significant share, such as 20\%.
If one trading country had no participating national line, its share was distributed among the remaining member lines in proportion to their existing shares.
\textcolor{black}{The Code's nationality-based participation principles do not supply the carrier-level service shares required by the empirical model. I therefore approximate each member's unobserved service share by its route-assigned carrying-capacity share, as described in Section \ref{sec:model}.}

Conference tariffs were publicly announced, but non-conference carriers could enter and serve the same trade routes at lower rates.
Conferences used loyalty arrangements to reduce shipper switching.
Under a dual-rate system, a shipper received a lower contract rate in exchange for using only conference carriers during the contract period.
Under a fidelity rebate, a shipper received a rebate after using only conference carriers for a specified period, commonly four to six months.
A deferred rebate added a further waiting period during which the shipper also had to avoid non-conference carriers; the rebate was commonly about 10\% of freight charges.
These arrangements are understood as entry-deterrence devices that supported conference tariffs \citep{fox1992empirical,marin2003exclusive}.

Conferences could also deploy ``fighting ships'' on schedules close to those of a non-conference carrier and charge lower rates until that carrier left the route.
Conference members shared the associated losses.
In the general cartel taxonomy, this practice is a coercive response to non-cartel supply \citep{marshall2014economics,harrington2018rent}.
Enforcement differed across markets: Asia--Europe conferences used stricter membership screening and sometimes pooled freight, whereas transpacific conferences were generally more open. \textcolor{black}{U.S. law did not categorically prohibit freight or traffic pooling; the Shipping Act of 1916 subjected such agreements to regulatory filing and approval \citep[secs.~14--15]{ShippingAct1916}.}

\subsection{Containerization and Market Structure}

International container service began on the transatlantic route in 1966 and on the transpacific route in 1967 with Sea-Land's service.
The first semi-container service on the Asia--Europe route also began in 1967, and full-container service followed in 1971.
Containerization standardized cargo handling and reduced the labor, time, and specialized knowledge required at ports.
At the same time, it made liner shipping more capital intensive because regular container service required substantial investment in ships, containers, and terminals.
The resulting combination of lower operational barriers and higher capital requirements helps explain both the entry of new carriers and the importance of shipbuilding investment in the model.

During the 1970s, developing and socialist countries increasingly promoted national and state-owned shipping lines, many of which initially entered as non-conference carriers.
The Convention sought to respond to demands for access by setting out participation rights for qualifying national lines and the trade-share principles described above in trades to which it applied.
In the 1980s, some Europe-related conferences also invited non-conference carriers to join as their own shares declined.
Containerization additionally encouraged operational consortia through space chartering, joint terminal use, and schedule coordination.
Unlike conferences, these consortia did not jointly set freight rates; their members retained individual control over pricing and sales.

\subsection{\textcolor{black}{Sea-Land's U.S.--Far East Conference Withdrawals in 1980}}

Competition on the transpacific routes intensified in the late 1970s.
\textcolor{black}{Sea-Land placed eight 33-knot SL-7 container ships in service in 1972--73 \citep[p.~157]{FMCReportsVol16}. Their fuel-intensive design became increasingly costly after the oil shocks.}
\textcolor{black}{In early 1980, Sea-Land withdrew from twelve eastbound conferences in the transpacific trade. The contemporaneous FMC report describes a market with increasing container capacity, shrinking eastbound cargo, and stronger independent competition \citep{FMC1980annual}.}
The dual-rate system initially limited its ability to attract conference shippers because switching could trigger a contractual penalty.
Sea-Land therefore focused on import and intermodal cargoes outside the port-to-port coverage of those contracts and absorbed the cost of returning empty containers from inland locations.
\textcolor{black}{I use 1980 as a common empirical boundary across the three markets because conference rates weakened broadly around this period, not because Sea-Land withdrew from every conference or route.}

\subsection{The Shipping Act of 1984}

The Shipping Act of 1984 was enacted on March 20, 1984, and took effect on June 18, 1984 \citep{ShippingAct1984,FMC1985annual}.
It changed conference enforcement through three main provisions.
First, conference agreements had to permit a member to take independent action on a rate or service item after no more than ten days' notice.
Second, the Act expressly authorized service contracts, under which a shipper committed a minimum quantity of cargo over a fixed period in exchange for a specified rate schedule and service level.
The minimum quantity was known as the Minimum Quantity Commitment.
The statutory independent-action right did not initially extend to service contracts, so conferences could still permit, restrict, or prohibit individual members from offering them.
Third, the Act removed antitrust immunity from conference loyalty contracts, effectively ending conference use of the dual-rate system in U.S. trades.
Together, mandatory independent action and the loss of loyalty-contract protection weakened common tariffs and increased rate competition on U.S.-related routes \citep{wilson1991some,JMC2008}.

After the traditional conference system weakened, carriers formed stabilization or discussion agreements \textcolor{black}{under which they exchanged information about demand and capacity and discussed} rate-restoration guidelines and surcharges without directly binding individual rates.
The Transpacific Stabilization Agreement, formed in 1989 by nine conference and four non-conference carriers, is one example.
These arrangements differed from the earlier conference regime because they lacked the same direct control over tariffs.

The U.S. legislation did not directly govern the Asia--Europe routes.
Nevertheless, Figure \ref{fg:log_freight_rate_year_coefficient_path} shows that Asia--Europe conference rates dropped sharply in 1983 and remained substantially below their 1979 level after 1984.
The paper therefore uses 1984 as a common empirical boundary for the competitive regime in all six routes, while interpreting the Shipping Act itself as a direct institutional shock only for the U.S.-related routes.

\section{Computation and Estimation Details}\label{sec:appendix}

\subsection{Dynamic Parameter Estimation}\label{sec:dynamic_estimation_details}

For each capacity level $l\in\{1,2,3,4\}$, let $X_{mt}^{l}$, $K_{mt}^{l}$, and $B_{mt}^{l}$ denote the numbers of incumbents that exit, continue without investment, and build ships. Thus, $N_{mt}^{l}=X_{mt}^{l}+K_{mt}^{l}+B_{mt}^{l}$, with $B_{mt}^{4}=0$. For potential entrants, let $O_{mt}^{pe}$ and $E_{mt}^{pe}$ denote the numbers that stay out and enter, so $N_{mt}^{pe}=O_{mt}^{pe}+E_{mt}^{pe}$.

Define the type-specific CCPs as
\begin{align*}
p_{lmt}^{a}(\theta_{\pi})
&\equiv \Pr(a_{imt}=a\mid s_{imt}=l,s_{mt},D_{mt};\theta_{\pi}),\\
p_{0mt}^{a}(\theta_{\pi})
&\equiv \Pr(a_{imt}=a\mid s_{imt}=0,s_{mt},D_{mt};\theta_{\pi}).
\end{align*}
Type symmetry makes these probabilities common across firms of the same type. \textcolor{black}{For $l\in\{1,2,3\}$, the likelihood contribution for incumbents of level $l$ is}
\begin{align*}
\textcolor{black}{P_{lmt}(X_{mt}^{l},K_{mt}^{l},B_{mt}^{l}\mid s_{mt},D_{mt};\theta_{\pi})}
&\textcolor{black}{=\frac{N_{mt}^{l}!}{X_{mt}^{l}!K_{mt}^{l}!B_{mt}^{l}!}}
\textcolor{black}{\left[p_{lmt}^{x}(\theta_{\pi})\right]^{X_{mt}^{l}}}
\textcolor{black}{\left[p_{lmt}^{k}(\theta_{\pi})\right]^{K_{mt}^{l}}}
\textcolor{black}{\left[p_{lmt}^{b}(\theta_{\pi})\right]^{B_{mt}^{l}}}.
\end{align*}
\textcolor{black}{For level 4, the contribution is the binomial likelihood}
\begin{align*}
\textcolor{black}{P_{4mt}(X_{mt}^{4},K_{mt}^{4}\mid s_{mt},D_{mt};\theta_{\pi})}
&\textcolor{black}{=\frac{N_{mt}^{4}!}{X_{mt}^{4}!K_{mt}^{4}!}}
\textcolor{black}{\left[p_{4mt}^{x}(\theta_{\pi})\right]^{X_{mt}^{4}}}
\textcolor{black}{\left[p_{4mt}^{k}(\theta_{\pi})\right]^{K_{mt}^{4}}}.
\end{align*}
The potential entrants' contribution is
\begin{align*}
P_{0mt}(O_{mt}^{pe},E_{mt}^{pe}\mid s_{mt},D_{mt};\theta_{\pi})
&=\frac{N_{mt}^{pe}!}{O_{mt}^{pe}!E_{mt}^{pe}!}
\left[p_{0mt}^{x}(\theta_{\pi})\right]^{O_{mt}^{pe}}
\left[p_{0mt}^{e}(\theta_{\pi})\right]^{E_{mt}^{pe}}.
\end{align*}
\textcolor{black}{For estimation, I omit the parameter-invariant multinomial and binomial coefficients from these count probabilities. The implemented market-specific log-likelihood is}
\begin{align*}
\textcolor{black}{ll_m(\theta_{\pi})}
={}&\sum_{t=1}^{T-1}\left\{
\textcolor{black}{O_{mt}^{pe}\log p_{0mt}^{x}+E_{mt}^{pe}\log p_{0mt}^{e}}\right.\\
&\quad+\sum_{l=1}^{3}\left(
\textcolor{black}{X_{mt}^{l}\log p_{lmt}^{x}
+K_{mt}^{l}\log p_{lmt}^{k}
+B_{mt}^{l}\log p_{lmt}^{b}}\right)\\
&\left.\quad+\textcolor{black}{X_{mt}^{4}\log p_{4mt}^{x}
+K_{mt}^{4}\log p_{4mt}^{k}}
\right\}.
\end{align*}
\textcolor{black}{The reported log-likelihood values are therefore defined up to data-only additive constants; omitting those constants does not affect the maximizer or likelihood differences.} The likelihood uses decisions observed in $t=1,\ldots,T-1$ (1973--1983). The terminal year $T=12$ (1984) supplies continuation values but no choice observation, consistent with the no-anticipation assumption.

Conditional on the static profit estimates, the entry, exit, continuation, and build frequencies across capacity and market states identify $\theta_{\pi}=(\kappa^e,\psi,\phi,\iota_1,\iota_2)$. I estimate the parameters separately by market to allow the dynamic costs to differ across the transpacific, transatlantic, and Asia--Europe markets. For each candidate $\theta_{\pi}$, the inner loop solves the finite-horizon game and its CCP fixed points; the outer loop maximizes \textcolor{black}{$ll_m(\theta_{\pi})$}. The following subsection gives the solution algorithm, and the subsequent subsections describe inference and computational performance.

\subsection{Algorithm}\label{sec:algorithm}

I adapt the algorithms of \textcolor{black}{\cite{igami2017estimating,igami2018industry}} to this model. For exposition, fix a candidate parameter vector $\theta_{\pi}$ and focus on market $m=1$.
For computation, I index each feasible market state $s_{mt}$ by an integer in $\{1,\ldots,S\}$.
As in the empirical exercise, I fix the number of potential entrants, $N_{mt}^{pe}$, at four, except in Asia--Europe, where it is five.
\textcolor{black}{Write $g(s,a)$ for the deterministic own-state transition: $g(s,x)=0$, $g(s,k)=s$, $g(s,b)=s+1$, and $g(0,e)=1$ for the feasible state-action pairs.}
\textcolor{black}{Let $P_m(\cdot\mid s_{imt},s_{mt},a_{imt})$ denote the induced market-state transition probability conditional on the focal firm's state and action.}
In equilibrium, a fixed point in CCPs determines the corresponding ex-ante value functions.
\textcolor{black}{Suppressing the deterministic demand-state argument for compactness,} the terminal values are initialized as $\mathcal V_{imT}\textcolor{black}{(s_{imT},s_{mT})}=\pi_{imT}\textcolor{black}{(s_{imT},s_{mT})}/(1-\beta)$, where $\pi_{imT}$ is the regime-continuation terminal profit consistent with the no-anticipation assumption in Section \ref{sec:model}.
The algorithm to compute ex-ante value functions via backward induction proceeds as follows.

\paragraph{Algorithm 1: Value Functions and CCPs}
\begin{enumerate}
    \item For $t=T-1,\ldots,1$, take the ex-ante value function \textcolor{black}{$\mathcal V_{imt+1}(s_{imt+1},s_{mt+1})$} as given and evaluate the exit, operating, and investment costs under $\theta_{\pi}$.
    \begin{enumerate}
        \item For each of the $S$ feasible values of $s_{mt}$, obtain the number of incumbent firms $N_{mt}$.
        \item \textcolor{black}{For each $i\in\mathcal N_{mt}$, initialize $\Pr(a_{imt}=a\mid s_{imt},s_{mt})$ for all $a\in\mathcal A(s_{imt})$ and the action-specific continuation expectations for $a\in\mathcal A(s_{imt})\setminus\{x\}$.}
        \item \textcolor{black}{For each $i\in\mathcal N_{mt}^{pe}$, initialize $\Pr(a_{imt}=a\mid s_{imt}=0,s_{mt})$ for all $a\in\mathcal A(0)$ and the continuation expectation for entry.}
        \item Find the CCP fixed point as follows.
        \begin{enumerate}[(i)]
            \item \textcolor{black}{For level $l=4$ and for all $a\in\mathcal{A}(l)$ and $i\in\mathcal{N}_{mt}^{l}$,}
            \begin{enumerate}

                \item Given the previous expectation and CCPs, together with $N_{mt}$ and $N_{mt}^{pe}$, compute the joint action-count likelihood defined in Section \ref{sec:dynamic_estimation_details} for all feasible exit, continuation, build, and entry counts.
                \item Compute \textcolor{black}{$P_m(s_{mt+1}\mid s_{imt},s_{mt},a_{imt}=a)$ for each own action $a$ and} all $S$ feasible next-period states from the joint likelihood and the incumbent and potential-entrant counts\textcolor{black}{, removing firm $i$ from its type's rival count and adding its own destination level deterministically}.
                \item \textcolor{black}{Given the joint likelihood, transition probabilities, and $\mathcal V_{imt+1}(g(s_{imt},a),s_{mt+1})$, compute}
                \begin{align*}
                &E\!\left[\textcolor{black}{\mathcal V_{imt+1}(g(s_{imt},a),s_{mt+1})}\right.\\
                &\left.\qquad\mid \textcolor{black}{s_{imt},s_{mt},a_{imt}=a}\right]\\
                &\quad=\sum_{s_{mt+1}=1}^{S}
                \textcolor{black}{P_m(s_{mt+1}\mid s_{imt},s_{mt},a_{imt}=a)}\\
                &\qquad\quad\times
                \textcolor{black}{\mathcal V_{imt+1}(g(s_{imt},a),s_{mt+1})},\\
                &\hspace{4em}\textcolor{black}{\text{for each }a\in\mathcal{A}(s_{imt})\setminus\{x\}}.
                \end{align*}
                \item Given the updated expectation, compute new CCPs. I use the log-sum-exp form to prevent numerical overflow.
                \item Compute the incumbent CCP gap as the sum of the absolute differences between the new and old choice probabilities.
                \item Replace each old incumbent CCP with the average of its new and old values.
                \item Set $l=l-1$ and repeat these steps until $l=1$.
            \end{enumerate}
            \item For potential entrants,
            \begin{enumerate}
                \item \textcolor{black}{Given the previous expectations and CCPs, together with $N_{mt}$ and $N_{mt}^{pe}$, compute the joint action-count likelihood for the other players.}
                \item \textcolor{black}{Compute $P_m(s_{mt+1}\mid s_{imt}=0,s_{mt},a_{imt}=a)$ for each $a\in\mathcal A(0)$, removing potential entrant $i$ from the rival count and adding its own destination state deterministically.}
                \item \textcolor{black}{Given the joint likelihood, transition probability, and $\mathcal V_{imt+1}(g(0,a),s_{mt+1})$, update the action-specific continuation expectation for entry using the old CCPs; staying out has no continuation value.}
                \item Given the updated expectation, compute new CCPs using the log-sum-exp form.
                \item Compute the potential-entrant CCP gap as the sum of the absolute differences between the new and old choice probabilities.

                \item Replace each old potential-entrant CCP with the average of its new and old values.
            \end{enumerate}
            \item Check whether $\text{(incumbent CCP gap)}+\text{(potential-entrant CCP gap)}<0.01$. If so, retain the new CCPs as the fixed point; otherwise, return to step (i).
            \item Given the fixed-point CCPs and updated expectations, compute the ex-ante value function \textcolor{black}{$\mathcal V_{imt}(s_{imt},s_{mt})$}.
        \end{enumerate}
    \end{enumerate}
\end{enumerate}

I nest Algorithm 1 as the inner loop and maximize the \textcolor{black}{market-specific log-likelihood $ll_m(\theta_{\pi})$} in the outer loop. The outer loop uses Nelder--Mead from five economically distinct starting values, followed by a polishing restart from the best point. Multiple starts address saturated likelihood plateaus, such as regions in which all incumbents exit with probability close to one. For the fixed point in step 1(d), I use a damped policy-function iteration similar to \cite{kasahara2012sequential}, with a one-half damping weight, a 0.01 tolerance for the summed CCP gap, and a maximum of 15 iterations.
I compute expectations over rivals' actions by sequentially convolving the capacity-level multinomial distributions of exit and investment counts with the binomial distribution of entries, rather than enumerating all joint action profiles. Unit tests confirm that the two methods agree to machine precision, while factorization reduces each likelihood evaluation to a few seconds.
\textcolor{black}{To limit the state space and computational burden while preserving each market's observed minimum presence in each capacity class, I treat the componentwise sample minima as fixed background carriers. These carriers enter the static capacity and profit calculations but do not make dynamic entry, exit, or investment choices; the dynamic state records carrier counts above this fixed background.}

\subsection{\textcolor{black}{Calculation of conditional likelihood-slice ranges}}

\textcolor{black}{I report conditional one-parameter likelihood-slice ranges rather than profile-likelihood confidence intervals.} Let $\hat{\theta}_{\pi}$ denote the maximum likelihood estimate and let $\tilde{\theta}_{\pi j}(u)$ equal $\hat{\theta}_{\pi}$ except that parameter $j$ is set to $\hat{\theta}_{\pi j}e^u$. For each \textcolor{black}{estimated parameter}, I evaluate the likelihood on an eleven-point grid with $u$ equally spaced on $[-1,1]$, holding all other parameters at their maximum likelihood values. \textcolor{black}{The range retains grid values} satisfying $2\{\textcolor{black}{ll_m(\hat{\theta}_{\pi})}-\textcolor{black}{ll_m(\tilde{\theta}_{\pi j}(u))}\}\le \chi^{2}_{1,0.90}=2.706$. \textcolor{black}{Because the nuisance parameters are held at their maximum likelihood values rather than re-optimized, the chi-squared cutoff serves as a descriptive likelihood threshold and the ranges do not claim nominal 90\% coverage.} I do not use Wald intervals because the likelihood has step-like regions, as in \textcolor{black}{\cite{igami2017estimating,igami2018industry,igami2020mergers}}.
When the initial grid yields a degenerate or boundary interval, the grid is recomputed with a narrower (0.35) or wider (2.5) log half-width and thirteen points.
\textcolor{black}{As in \cite{igami2017estimating}, the reported ranges understate overall uncertainty because they do not incorporate the estimation error of the first-stage static parameters.}

\subsection{Computational time}
\textcolor{black}{The main computational burden comes from Algorithm 1, in particular, finding a CCP fixed point at every feasible market state and year for each candidate parameter vector.} With the factorized computation of rivals' action distributions described above, one evaluation of the joint log-likelihood takes a few seconds per market on a laptop.
Using eight Julia threads, the multistart estimation of $\theta_{\pi}$ takes about two minutes for the transatlantic market, ten minutes for the transpacific market, and twenty-two minutes for the Asia--Europe market, whose state space is the largest.
Computing the conditional likelihood-slice ranges adds several minutes per market. Simulating 1,000 equilibrium paths takes less than a minute per market.

\subsection{Discretization details}

\begin{figure}[!htbp]
  \begin{center}
  \subfloat[Transpacific: price]{\includegraphics[width = 0.25\textwidth]
  {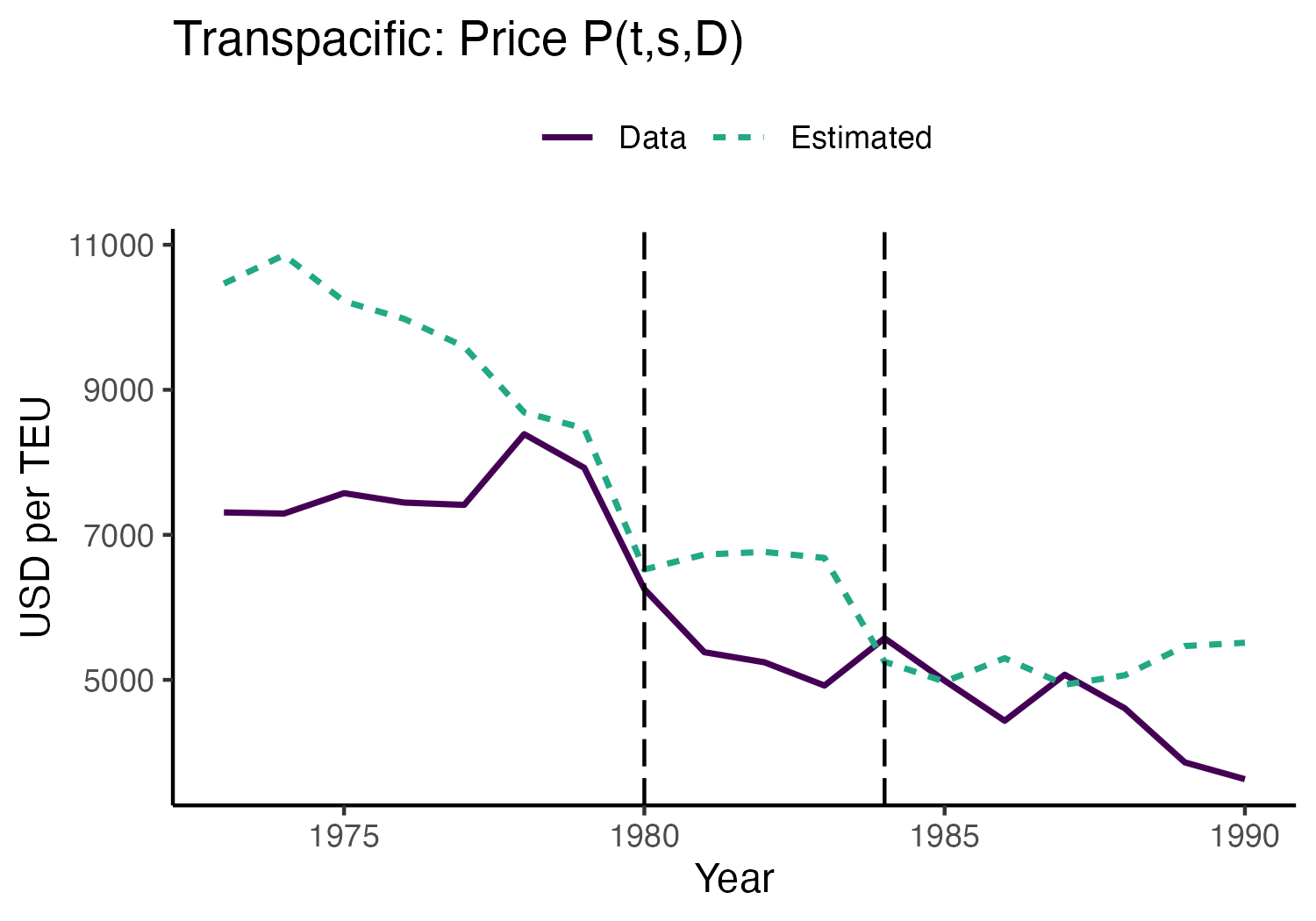}}
  \subfloat[Transatlantic: price]{\includegraphics[width = 0.25\textwidth]
  {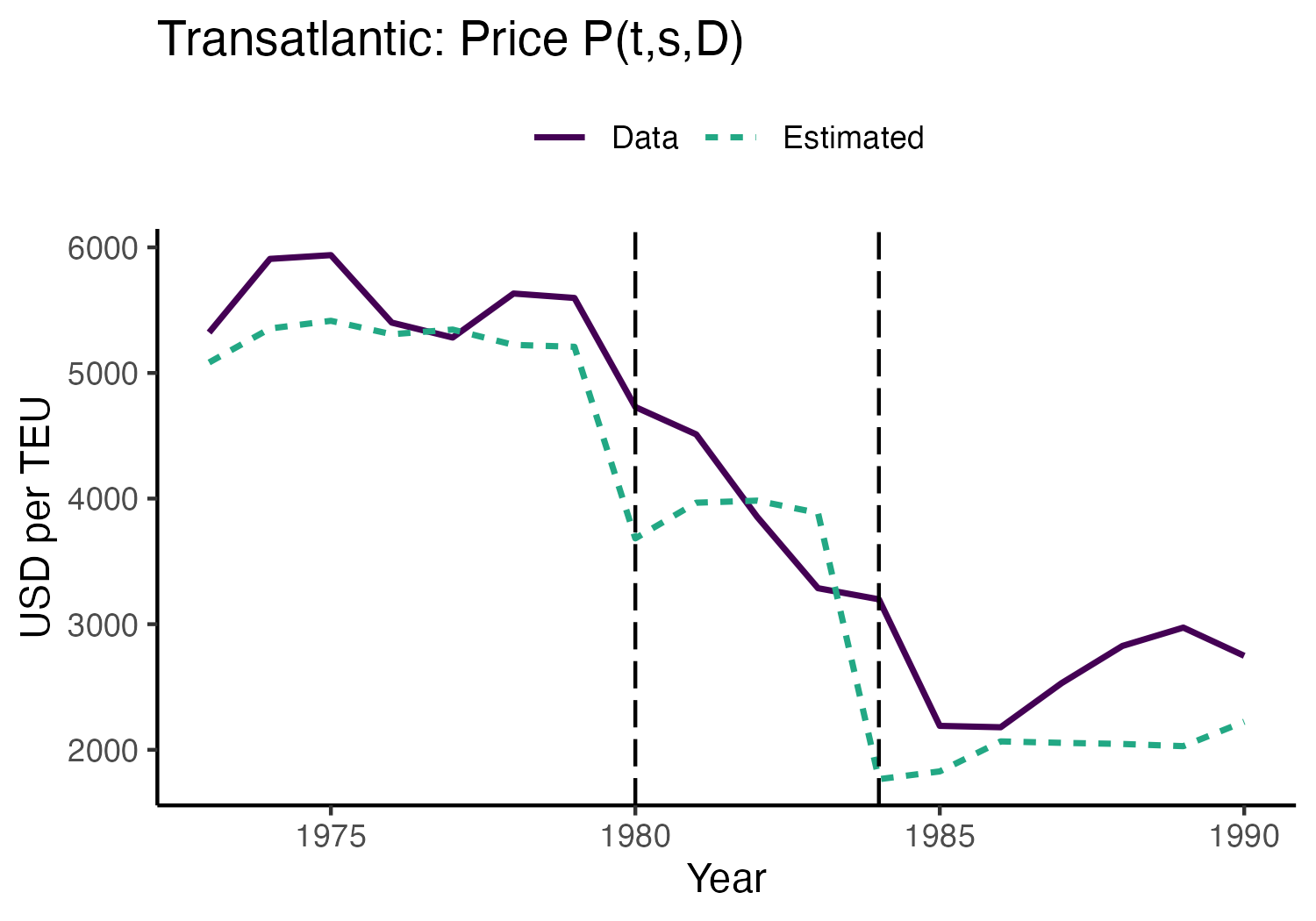}}
  \subfloat[Asia--Europe: price]{\includegraphics[width = 0.25\textwidth]
  {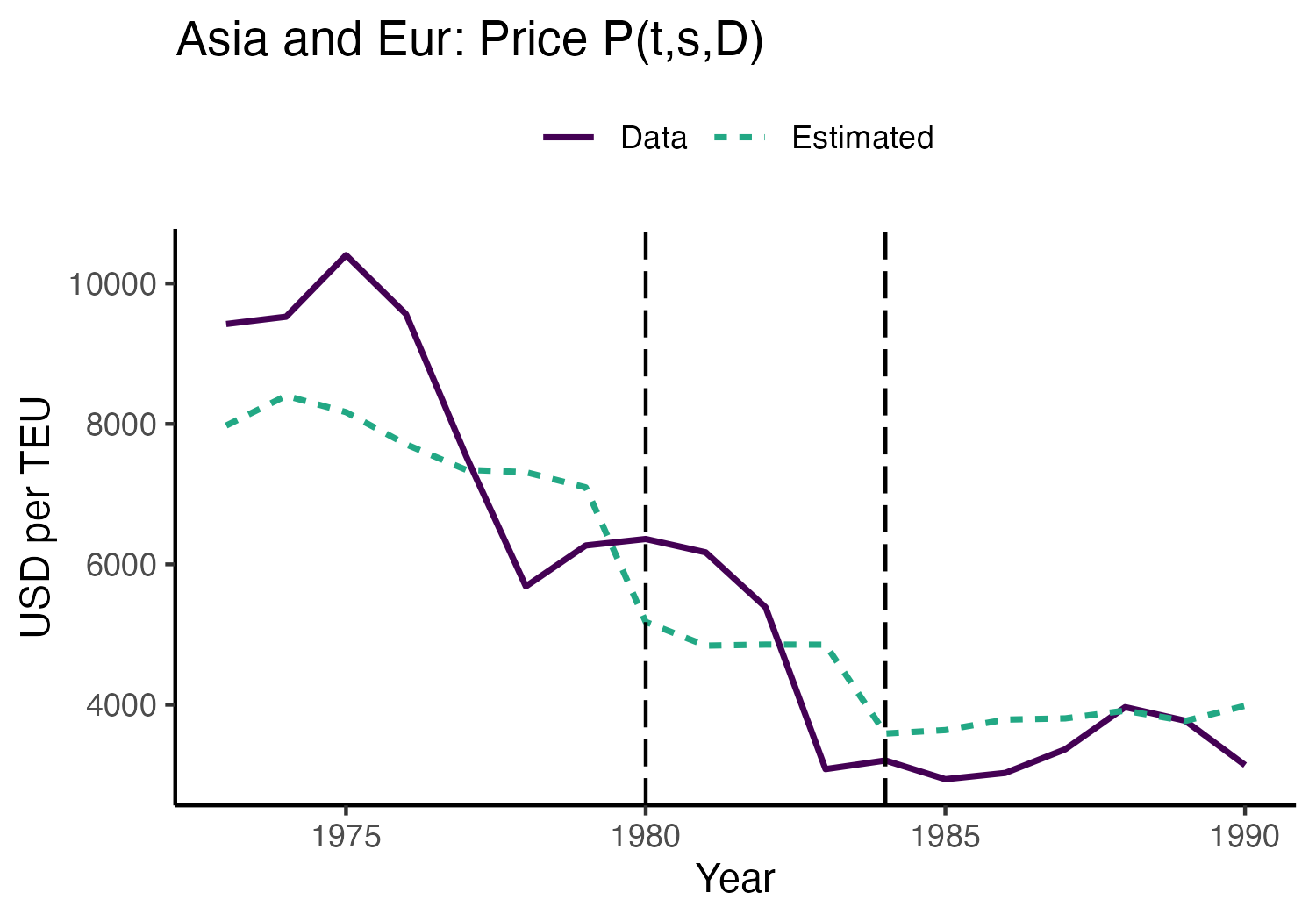}}\\
  \subfloat[Transpacific: quantity]{\includegraphics[width = 0.25\textwidth]
  {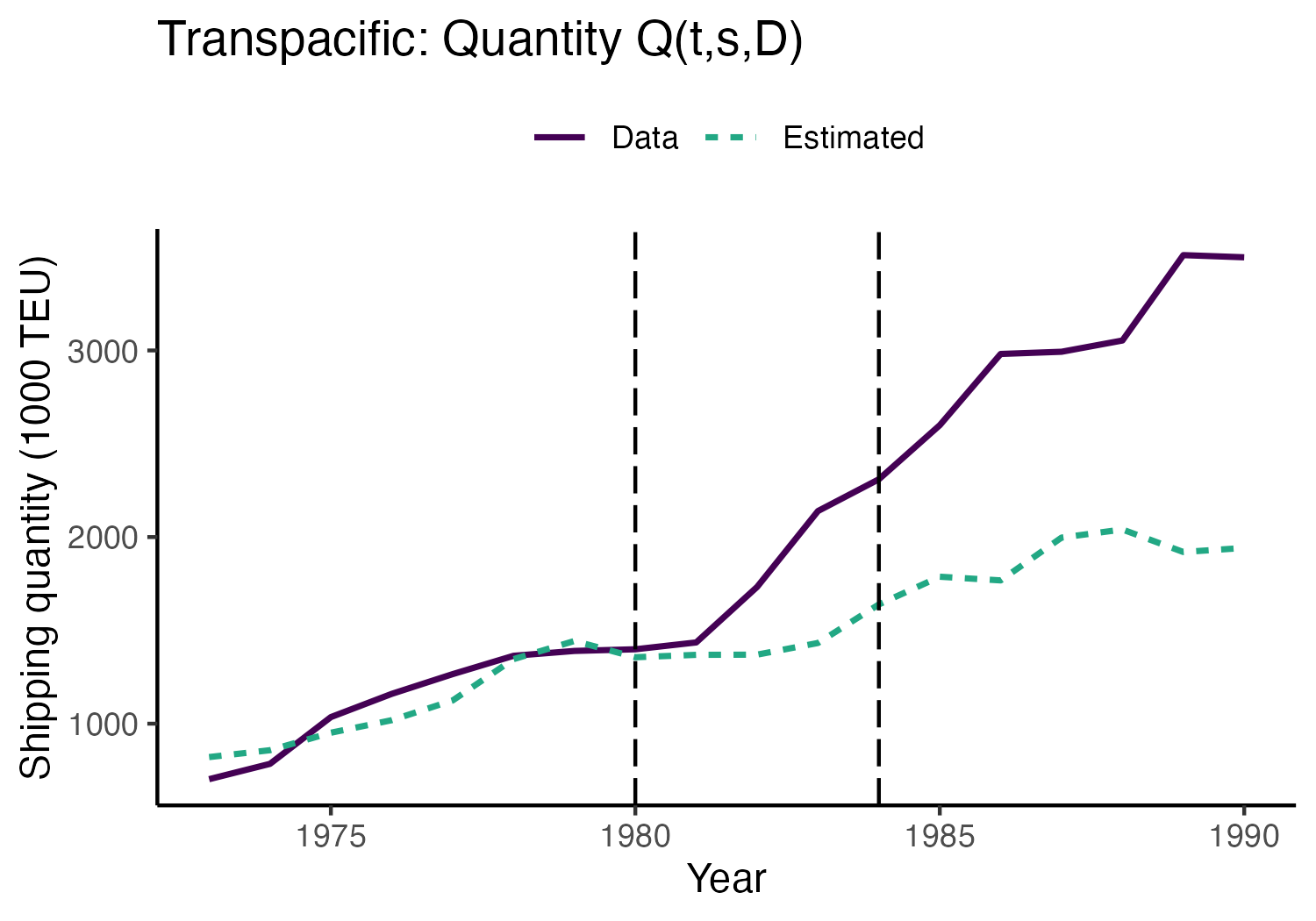}}
  \subfloat[Transatlantic: quantity]{\includegraphics[width = 0.25\textwidth]
  {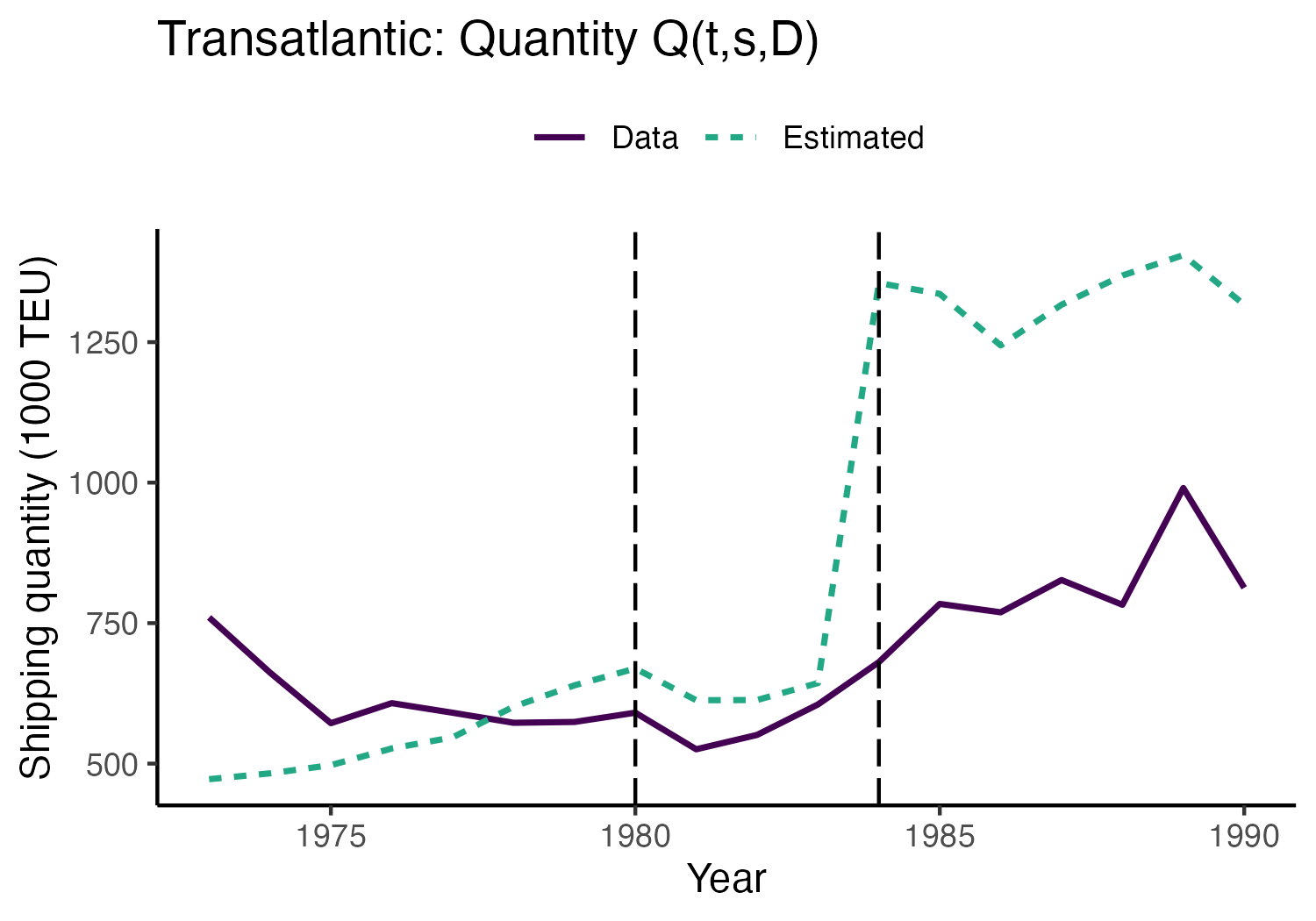}}
  \subfloat[Asia--Europe: quantity]{\includegraphics[width = 0.25\textwidth]
  {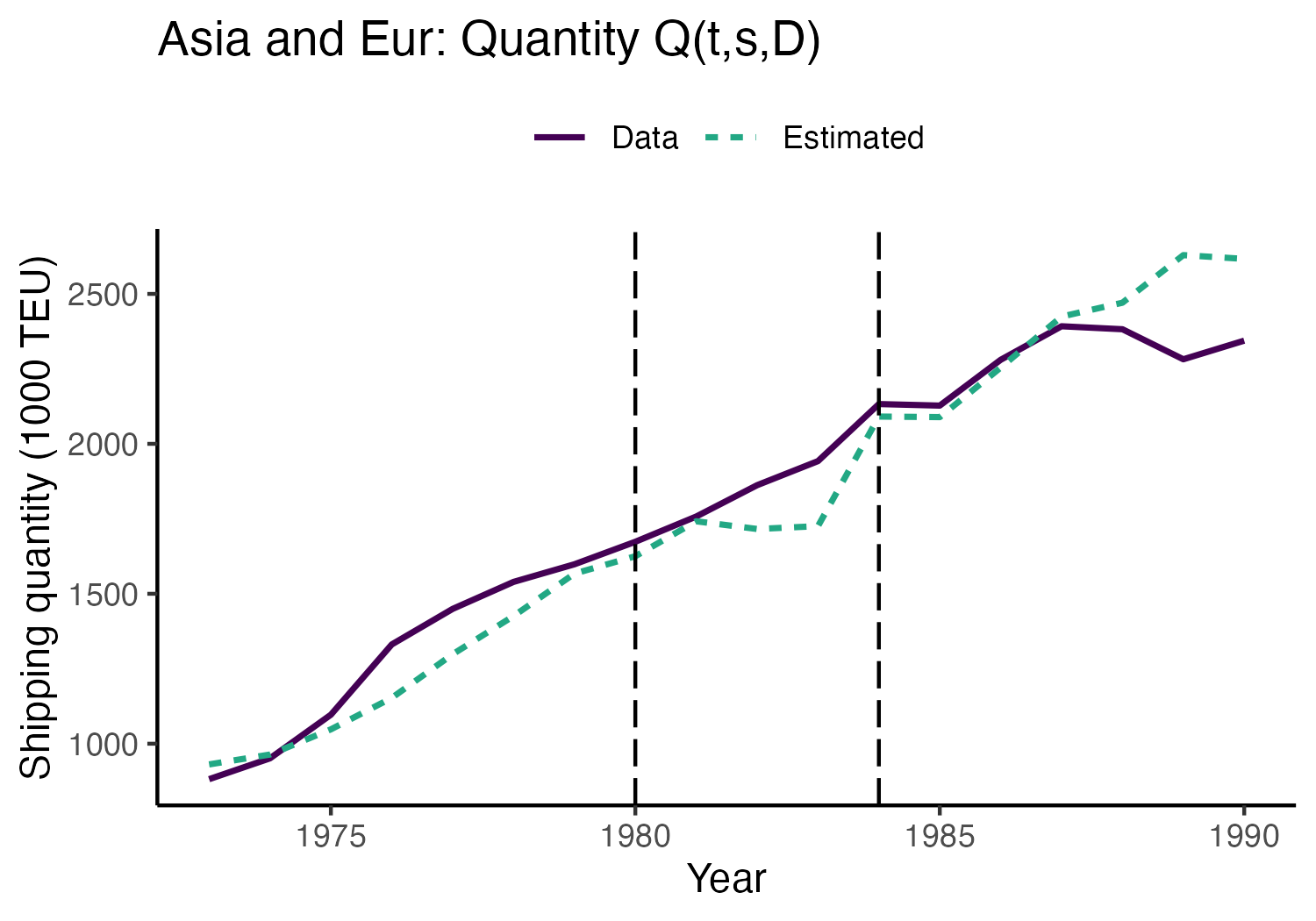}}\\
  \subfloat[Transpacific: total tonnage]{\includegraphics[width = 0.25\textwidth]
  {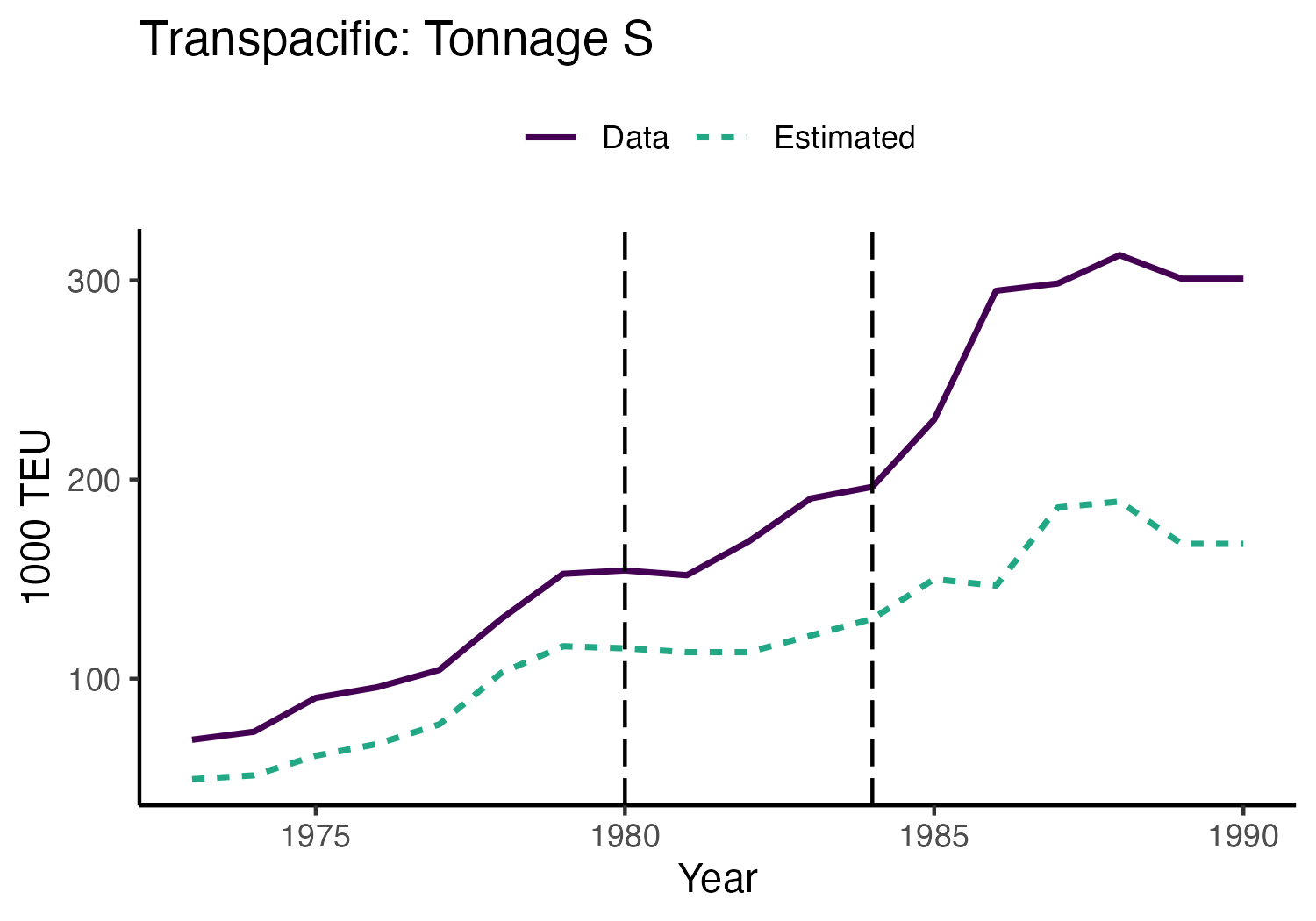}}
  \subfloat[Transatlantic: total tonnage]{\includegraphics[width = 0.25\textwidth]
  {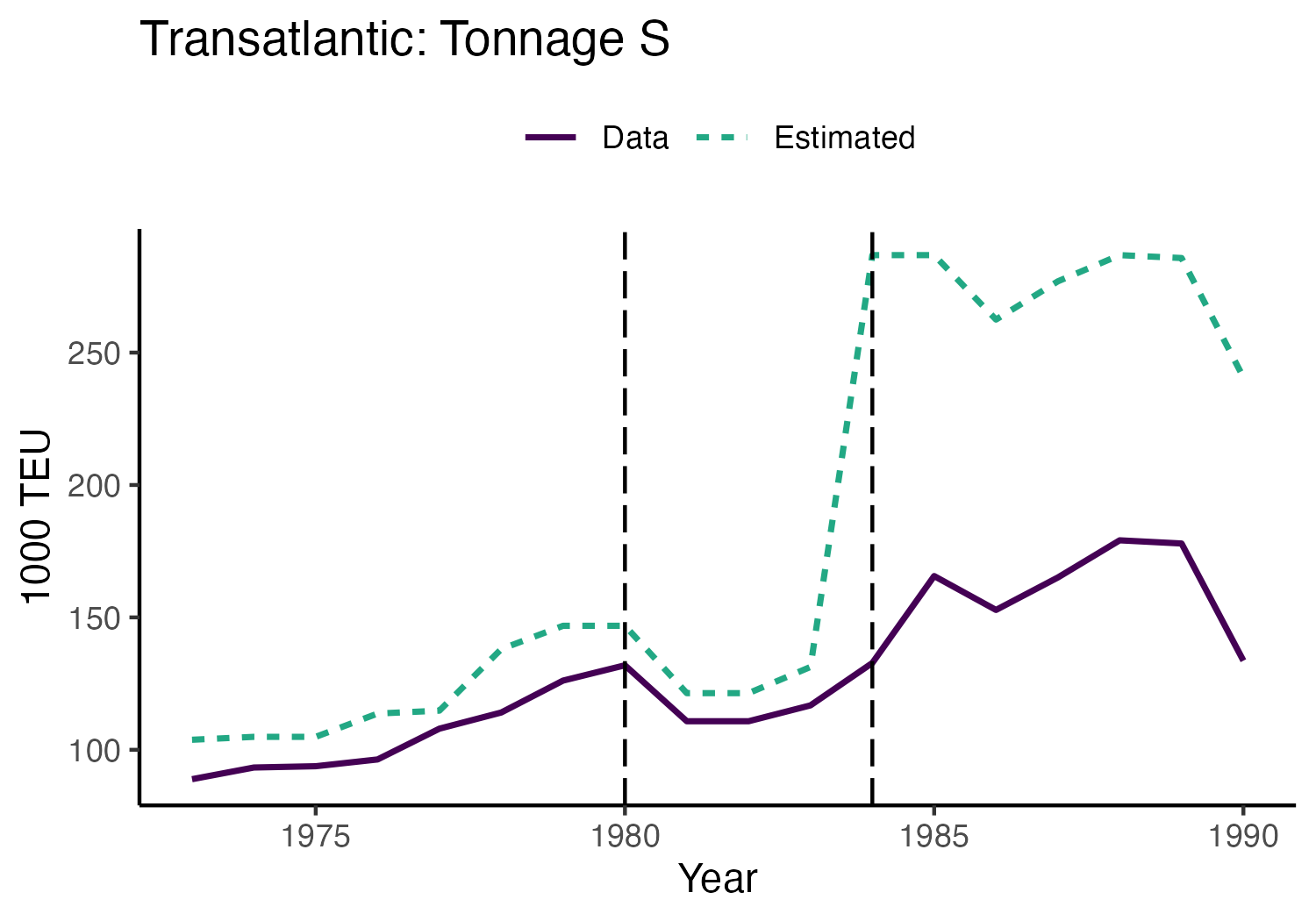}}
  \subfloat[Asia--Europe: total tonnage]{\includegraphics[width = 0.25\textwidth]
  {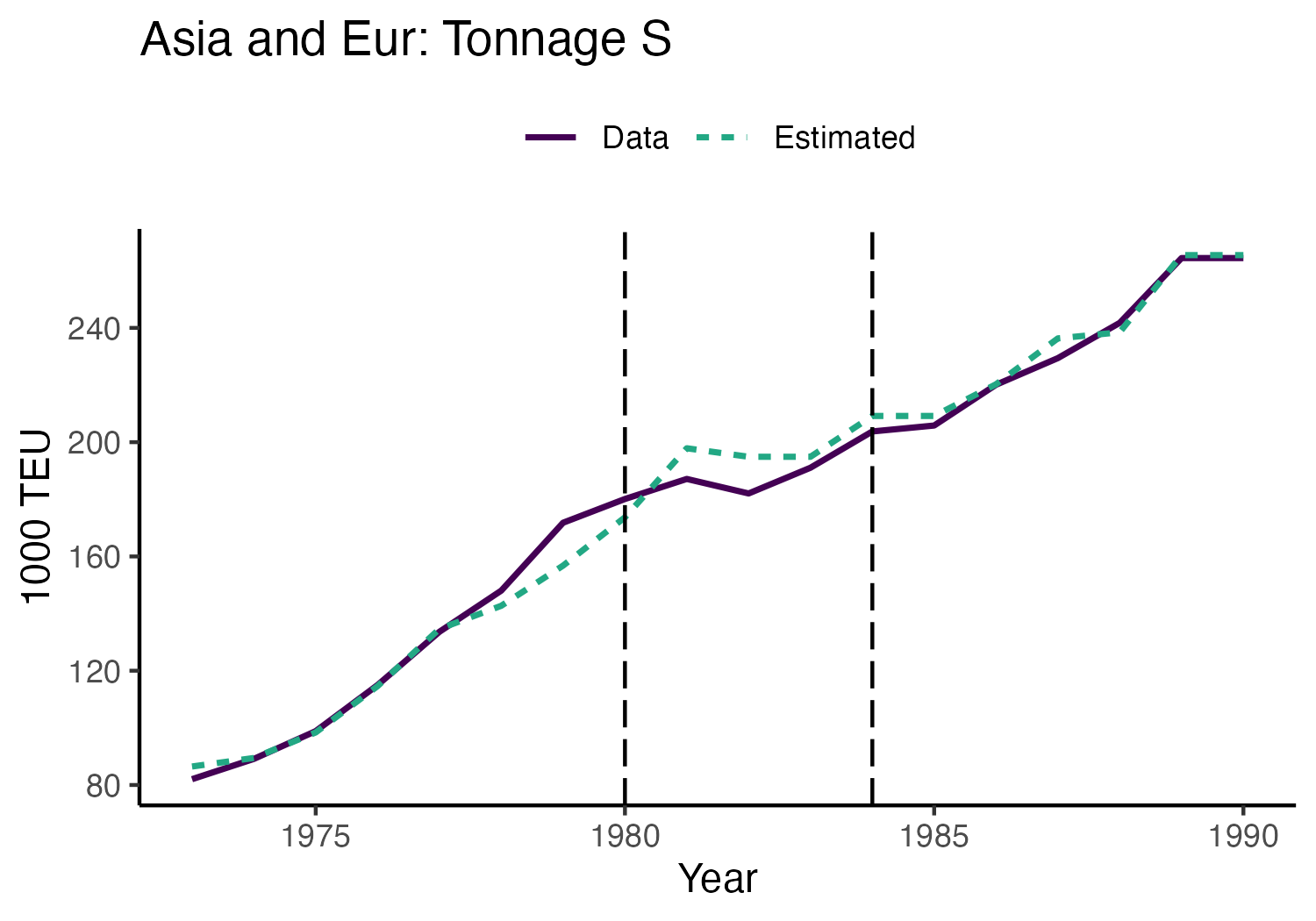}}\\
  \subfloat[Transpacific: firm profit]{\includegraphics[width = 0.25\textwidth]
  {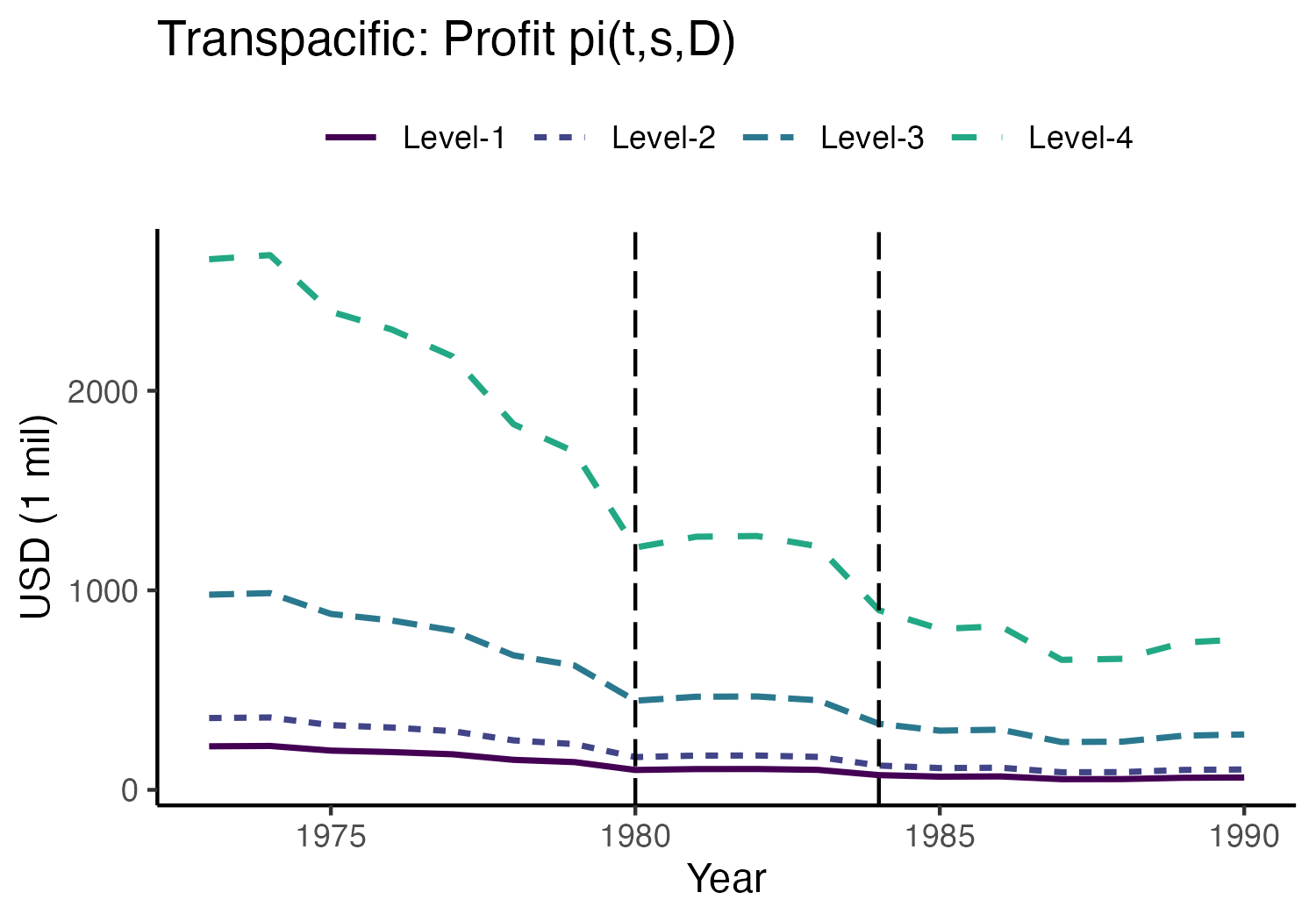}}
  \subfloat[Transatlantic: firm profit]{\includegraphics[width = 0.25\textwidth]
  {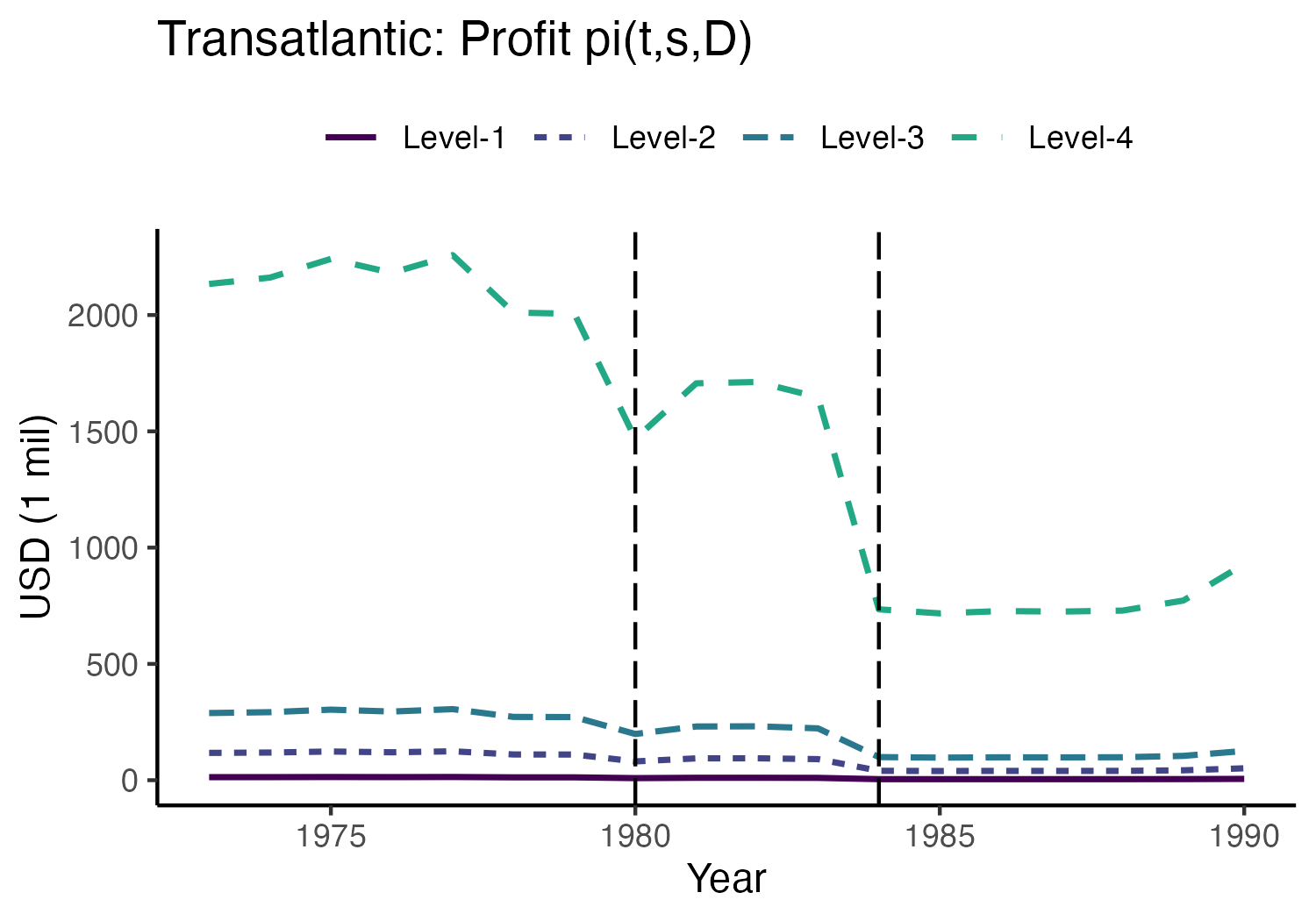}}
  \subfloat[Asia--Europe: firm profit]{\includegraphics[width = 0.25\textwidth]
  {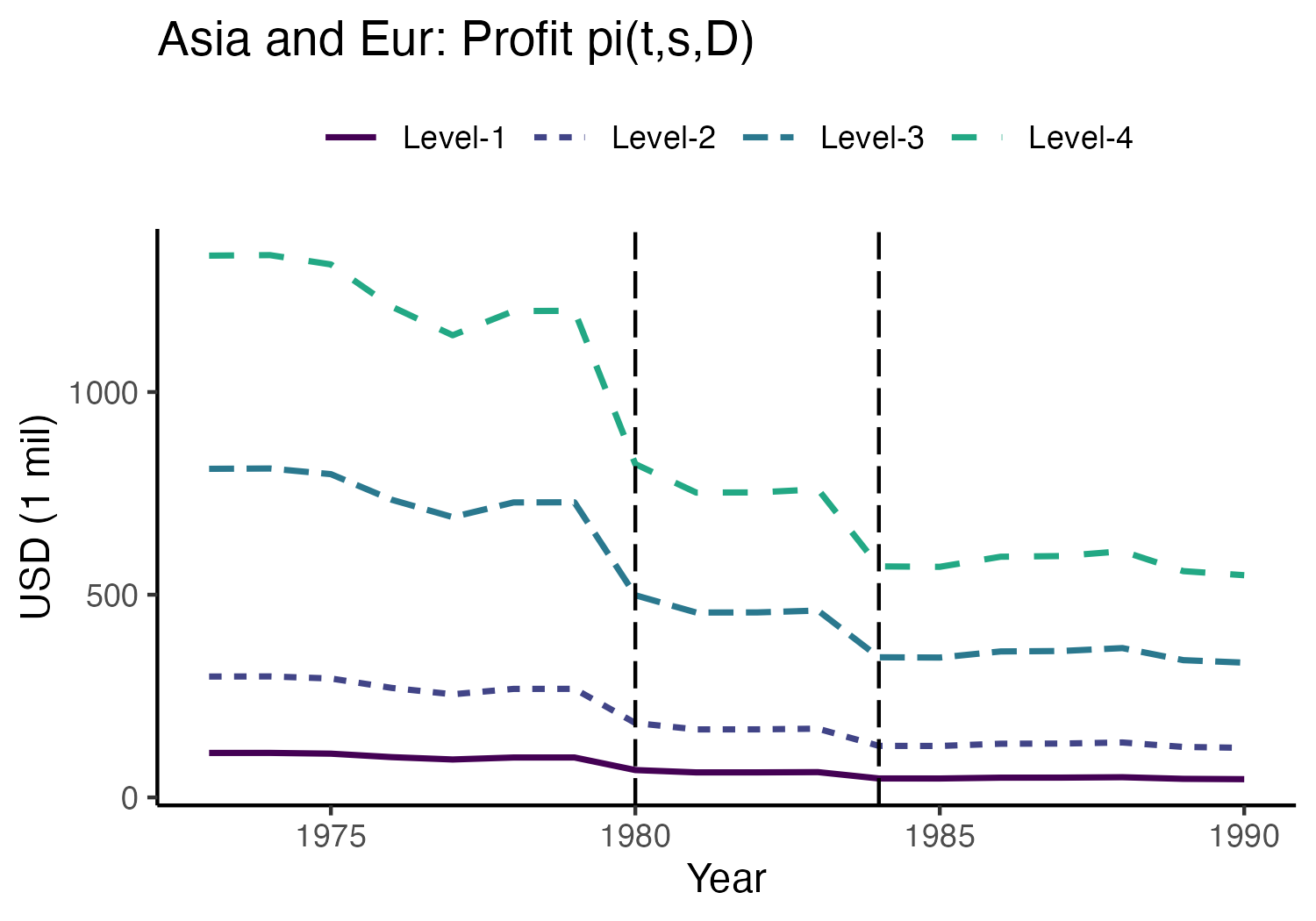}}\\
  \caption{Estimated market-year prices, quantities, total tonnage, and firm profits by capacity level}
  \label{fg:market_level_estimated_and_actual_profit_asia_and_eur}
  \end{center}
  \footnotesize
  Note: Each outcome is evaluated at the discretized state corresponding to the observed market state. Price and quantity are sums over the two directed routes in a market, whereas total tonnage counts the market's capacity once.
\end{figure}

\textcolor{black}{To compute static equilibrium outcomes from the discrete capacity states, I let $\bar s_{imt}$ denote the market-specific representative TEU capacity assigned to state $s_{imt}$ and specify its logarithm below.}
The state thresholds are common across markets, but representative capacities may differ because the capacity distributions are heterogeneous.
\textcolor{black}{I selected the following representative log capacities manually, before dynamic estimation, to approximate the levels and time paths of equilibrium prices, quantities, and total tonnage; no formal objective function or sampling weights were used, and the values are held fixed throughout estimation and the counterfactuals. Figure \ref{fg:market_level_estimated_and_actual_profit_asia_and_eur} shows that the approximation is weaker in the transpacific market, especially for prices.}
For the Asia--Europe market,
\begin{align*}
    \log \textcolor{black}{\bar{s}_{imt}}= \begin{cases}
        10.5 \quad &\text{if } s_{imt} = 4\\
        10.0 \quad &\text{if } s_{imt} = 3\\
        9.0  \quad &\text{if } s_{imt}=2\\
        8.0 \quad &\text{if } s_{imt}=1.
    \end{cases}
\end{align*}
For the transpacific market,
\begin{align*}
    \log \textcolor{black}{\bar{s}_{imt}}= \begin{cases}
        10.5 \quad &\text{if } s_{imt} = 4\\
        9.5 \quad &\text{if } s_{imt} = 3\\
        8.5  \quad &\text{if } s_{imt}=2\\
        8.0 \quad &\text{if } s_{imt}=1.
    \end{cases}
\end{align*}
For the transatlantic market,
\begin{align*}
    \log \textcolor{black}{\bar{s}_{imt}}= \begin{cases}
        12.1 \quad &\text{if } s_{imt} = 4\\
        10.1 \quad &\text{if } s_{imt} = 3\\
        9.2  \quad &\text{if } s_{imt}=2\\
        7.0 \quad &\text{if } s_{imt}=1.
    \end{cases}
\end{align*}

Figure \ref{fg:market_level_estimated_and_actual_profit_asia_and_eur} compares equilibrium outcomes evaluated at the discretized observed states with the data. The model captures the broad patterns, with some lags and deviations due to discretization.

\section{Data Details}\label{sec:data_details}

\subsection{Route-Year Prices and Quantities}

The route-year price and quantity series follow the construction in \cite{matsuda2022unified}.
For each route-year, total quantity is decomposed as $Q_{rt}^{whole}=Q_{rt}+Q_{rt}^{non}$, where $Q_{rt}$ and $Q_{rt}^{non}$ are conference and non-conference quantities.
The two directional route quantities sum to market quantity $Q_{mt}^{whole}$.
I recover conference and non-conference quantities by applying their capacity shares to total quantity, which assumes that utilization does not differ systematically between the two segments within a market-year.
Historical industry utilization rates were 80--95\%, and I found no evidence of systematically lower utilization in either segment.
The model treats $Q_{rt}^{non}$ as exogenous and does not model customers' choices between conference and non-conference services.
I found no continuous non-conference freight-rate series in the \textit{Containerisation International Yearbook} or \textit{The Japan Maritime Daily}.
Practitioner interviews suggest that non-conference rates were often 20--30\% below conference rates, but they do not provide a consistent time series.

\subsection{Ship-Level Records and Firm Histories}

The ship-level records assign each vessel a main route.
When a vessel is reported on multiple main routes and its capacity cannot be allocated across them, I classify it as ``Multiple routes'' and omit it from the market-level analysis.
I construct the entry, exit, and conference-status histories from several archival sources \citep{NYTIMES1981,NYTIMES1983,DOT1992conference,kaiun_ni_okeru2004,VANHAM2012} and the Port of Long Beach timeline (\url{https://polb.com/port-info/timeline/}).
Firm shipment shares are proxied by capacity shares, which assumes that utilization does not differ systematically across firms within a market-year.

\subsection{Mergers}

Industry-level merger patterns between 1966 and 2022 are documented by \cite{otani2025unified}; \cite{otani2021estimating} studies the 1964 consolidation of Japanese shipping firms before global containerization.
The sample contains two mergers in 1986, three in 1988, and two in 1990. Because the dynamic likelihood uses firm decisions only through 1983, these mergers do not enter the dynamic estimation. In the descriptive post-1984 firm histories, I code an acquired firm as exiting. I code an acquirer as investing if its capacity moves to a higher level in the following year.

\section{Model Details}\label{sec:model_details}

\subsection{Cartel-Effect Specification and Alternative Conduct Models}

The main model and \cite{porter1983study}, who studies the Joint Executive Committee railroad cartel, differ in how they represent collusive conduct. The main model does not specify a static strategic game: it recovers competitive marginal cost from the 1984--1990 regime and represents earlier conference pricing with regime-specific wedges. By contrast, \cite{porter1983study} embeds a conduct parameter in a quantity-setting game. Both approaches imply that price equals marginal cost under competition, but they interpret collusive prices differently.

To make the comparison explicit, consider the constant-elasticity demand equation \eqref{eq:log_demand_model}. Under Porter's marginal-cost specification, \textcolor{black}{$mc_{irt}(q_{irt})=\nu a_{ir}q_{irt}^{\nu-1}$, where $a_{ir}$ is a firm-specific cost shifter and $\nu>1$}. The firm-level first-order condition can be written as
\begin{align*}
P_{rt}\left(1+\frac{\theta_{irt}}{\alpha_1}\right)
=mc_{irt}(q_{irt}),
\end{align*}
where the conduct parameter $\theta_{irt}$ equals zero under competition, one under joint-profit maximization, and $q_{irt}/Q_{rt}$ under Cournot competition \citep{bresnahan1982oligopoly}. Weighting these conditions by quantity shares gives
\begin{align*}
P_{rt}\left(1+\frac{\theta_{rt}}{\alpha_1}\right)
&=\sum_{i\in\mathcal N_{rt}}\frac{q_{irt}}{Q_{rt}}mc_{irt}(q_{irt}),\\
\theta_{rt}
&=\sum_{i\in\mathcal N_{rt}}\frac{q_{irt}}{Q_{rt}}\theta_{irt}.
\end{align*}
\textcolor{black}{Conditional on a fixed active set and a conduct factor common across firms---as under competition or joint-profit maximization---}Porter's cost specification implies output shares that are constant over time and invariant to changes in conduct\textcolor{black}{; the formula does not generally apply under Cournot conduct, where $\theta_{irt}=q_{irt}/Q_{rt}$ varies across firms}:
\begin{align*}
\frac{q_{irt}}{Q_{rt}}
=\textcolor{black}{\frac{a_{ir}^{1/(1-\nu)}}
{\sum_{j\in\mathcal N_{rt}}a_{jr}^{1/(1-\nu)}}}.
\end{align*}

Under the marginal-cost specification in \eqref{eq:individual_marginal_cost}, and using the capacity-proportional allocation $q_{irt}/Q_{rt}=s_{irt}/S_{rt}$, the corresponding aggregate condition is
\begin{align*}
P_{rt}\left(1+\frac{\theta_{rt}}{\alpha_1}\right)
&=\sum_{i\in\mathcal N_{rt}}\frac{s_{irt}}{S_{rt}}
\left(c_{rt}+\gamma_1\frac{q_{irt}}{s_{irt}}\right)\\
&=c_{rt}+\gamma_1\frac{Q_{rt}}{S_{rt}}.
\end{align*}
When $\theta_{rt}=0$, this condition coincides with the competitive supply equation in the main model. A conduct interpretation of collusive periods would additionally require conferences to choose total quantity under a specified static game. The implied markup, $\theta_{rt}P_{rt}/(-\alpha_1)$, varies with price; the main model instead represents the observed collusive price as competitive marginal cost plus an additive regime-specific wedge.

I use the wedge specification for four reasons. First, fixed liner schedules limit conferences' ability to adjust service quantity to a static joint-profit target. Second, conferences used prices to compete with non-conference carriers, whose prices are unobserved and whose strategic interaction lies outside the model. Third, capacity, entry, exit, and investment changed substantially over the sample, making a stable conduct index and time-invariant member output shares restrictive. The capacity term $s_{irt}$ instead links current marginal cost directly to the dynamic investment state. Finally, the short route-year panel does not support a flexible time-varying conduct parameter \citep{matsumura2023test,matsumura2023revisiting,matsumura2023mpec}. Following \cite{igami2015market}, I recover marginal cost from a competitive regime and measure collusive markups as the difference between observed collusive prices and recovered competitive marginal cost, without specifying the cartel's static game.

\subsection{Dynamic State, Payoffs, and Timing}\label{sec:dynamic_setup_details}

This subsection provides the full setup summarized in Section \ref{sec:dynamics}. Let $\tilde{s}_{imt}\in\mathbb{R}_{+}$ denote firm $i$'s actual tonnage in market $m$ and year $t$. The capacity state used in estimation is
\begin{align*}
s_{imt}= \begin{cases}
1 &\textcolor{black}{\text{if } \tilde{s}_{imt}< \exp(8.5)},\\
2 &\textcolor{black}{\text{if } \exp(8.5) \le \tilde{s}_{imt} < \exp(9.5)},\\
3 &\textcolor{black}{\text{if } \exp(9.5) \le \tilde{s}_{imt} < \exp(10.5)},\\
4 &\textcolor{black}{\text{if } \exp(10.5) \le \tilde{s}_{imt}}.
\end{cases}
\end{align*}
Potential entrants have the null state $s_{imt}=0$. The market state is
\begin{align*}
s_{mt}=\left(N_{mt}^{1},N_{mt}^{2},N_{mt}^{3},N_{mt}^{4}\right),
\qquad
N_{mt}^{l}=\sum_{i\in\mathcal{N}_{mt}}1(s_{imt}=l),
\end{align*}
and $\mathcal{N}_{mt}^{l}$ denotes the set of level-$l$ incumbents. The model has four potential entrants in each market-year, except in Asia--Europe, where it has five to accommodate the maximum number of entries observed in a year.

The action-specific private payoff shocks are
\begin{align*}
\varepsilon_{imt}=
\begin{cases}
(\varepsilon_{imt}^{a})_{\textcolor{black}{a\in\mathcal{A}(s_{imt})}} & \text{if } i\in\mathcal{N}_{mt},\\
(\varepsilon_{imt}^{a})_{\textcolor{black}{a\in\mathcal{A}(0)}} & \text{if } i\in\mathcal{N}_{mt}^{pe}.
\end{cases}
\end{align*}
The components are private information and are independently distributed type-I extreme-value shocks across actions, firms, and periods. Thus, each firm observes $(t,s_{imt},s_{mt},D_{mt},\varepsilon_{imt})$ at its turn, whereas the researcher observes all components except $\varepsilon_{imt}$.

For an incumbent or potential entrant, the individual capacity transition is
\begin{align*}
s_{imt+1}=\begin{cases}
0 & \text{if }s_{imt}\in\{1,2,3,4\},\ a_{imt}=x,\\
s_{imt} & \text{if }s_{imt}\in\{1,2,3,4\},\ a_{imt}=k,\\
s_{imt}+1 & \text{if }s_{imt}\in\{1,2,3\},\ a_{imt}=b,\\
1 & \text{if }s_{imt}=0,\ a_{imt}=e,\\
0 & \text{if }s_{imt}=0,\ a_{imt}=x.
\end{cases}
\end{align*}
The build action is unavailable to level-4 firms. Entry therefore begins at level 1, and the induced market-state transitions are
\begin{align*}
N_{mt+1}^{1}&=N_{mt}^{1}+\sum_{i\in\mathcal{N}_{mt}^{pe}}1(a_{imt}=e)
-\sum_{i\in\mathcal{N}_{mt}^{1}}1(a_{imt}=b)-\sum_{i\in\mathcal{N}_{mt}^{1}}1(a_{imt}=x),\\
N_{mt+1}^{2}&=N_{mt}^{2}+\sum_{i\in\mathcal{N}_{mt}^{1}}1(a_{imt}=b)
-\sum_{i\in\mathcal{N}_{mt}^{2}}1(a_{imt}=b)-\sum_{i\in\mathcal{N}_{mt}^{2}}1(a_{imt}=x),\\
N_{mt+1}^{3}&=N_{mt}^{3}+\sum_{i\in\mathcal{N}_{mt}^{2}}1(a_{imt}=b)
-\sum_{i\in\mathcal{N}_{mt}^{3}}1(a_{imt}=b)-\sum_{i\in\mathcal{N}_{mt}^{3}}1(a_{imt}=x),\\
N_{mt+1}^{4}&=N_{mt}^{4}+\sum_{i\in\mathcal{N}_{mt}^{3}}1(a_{imt}=b)
-\sum_{i\in\mathcal{N}_{mt}^{4}}1(a_{imt}=x).
\end{align*}
The demand state $D_{mt}$ follows the path estimated in the static model. The econometric residuals $\zeta_{rt}$ and $\eta_{rt}$ from the static estimation do not enter the dynamic state or payoff function; $\varepsilon_{imt}$ contains the only private shocks in the dynamic game. Because these shocks are i.i.d., the model satisfies conditional independence: conditional on the current public state and action, next period's capacity does not depend on the current private shock \citep{rust1987optimal}.

For incumbent $i\in\mathcal{N}_{mt}$ and action $a\in\mathcal{A}(s_{imt})$, the action-specific per-period payoff is
\begin{align*}
\pi_{imt}^{a}(s_{imt},s_{mt},D_{mt},\varepsilon_{imt}^{a})
&=\pi_{imt}(s_{imt},s_{mt},D_{mt})-\psi 1(a=x)\\
&\quad-\phi 1(a=k)
-\left[\phi+I(s_{imt};\iota_1,\iota_2)\right]1(a=b)
+\varepsilon_{imt}^{a}.
\end{align*}
Thus, the current static profit is common to all incumbent choices. For potential entrant $i\in\mathcal{N}_{mt}^{pe}$ and action $a\in\mathcal{A}^{pe}$,
\begin{align*}
\pi_{imt}^{a}(s_{imt},s_{mt},D_{mt},\varepsilon_{imt}^{a})
=-\kappa^{e}1(a=e)+\varepsilon_{imt}^{a}.
\end{align*}

The within-year game proceeds as follows. Static market outcomes first determine the profits of all current incumbents. Level-4 incumbents then draw their private payoff shocks and move simultaneously. Level-3, level-2, and level-1 incumbents move in that order; each group observes the actions of earlier groups, draws its shocks, and moves simultaneously. Potential entrants move last after observing all incumbent actions. The capacity and market states are then updated. This deterministic order follows \cite{igami2017estimating}; annual data do not reveal the shorter-period stochastic order used by \cite{igami2020mergers}. At the terminal date,
\begin{align*}
\mathcal V_{imT}\textcolor{black}{(s_{imT},s_{mT},D_{mT})}
=\frac{\pi_{imT}\textcolor{black}{(s_{imT},s_{mT},D_{mT})}}{1-\beta},
\end{align*}
where $\pi_{imT}$ is evaluated under firms' belief that the 1980--1983 regime continues.

\subsection{Dynamic Optimization and Equilibrium Details}\label{sec:dynamic_optimization_details}

This subsection expands the dynamic programming problem in Section \ref{sec:dynamic_optimization}. For incumbents at levels $l\in\{1,2,3\}$, the choice-specific values are
\begin{align*}
\bar V_{imt}^{x}&=\pi_{imt}(l,s_{mt},D_{mt})-\psi+\varepsilon_{imt}^{x},\\
\bar V_{imt}^{k}&=\pi_{imt}(l,s_{mt},D_{mt})-\phi+\varepsilon_{imt}^{k}
+\beta E\!\left[\mathcal V_{imt+1}(l,s_{mt+1},D_{mt+1})\mid l,s_{mt},a_{imt}=k\right],\\
\bar V_{imt}^{b}&=\pi_{imt}(l,s_{mt},D_{mt})-\phi-I(l;\iota_1,\iota_2)+\varepsilon_{imt}^{b}
+\beta E\!\left[\mathcal V_{imt+1}(l+1,s_{mt+1},D_{mt+1})\mid l,s_{mt},a_{imt}=b\right].
\end{align*}
Level-4 incumbents have the first two values but no build option. For a potential entrant,
\begin{align*}
\bar V_{imt}^{x}&=\varepsilon_{imt}^{x},\\
\bar V_{imt}^{e}&=-\kappa^e+\varepsilon_{imt}^{e}
+\beta E\!\left[\mathcal V_{imt+1}(1,s_{mt+1},D_{mt+1})\mid 0,s_{mt},a_{imt}=e\right].
\end{align*}
All values condition on the actions of groups that have already moved and integrate over the strategies and private shocks of later movers. This conditioning is suppressed in the notation.

To expand the expectation in equation \eqref{eq:representative_csvf}, let $h_{imt}$ collect the actions observed before firm $i$ moves, let $g(s_{imt},a_{imt})$ denote its deterministic capacity transition, and \textcolor{black}{let $P_m(\cdot\mid s_{imt},s_{mt},h_{imt},a_{imt})$ denote the induced transition probability for the market state}. Conditional independence gives
\begin{align*}
&E\!\left[\mathcal V_{imt+1}(s_{imt+1},s_{mt+1},D_{mt+1})
\mid s_{imt},s_{mt},h_{imt},a_{imt}\right]\\
&\quad=\sum_{s_{mt+1}}
\textcolor{black}{P_m(s_{mt+1}\mid s_{imt},s_{mt},h_{imt},a_{imt})}
\mathcal V_{imt+1}(g(s_{imt},a_{imt}),s_{mt+1},D_{mt+1}).
\end{align*}
For example, a level-4 incumbent that continues has $g(4,k)=4$, so the expression reduces to
\begin{align*}
\sum_{s_{mt+1}}\textcolor{black}{P_m(s_{mt+1}\mid s_{imt}=4,s_{mt},h_{imt},a_{imt}=k)}
\mathcal V_{imt+1}(4,s_{mt+1},D_{mt+1}).
\end{align*}

The generic integrated value in equation \eqref{eq:integrated_value_generic} specializes to sums over $(x,k,b)$ for levels 1--3, $(x,k)$ for level 4, and $(x,e)$ for potential entrants. \textcolor{black}{Under type symmetry, let $\widetilde V_{lmt}^{a}$ denote the common deterministic choice-specific value for a level-$l$ incumbent and $\widetilde V_{0mt}^{a}$ the corresponding value for a potential entrant.} Equation \eqref{eq:ccp_generic} then implies
\begin{align*}
p_{lmt}^{a}
\equiv\Pr(a_{imt}=a\mid s_{imt}=l,s_{mt},D_{mt})
&=\frac{\exp(\widetilde V_{lmt}^{a}/\sigma)}
{\sum_{a'\in\mathcal A(l)}\exp(\widetilde V_{lmt}^{a'}/\sigma)},
&&a\in\mathcal A(l),\\
p_{0mt}^{a}
\equiv\Pr(a_{imt}=a\mid s_{imt}=0,s_{mt},D_{mt})
&=\frac{\exp(\widetilde V_{0mt}^{a}/\sigma)}
{\sum_{a'\in\{x,e\}}\exp(\widetilde V_{0mt}^{a'}/\sigma)},
&&a\in\{x,e\}.
\end{align*}
Type symmetry makes these probabilities independent of firm identity within a capacity level.

I next construct the market-state transition from these CCPs. Let $(X_{mt}^{l},K_{mt}^{l},B_{mt}^{l})$ be the numbers of level-$l$ incumbents that exit, continue, and build, respectively, with $N_{mt}^{l}=X_{mt}^{l}+K_{mt}^{l}+B_{mt}^{l}$ and $B_{mt}^{4}=0$. Let $(O_{mt}^{pe},E_{mt}^{pe})$ be the numbers of potential entrants that stay out and enter. \textcolor{black}{Write $X=(X_{mt}^{l})_{l=1}^{4}$, $K=(K_{mt}^{l})_{l=1}^{4}$, and $B=(B_{mt}^{l})_{l=1}^{4}$. Each probability below is a sequential conditional probability evaluated at the realized actions of groups that moved earlier; the corresponding histories are suppressed.} Conditional on the public state, type symmetry implies the multinomial probabilities
\begin{align*}
P_{lmt}(X_{mt}^{l},K_{mt}^{l},B_{mt}^{l}\mid s_{mt})
&=\frac{N_{mt}^{l}!}{X_{mt}^{l}!K_{mt}^{l}!B_{mt}^{l}!}
[p_{lmt}^{x}]^{X_{mt}^{l}}[p_{lmt}^{k}]^{K_{mt}^{l}}[p_{lmt}^{b}]^{B_{mt}^{l}},\\
P_{0mt}(O_{mt}^{pe},E_{mt}^{pe}\mid s_{mt})
&=\frac{N_{mt}^{pe}!}{O_{mt}^{pe}!E_{mt}^{pe}!}
[p_{0mt}^{x}]^{O_{mt}^{pe}}[p_{0mt}^{e}]^{E_{mt}^{pe}}.
\end{align*}
\textcolor{black}{The first expression applies to $l\in\{1,2,3\}$; for level 4 it is the corresponding binomial over $(x,k)$.} The realized action counts map into the next state according to
\begin{align*}
\Psi(s_{mt},X,K,B,O^{pe},E^{pe})
=\left(K_{mt}^{1}+E_{mt}^{pe},\ K_{mt}^{2}+B_{mt}^{1},\ K_{mt}^{3}+B_{mt}^{2},\ K_{mt}^{4}+B_{mt}^{3}\right).
\end{align*}
Therefore,
\begin{align*}
P_m(s_{mt+1}\mid s_{mt})
=\sum_{(X,K,B,O^{pe},E^{pe}):\,\Psi(\cdot)=s_{mt+1}}
P_{0mt}(O_{mt}^{pe},E_{mt}^{pe}\mid s_{mt})
\prod_{l=1}^{4}P_{lmt}(X_{mt}^{l},K_{mt}^{l},B_{mt}^{l}\mid s_{mt}).
\end{align*}
At each within-year turn, actions already observed are held fixed and the sum is taken only over firms that have not yet moved.
When computing firm $i$'s continuation value, \textcolor{black}{its own state and chosen action are also held fixed, which gives $P_m(s_{mt+1}\mid s_{imt},s_{mt},h_{imt},a_{imt})$ above}.

Formally, a strategy maps the public state, a firm's private shocks, and earlier movers' actions to a feasible action. A perfect Bayesian equilibrium requires sequential optimality at every information set and beliefs consistent with these strategies and Bayes' rule. Type symmetry restricts firms at the same capacity level to use the same strategy. As in \cite{igami2017estimating}, private shocks affect rivals only through observed choices, capacity types move sequentially, and the finite horizon permits recursive solution by backward induction. Continuous private shocks make ties occur with probability zero.

\subsection{Welfare Measurement}\label{sec:welfare_measurement}

This subsection defines the welfare measures used in Tables \ref{tb:welfare_benchmark}--\ref{tb:welfare_counterfactual_inner_allocation_rule_1} and Figures \ref{fg:optimal_allocation_rule_lambda}--\ref{fg:optimal_allocation_rule_lambda_components}. Given the estimated demand elasticity $\widehat{\alpha}_1<-1$, consumer surplus, producer surplus, and social welfare in market $m$ and year $t$ are
\begin{align}
CS_{mt}
&=\sum_{r\in\textcolor{black}{\mathcal R_m}}\frac{P_{rt}^{*}Q_{rt}^{*}}
{|1+\widehat{\alpha}_1|},\label{eq:welfare_cs}\\
PS_{mt}
&=\sum_{i\in\mathcal N_{mt}}\pi_{imt},\label{eq:welfare_ps}\\
SW_{mt}
&=CS_{mt}+PS_{mt}.\label{eq:welfare_sw}
\end{align}
Let $N_{mt}=\sum_{l=1}^{4}N_{mt}^{l}$ and $X_{mt}=\sum_{l=1}^{4}X_{mt}^{l}$ denote the numbers of current and exiting incumbents. Define operating costs, exit costs, and entry and shipbuilding costs, respectively, as
\begin{align}
FC_{mt}&=\phi(N_{mt}-X_{mt}),\label{eq:welfare_fc}\\
XC_{mt}&=\psi X_{mt},\label{eq:welfare_xc}\\
SC_{mt}&=\kappa^{e}E_{mt}^{pe}
+\sum_{l=1}^{3}I(l;\iota_1,\iota_2)B_{mt}^{l}.\label{eq:welfare_sc}
\end{align}
Thus, continuing and building incumbents pay the operating cost, exiting incumbents pay the exit cost, entrants pay the entry cost, and builders pay the capacity-expansion cost. Net social welfare is
\begin{align}
\operatorname{NetSW}_{mt}
=SW_{mt}-FC_{mt}-XC_{mt}-SC_{mt}.\label{eq:welfare_net_sw}
\end{align}
\textcolor{black}{This calculation treats the estimated operating, exit, entry, and shipbuilding payoff costs dollar for dollar as real resource costs. This is a maintained accounting assumption: because these parameters are reduced-form private costs, any components reflecting transfers, financing premia, or other private wedges would not be social resource costs. I therefore report $CS+PS$ separately and interpret net social welfare as conditional on the resource-cost assumption and the logit-scale normalization.}
For any reported set of years $\mathcal T$, the present value of measure $Z_{mt}$ is
\begin{align}
PV_{\mathcal T}(Z_m)
=\sum_{t\in\mathcal T}\beta^{t-1973}Z_{mt}.\label{eq:welfare_pv}
\end{align}
The counterfactual tables and figures use $\mathcal T=\{1973,\ldots,1983\}$; Table \ref{tb:welfare_benchmark} also reports the two conference subperiods separately.

\subsection{Capacity, Revenue, Total Cost, and Profit}\label{sec:qualitative_analysis_static_dynamic_link}
This subsection gives comparative-static intuition for how a change in firm capacity affects static profit. I hold rivals' capacities fixed and treat own capacity as continuous, although capacity choices are discrete in the dynamic model.

In equation \eqref{eq:market_level_profit}, firm $i$ on route $r$ in year $t$ has static profit
\begin{align}
        &\pi_{irt}(s_{irt},s_{rt},D_{rt})=\underbrace{P_{rt}^{*}(s_{rt},D_{rt})q_{irt}}_{=\text{Revenue}}-\underbrace{\int_{0}^{q_{irt}}mc_{irt}(q,s_{irt})dq}_{=\text{Total cost}},\nonumber\\
        &\text{s.t.}\quad  q_{irt}=\begin{cases}
        Q_{rt}^{*}(s_{rt},D_{rt})\omega_{irt} \quad\quad  \text{ if } t \le 1983\\
        \frac{P_{rt}^{*}(s_{rt},D_{rt}) - c_{rt}}{\gamma_1}s_{irt}\quad  \text{ otherwise },
        \end{cases}\quad\omega_{irt}\in[0,1],\forall i, \sum_{i=1}^{N_{rt}} \omega_{irt}=1.\nonumber
\end{align}

\begin{figure}[!htbp]
  \begin{center}
  \subfloat[Transpacific $\pi_{imt}$]{\includegraphics[width = 0.33\textwidth]
  {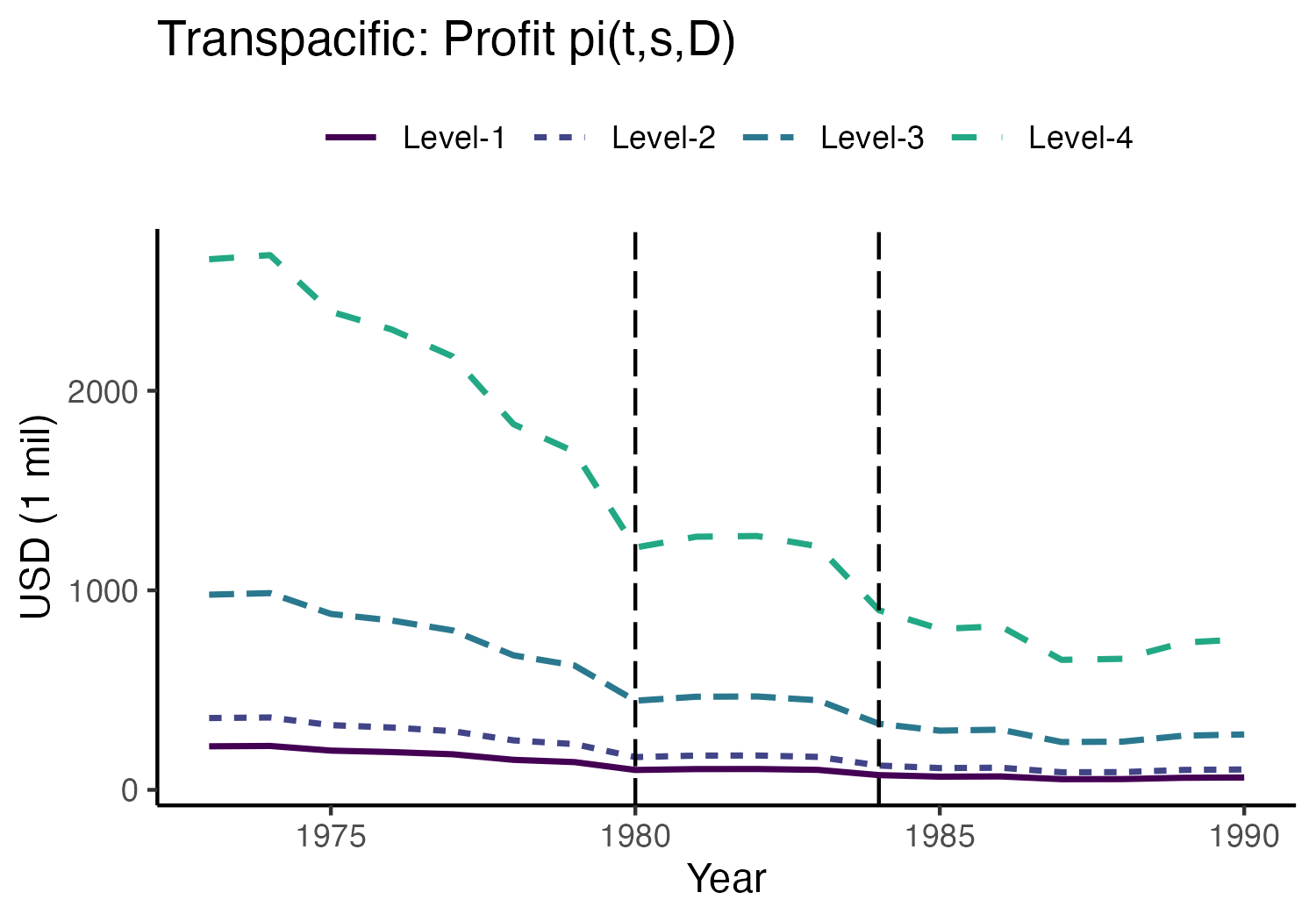}}
  \subfloat[Transatlantic $\pi_{imt}$]{\includegraphics[width = 0.33\textwidth]
  {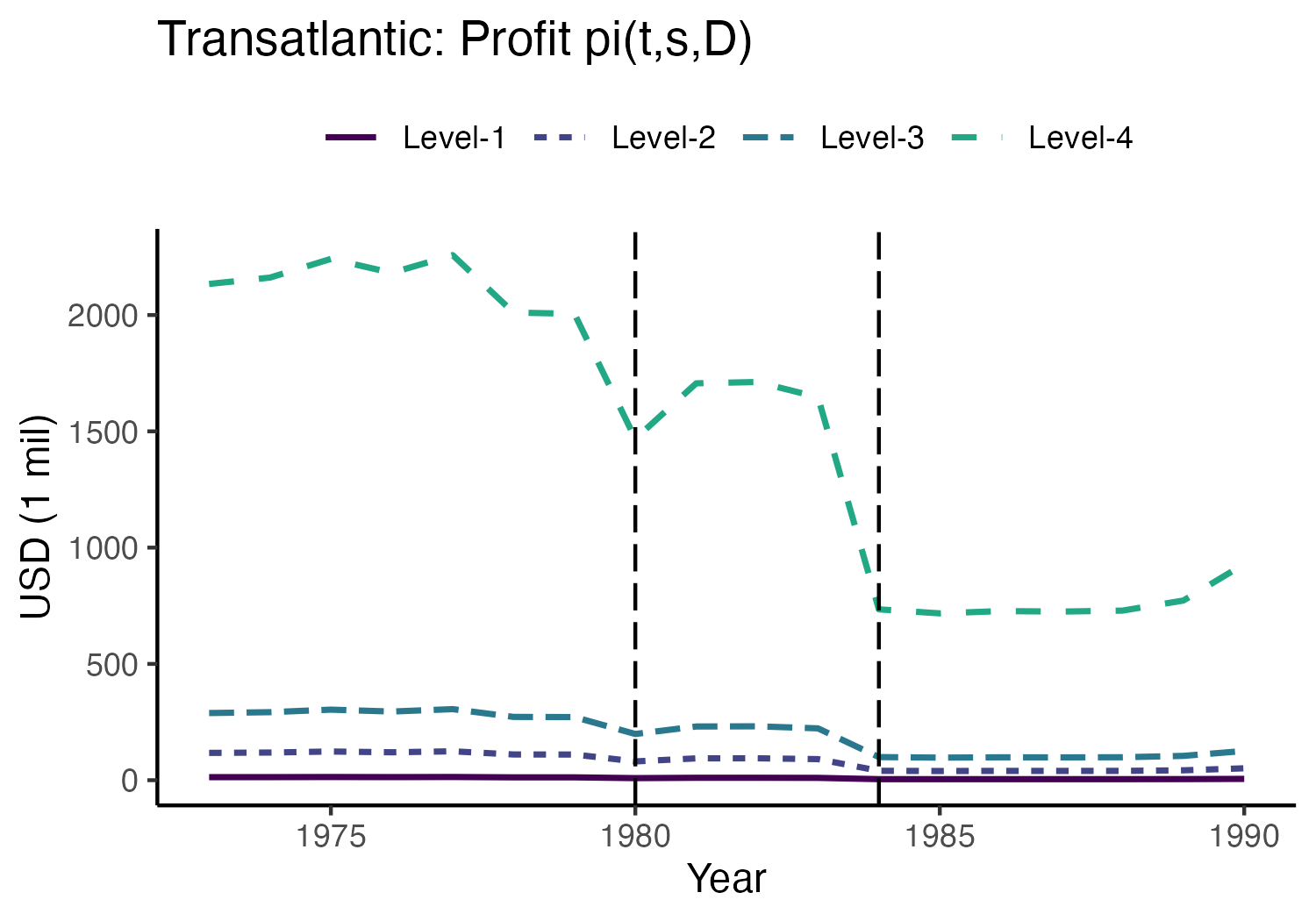}}
  \subfloat[Asia--Europe $\pi_{imt}$]{\includegraphics[width = 0.33\textwidth]
  {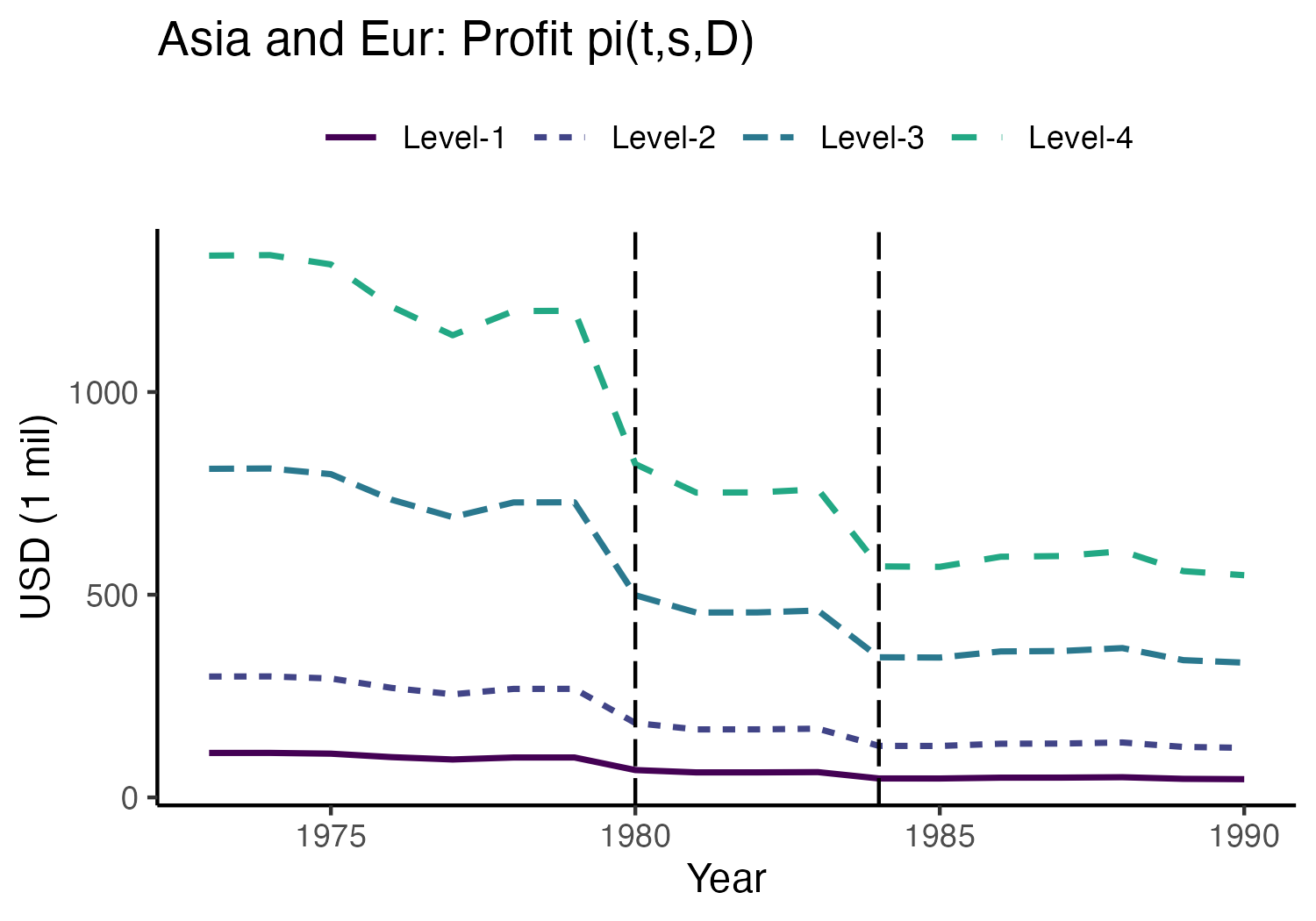}}\\
  \caption{Estimated static profits by capacity level and market}
  \label{fg:market_level_estimated_and_actual_profit_transpacific}
  \end{center}
  \footnotesize
  Note: Each series reports estimated static profit at a discrete capacity level, conditional on the observed market states. Level 1 is the smallest capacity state and level 4 is the largest.
\end{figure}

For compactness, write $P=P_{rt}^{*}(s_{rt},D_{rt})$, $Q=Q_{rt}^{*}(s_{rt},D_{rt})$, $q_i=q_{irt}$, and $s_i=s_{irt}$. The revenue derivative is
\begin{align*}
    \frac{\partial(Pq_i)}{\partial s_i}
    =q_i\frac{\partial P}{\partial s_i}
    +P\frac{\partial q_i}{\partial s_i}.
\end{align*}
The quantity response depends on the regime:
\begin{align*}
\frac{\partial q_i}{\partial s_i}
&=\omega_{irt}\frac{\partial Q}{\partial s_i}
  +Q\frac{\partial\omega_{irt}}{\partial s_i}, && t\leq 1983,\\
\frac{\partial q_i}{\partial s_i}
&=\frac{s_i}{\gamma_1}\frac{\partial P}{\partial s_i}
  +\frac{P-c_{rt}}{\gamma_1}, && t\geq 1984.
\end{align*}
Total cost is $TC_i=c_{rt}q_i+\gamma_1q_i^2/(2s_i)$. Because equilibrium quantity also changes with capacity, its derivative is
\begin{align*}
\frac{\partial TC_i}{\partial s_i}
&=\left(c_{rt}+\gamma_1\frac{q_i}{s_i}\right)
  \frac{\partial q_i}{\partial s_i}
  -\gamma_1\frac{q_i^2}{2s_i^2}.
\end{align*}
Hence,
\begin{align*}
\frac{\partial\pi_{irt}}{\partial s_i}
&=q_i\frac{\partial P}{\partial s_i}
 +\left[P-mc_{irt}(q_i,s_i)\right]\frac{\partial q_i}{\partial s_i}
 +\gamma_1\frac{q_i^2}{2s_i^2}.
\end{align*}
The sign is not determined analytically because price, total quantity, and the quota all respond to capacity. I therefore evaluate these responses at the numerical equilibrium. Figure \ref{fg:market_level_estimated_and_actual_profit_transpacific} reports equilibrium profits conditional on the observed states; static profit increases with capacity in all three markets.

\section{Additional Estimation Results}\label{sec:data_fitting_and_additional_results}

\subsection{Route-Specific Conference Freight-Rate Trends}\label{subsec:route_specific_freight_rate_trends}

\begin{figure}[!htbp]
\begin{center}
  \includegraphics[width = \textwidth]{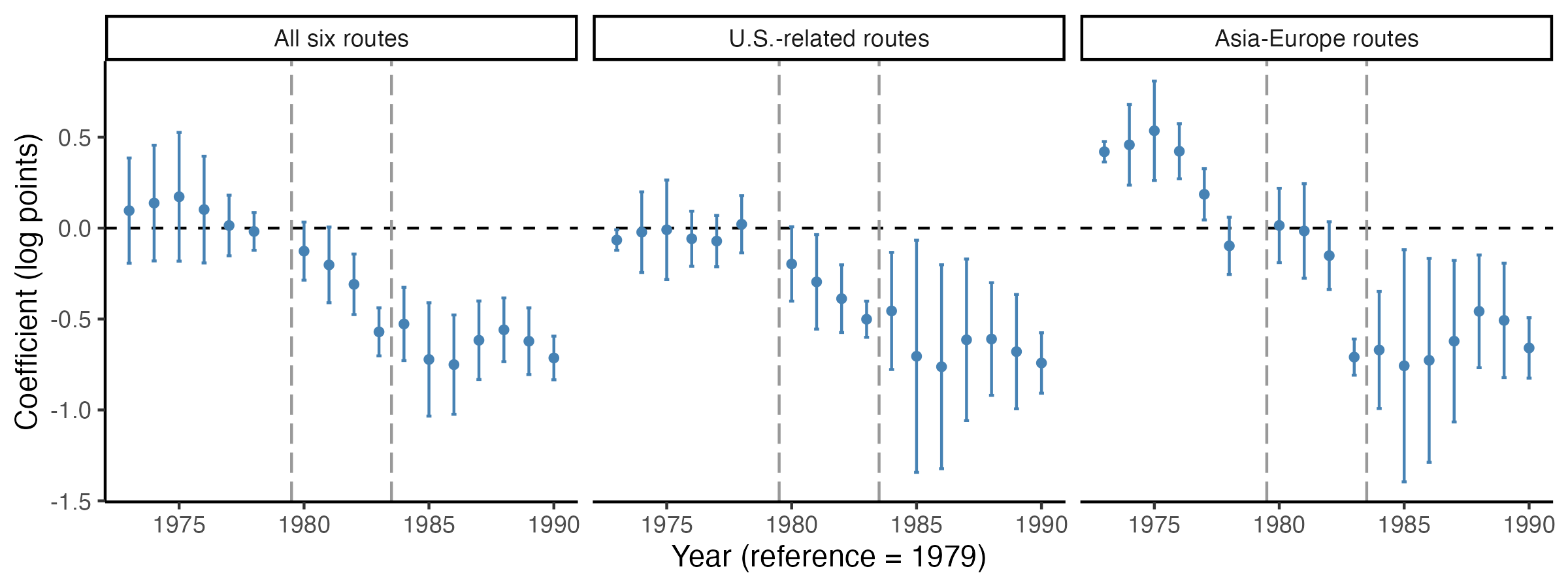}
  \caption{Log conference freight rates relative to 1979}
  \label{fg:log_freight_rate_year_coefficient_path}
\end{center}
\footnotesize
\textcolor{black}{Note: Each panel reports year coefficients from a regression of log conference freight rates on year and route fixed effects, normalized to 1979. The sample covers 1973--1990; dashed lines mark the common empirical regime boundary in 1980 and the Shipping Act of 1984. In the first two panels, error bars are 95\% confidence intervals based on route-clustered standard errors. Because the eastbound and westbound Asia--Europe rates are obtained in part by applying proportional conversion factors to the same raw-data series, they provide effectively one independent price series. The third panel therefore shows an illustrative reference band obtained by applying the year-specific standard errors from the four U.S.-related routes to the Asia--Europe coefficients; it is not an estimated confidence interval for those routes.}
\end{figure}

This subsection documents the timing of the price decline around the 1980 and 1984 empirical regime boundaries.
The exercise is descriptive: it does not by itself separate changes in conference conduct from changes in costs, demand, or capacity.
Figure \ref{fg:log_freight_rate_year_coefficient_path} plots year coefficients from regressions of log conference freight rates on year and route fixed effects, normalized to 1979.
On the U.S.-related routes, rates were close to their 1979 level throughout the 1970s, fell from 1980 through 1983, and remained lower after 1984.
Asia--Europe rates had already declined before 1980 and then dropped sharply in 1983; they also remained well below their 1979 level after 1984.

\subsection{\textcolor{black}{Fit of the Main Models}}

\begin{figure}[!htbp]
  \begin{center}
  \subfloat[$P_{rt}$: Estimated (solid) and data (dotted)]{\includegraphics[width = 0.45\textwidth]
  {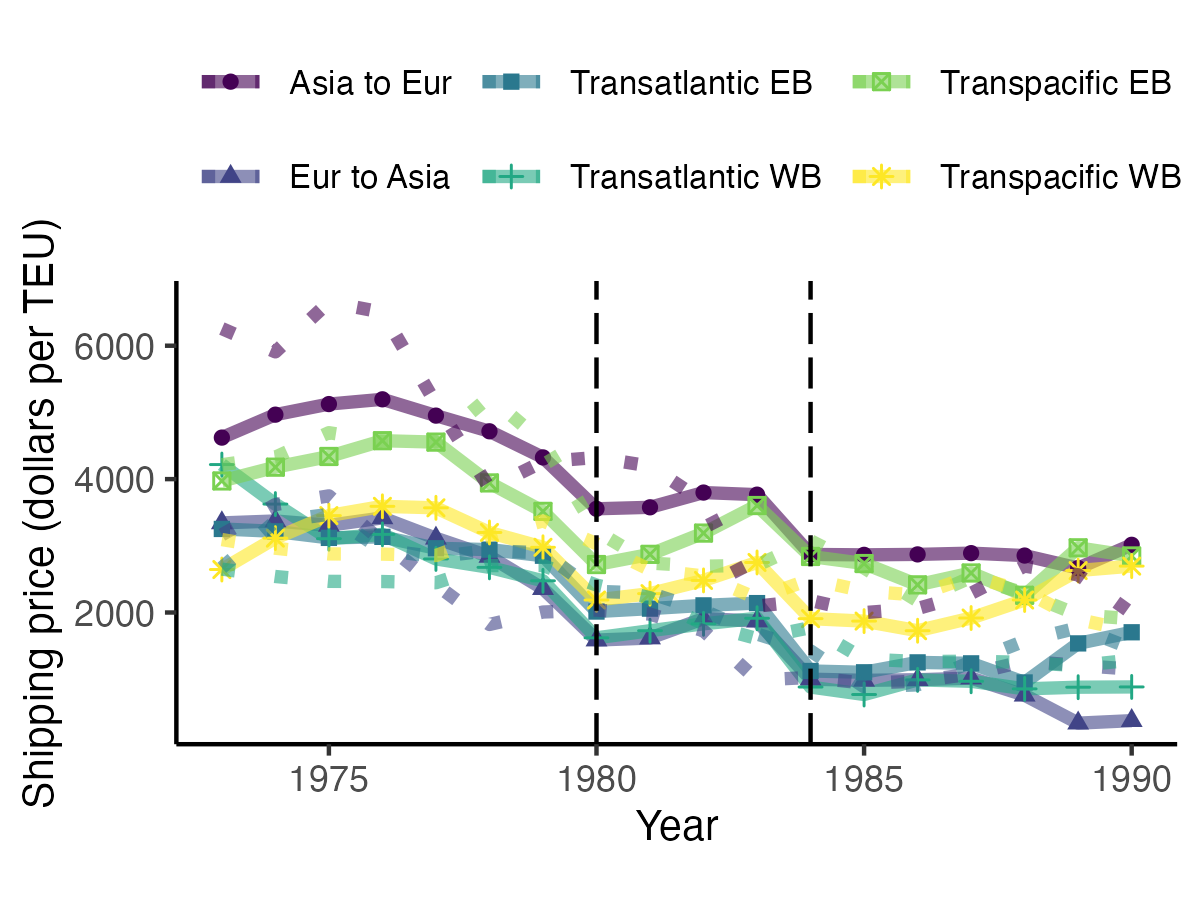}}
  \subfloat[$Q_{rt}$: Estimated (solid) and data (dotted)]{\includegraphics[width = 0.45\textwidth]
  {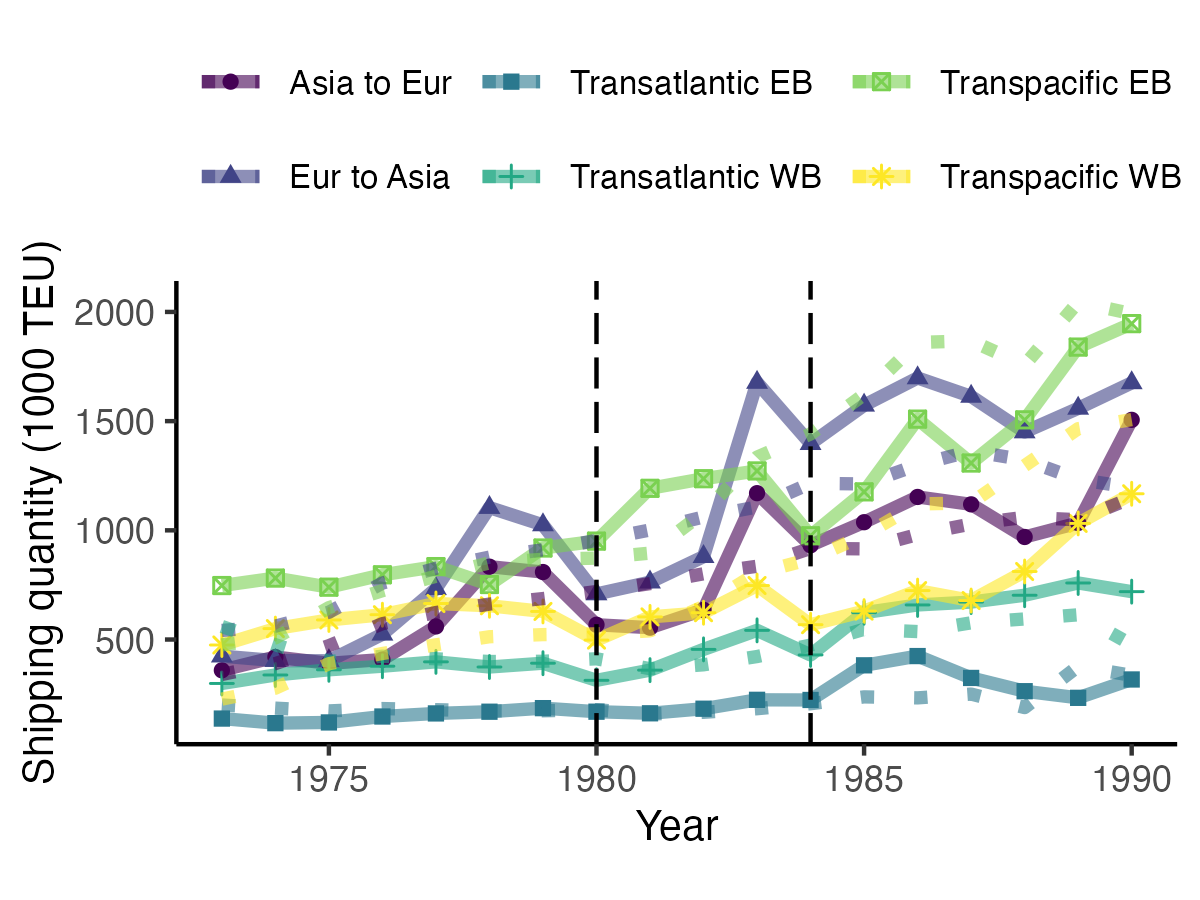}}
  \caption{Fitted and observed route-year prices and quantities}
  \label{fg:estimated_and_actual_quantity_comparison}
  \end{center}
\end{figure}

Panels (a) and (b) of Figure \ref{fg:estimated_and_actual_quantity_comparison} compare fitted route-year prices and quantities with the data. The fitted series reproduce the broad regime shifts and trends but are smoother than the data.
\textcolor{black}{Table \ref{tb:datafit_for_supply_side} evaluates the supply-side fit by market and regime. The ratio of the fitted mean price to the observed mean price ranges from 0.812 to 1.115.}

\begin{table}[!htbp]
  \begin{center}
      \caption{Fit of shipping prices by regime}
      
\begin{tabular}[t]{llrrr}
\toprule
Regime & Market & (1) Mean $\hat{P}_{rt}$ & (2) Mean $P_{rt}$ & (1)/(2)\\
\midrule
1973-1979 & Asia and Eur & 3975.553 & 4172.739 & 0.953\\
 & Transatlantic & 3111.218 & 2791.502 & 1.115\\
 & Transpacific & 3687.998 & 3810.528 & 0.968\\
1980-1983 & Asia and Eur & 2710.169 & 2625.379 & 1.032\\
 & Transatlantic & 1926.517 & 2047.711 & 0.941\\
 & Transpacific & 2760.922 & 2724.519 & 1.013\\
1984-1990 & Asia and Eur & 1821.684 & 1672.949 & 1.089\\
 & Transatlantic & 1081.437 & 1331.900 & 0.812\\
 & Transpacific & 2399.493 & 2297.765 & 1.044\\
\bottomrule
\end{tabular}
      \label{tb:datafit_for_supply_side}
  \end{center}
  \footnotesize
  \textcolor{black}{Note: Each row pools the two directed-route observations within a market and regime. Fitted route-year prices are calculated from the estimates in Table \ref{tb:cost_estimate_results_after_1973_before_1990}.}
\end{table}

\subsection{Specification Sensitivity and Instrument Selection for Demand and Cost}\label{sec:iv_specification_search}

This subsection documents sensitivity to the demand instruments and cost shifters used in Section \ref{sec:estimation_results}.
I explored a broad grid of demand and cost specifications and then compared economically admissible specifications by their alignment with the model, first-stage relevance, and overidentification diagnostics where available.
Model admissibility requires $\alpha_1<-1$ for finite consumer surplus under constant-elasticity demand and $\gamma_1>0$ for upward-sloping supply. Because the reported specification was selected from the same 108 observations using these criteria, the coefficient $p$-values and confidence sets are conditional on specification selection and are not selection-adjusted. The comparisons below should therefore be read as sensitivity analysis rather than independent confirmation of the preferred estimates.
Tables \ref{tb:demand_iv_specification_comparison} and \ref{tb:cost_iv_specification_comparison} report the main candidates for the common 1973--1990 sample.

On the demand side, column (1) instruments $\log(P_{rt})$ with average ship age and the contemporaneous share of tonnage over 20 years old; it yields $\hat{\alpha}_1=-0.842$.
The preferred specification in column (2) uses average ship age and the share of tonnage over 15 years old. It yields $\hat{\alpha}_1=-1.111$ and a route-clustered first-stage F of 19.0. Both instruments are contemporaneous functions of the fleet state, which is determined by past entry and shipbuilding decisions and is therefore predetermined with respect to the current i.i.d. demand shock; the lagged variant in column (3) gives a nearly identical $\hat{\alpha}_1=-1.095$.
Columns (4) and (5) show that each instrument alone is either too weak or yields an imprecise inelastic estimate.
The instruments derive most of their power from the low-frequency aging of the fleet; the elasticity estimated from the detrended, within-route variation alone is $-0.93$, close to the preferred estimate, although the detrended instruments are weak.
No examined specification delivers $\alpha_1$ significantly below $-1$: the preferred estimate is statistically different from zero but not from unit elasticity.
In leave-one-route-out estimates, $\hat{\alpha}_1$ ranges from $-1.211$ to $-0.482$; omitting the Europe-to-Asia route produces the upper endpoint. Thus, the sign is stable but the classification as elastic is sensitive to individual routes.

On the cost side, the preferred specification in column (1) includes average ship size and the share of tonnage over 20 years old \textcolor{black}{as fleet-technology cost shifters} and yields $\hat{\gamma}_1=597$.
Its route-clustered first-stage F is 8.5, below the conventional threshold of ten, although its Anderson--Rubin 95\% confidence set is bounded and positive.
Columns (2)--(5) vary the age threshold and, in columns (4) and (5), omit average ship size; the supply slope is positive in columns (1)--(4) but negative in column (5), and the first-stage F ranges from 2.5 to 8.5.
The \textcolor{black}{1973--1979 wedge} is positive across the reported specifications, whereas the 1980--1983 wedge is more sensitive to the treatment of fleet variables.
Across leave-one-route-out estimates, $\hat{\gamma}_1$ ranges from 337 to 1,115, the pre-1980 wedge from \$1,385 to \$1,975, and the 1980--1983 wedge from \$596 to \$1,220.

\begin{table}[!htbp]
  \begin{center}
      \caption{Demand specifications and instrument choice}
      \resizebox{\textwidth}{!}{\begin{tabular}{llllll}
\hline
& (1) & (2) & (3) & (4) & (5) \\ \hline
$\log(P_{rt})$: $\alpha_1$ & -0.842*** & -1.111*** & -1.095*** & -0.691 & -4.893 \\
& (0.274) & (0.354) & (0.339) & (0.476) & (23.917) \\
log GDP & 0.419** & 0.410* & 0.411* & 0.425** & 0.279 \\
& (0.197) & (0.217) & (0.216) & (0.190) & (0.634) \\
1(t $\le$ 1979) & 0.369** & 0.552* & 0.541** & 0.266 & 3.129 \\
& (0.158) & (0.304) & (0.268) & (0.287) & (16.297) \\
1(1980 $\le$ t $\le$ 1983) & 0.080** & 0.168 & 0.162* & 0.030 & 1.405 \\
& (0.039) & (0.111) & (0.090) & (0.153) & (7.812) \\
Num.Obs. & 108 & 108 & 108 & 108 & 108 \\
R2 Adj. & 0.850 & 0.804 & 0.807 & 0.869 & -1.173 \\
IV: mean ship age & X & X & X & X &  \\
IV: share of tonnage $>$20yo & X &  &  &  &  \\
IV: share of tonnage $>$15yo &  & X &  &  &  \\
IV: lag share of tonnage $>$15yo &  &  & X &  & X \\
First-stage F & 36.6 & 19.0 & 25.0 & 8.3 & 0.1 \\
\hline
\end{tabular}
}
      \label{tb:demand_iv_specification_comparison}
  \end{center}
  \footnotesize
  Note: The sample covers six routes in 1973--1990. All columns include the demand shifters of Table \ref{tb:demand_estimate_results_after_1973_before_1990} and vary only the excluded instruments; standard errors in parentheses are clustered at the route level. The rows ``IV'' indicate the excluded instruments for $\log(P_{rt})$. The first-stage F statistics are route-clustered Wald statistics. Column (2) is the preferred specification used in Table \ref{tb:demand_estimate_results_after_1973_before_1990}. Significance levels are denoted by $^{*}p<0.1$, $^{**}p<0.05$, and $^{***}p<0.01$.
\end{table}

\begin{table}[!htbp]
  \begin{center}
      \caption{Cost specifications and instrument choice}
      \resizebox{\textwidth}{!}{\begin{tabular}{llllll}
\hline
& (1) & (2) & (3) & (4) & (5) \\ \hline
$Q_{rt}/S_{rt}$: $\gamma_1$ & 596.922* & 103.019 & 901.129* & 433.584** & -234.438 \\
& (327.028) & (129.756) & (514.259) & (195.173) & (639.550) \\
share of tonnage $>$10yo &  & -1976.313** &  &  & -1697.562** \\
&  & (913.555) &  &  & (778.658) \\
share of tonnage $>$20yo & -1491.488 &  &  & 1424.614 &  \\
& (2927.073) &  &  & (3464.868) &  \\
share of tonnage $>$15yo &  &  & 990.437 &  &  \\
&  &  & (1481.399) &  &  \\
mean ship size (1,000 TEU) & 1817.274 & 2226.129 & 1209.211 &  &  \\
& (1827.054) & (2016.352) & (1465.583) &  &  \\
1(t $\ge$ 1974) & 113.941 & 3.965 & 202.355 & 48.200 & -130.107 \\
& (320.233) & (160.761) & (446.873) & (296.061) & (266.190) \\
1(t $\le$ 1979): $\tilde{\gamma}_1$ & 1807.343*** & 1229.007*** & 1891.642*** & 1743.573*** & 1187.760*** \\
& (411.167) & (309.217) & (475.685) & (386.512) & (333.911) \\
1(1980 $\le$ t $\le$ 1983): $\tilde{\gamma}_2$ & 997.543*** & 588.864*** & 1192.165** & 823.640*** & 315.925 \\
& (368.733) & (210.387) & (512.491) & (244.032) & (245.232) \\
Num.Obs. & 108 & 108 & 108 & 108 & 108 \\
R2 Adj. & 0.714 & 0.800 & 0.630 & 0.732 & 0.769 \\
First-stage F & 8.5 & 7.5 & 4.3 & 3.9 & 2.5 \\
\hline
\end{tabular}
}
      \label{tb:cost_iv_specification_comparison}
  \end{center}
  \footnotesize
  Note: The sample covers six routes in 1973--1990. All columns include the regime dummies and route fixed effects and instrument $Q_{rt}/S_{rt}$ with log GDP; standard errors in parentheses are clustered at the route level. The columns vary the cost shifters as indicated by the coefficient rows. Column (1) is the preferred specification used in Table \ref{tb:cost_estimate_results_after_1973_before_1990}. Significance levels are denoted by $^{*}p<0.1$, $^{**}p<0.05$, and $^{***}p<0.01$.
\end{table}

\subsection{Robustness of the Dynamic Estimates to Anticipating the Cartel Breakdown}\label{sec:pf_robustness}

The benchmark model assumes that firms do not anticipate the 1984 cartel breakdown, so the terminal ex-ante value $\mathcal V_{imT}=\pi_{imT}/(1-\beta)$ uses the 1984 profits evaluated under the belief that the 1980--1983 conference regime continues.
As a robustness check, I re-estimate the dynamic parameters under the alternative assumption that firms perfectly foresaw the breakdown: the terminal value uses realized competitive profits in 1984, so backward induction incorporates the end of the cartel into every pre-1984 decision.
Table \ref{tb:dynamic_estimate_perfect_foresight} reports the results.
\textcolor{black}{Entry costs fall by 0.9--3.2\% across markets. Exit and operating costs remain similar in magnitude, with the largest proportional movement being a 27\% rise in the imprecisely estimated transpacific exit cost. The identified investment costs fall by 2.1--6.0\%, and every perfect-foresight cost estimate lies within its benchmark likelihood-slice range.}
\textcolor{black}{The stability of the estimates is an empirical robustness result rather than an exact logit-invariance result: the terminal-value change enters continuation, build, and entry values but not exit or stay-out values, so it can affect the corresponding odds, and under the capacity-proportional allocation the change need not be uniform across capacity levels.} \textcolor{black}{With the moderate continuing 1980--1983 wedge ($\tilde{\gamma}_2 = \$998$ per TEU), the re-estimated costs remain within the benchmark likelihood-slice ranges.}
The no-anticipation assumption is therefore maintained because the timing and effectiveness of the breakdown were uncertain before 1984; \textcolor{black}{the robustness exercise suggests that this terminal-value assumption is not driving the estimated dynamic costs}.

\begin{table}[!htbp]
  \begin{center}
      \caption{Dynamic parameters under perfect foresight of the 1984 breakdown}
      \begin{tabular}[t]{lccc}
\toprule
Parameter & Transpacific & Transatlantic & Asia--Europe\\
\midrule
Entry cost: $\kappa^{e}$ & 4.018 & 2.91 & 3.804\\
 & {}[3.575,4.515] & {}[2.382,3.554] & {}[3.385,4.275]\\
Exit cost: $\psi$ & 0.85 & 3.419 & 2.843\\
 & {}[0.07,1.29] & {}[2.799,4.176] & {}[1.56,4.241]\\
Operation cost: $\phi$ & 0.224 & 0.505 & 0.052\\
 & {}[0.097,0.339] & {}[0.413,0.617] & {}[0.004,0.181]\\
Investment cost: $\iota_1$ & 3.211 & 3.572 & 3.709\\
 & {}[2.696,3.825] & {}[2.925,4.363] & {}[3.3,4.168]\\
Investment cost: $\iota_2$ & Not identified & 9.185 & 3.913\\
 &  & {}[7.71,10.942] & {}[2.623,5.837]\\
Logit scale: $\sigma$ (normalized) & 1.000 & 1.000 & 1.000\\
 &  &  & \\
Log Likelihood & -67.652 & -53.796 & -67.571\\
\bottomrule
\end{tabular}

      \label{tb:dynamic_estimate_perfect_foresight}
  \end{center}
  \footnotesize
  Note: Each market aggregates its eastbound and westbound routes. The terminal value uses realized competitive profits in 1984 instead of the regime-continuation beliefs of the benchmark model. Brackets report conditional likelihood-slice ranges based on the same 90\% chi-squared cutoff as in Table \ref{tb:dynamic_estimate}. As in Table \ref{tb:dynamic_estimate}, no level-3-to-4 investment is observed during the 1973--1983 transpacific estimation sample, so $\iota_2$ is not finitely identified there. Cost parameters are measured in billions of U.S. dollars, and the logit scale $\sigma$ is normalized to one.
\end{table}

\renewcommand{\refname}{References for the Appendix}
\putbib[ship_cartel]
\end{bibunit}

\end{document}